\documentclass[useAMS,usenatbib]{mn2e}
\usepackage{epsfig}

\title[Gas stripping and mixing in galaxy clusters]
  {Gas stripping and mixing in galaxy clusters: A numerical comparison study}
\author[S.~He{\ss} \& V.~Springel]
  {Steffen~He{\ss}$^{1,2}$ and
  Volker~Springel$^{3,4}$ \vspace{0.2cm}\\ 
  $^1$Max-Planck-Institut
f\"{u}r Astrophysik, Karl-Schwarzschild-Stra\ss{}e 1, 85740 Garching
bei M\"{u}nchen, Germany\\
  $^2$Leibniz Institute for Astrophysics, An der Sternwarte 16,
14482 Potsdam, Germany\\
$^3$Heidelberg Institute for Theoretical Studies,
Schloss-Wolfsbrunnenweg 35, 69118 Heidelberg, Germany\\
$^4$Zentrum f\"ur Astronomie der Universit\"at Heidelberg,
Astronomisches
Recheninstitut,
    M\"{o}nchhofstr. 12-14, 69120 Heidelberg, Germany}

\renewcommand{\vec}[1]{ {\bmath #1} }

\begin{document}
\label{firstpage}
\maketitle

\begin{abstract}
  The ambient hot intra-halo gas in clusters of galaxies is constantly
  fed and stirred by in-falling galaxies, a process that can be
  studied in detail with cosmological hydrodynamical
  simulations. However, different numerical methods yield discrepant
  predictions for crucial hydrodynamical processes,
  leading for example to different entropy profiles in clusters of
  galaxies. In particular, the widely used Lagrangian smoothed
  particle hydrodynamics (SPH) scheme is suspected to strongly damp
  fluid instabilities and turbulence, which are both crucial to
  establish the thermodynamic structure of clusters. In this study, we
  test to which extent our recently developed Voronoi particle
  hydrodynamics (VPH) scheme yields different results for the
  stripping of gas out of infalling galaxies, and for the bulk gas
  properties of cluster. We consider both the evolution of isolated
  galaxy models that are exposed to a stream of intra cluster medium
  or are dropped into
  cluster models, as well as non-radiative cosmological simulations of
  cluster formation. We also compare our particle-based method with
  results obtained with a fundamentally different discretisation
  approach as implemented in the moving-mesh code {\small AREPO}. We
  find that VPH leads to noticeably faster stripping of gas out of
  galaxies than SPH, in better agreement with the mesh-code than with
  SPH. We show that despite the fact that VPH in its present form
is not as accurate as the moving mesh code in our investigated cases,
its improved accuracy of gradient estimates makes VPH an attractive
alternative to SPH.

\end{abstract}


\section{Introduction}

In the hierarchical structure formation model \citep{White1991},
interactions or mergers of galaxies with other galaxies are common
processes, and depending on the environment, a diverse set of physical
processes can become important.  For example, when a galaxy falls into
a cluster of galaxies, gravity may affect its integrity through
gravitational tidal forces. The closer the galaxy gets to the cluster
the more it is exposed to the hot intra-cluster medium (ICM). This is
experienced as a headwind by the galaxy that compresses and removes
the gas of the galaxy \citep{GunnGott}. This may lead to the formation
of a bow shock in front of the galaxy if it supersonically ploughs
through the ICM.

Whereas the gas at the front of the galaxy is primarily compressed by
ram pressure in this situation, the sides are exposed to strong shear
flows which may trigger fluid instabilities, leading to stripping and
mixing of the gas with the ICM. This in turn crucially determines how
rapidly star formation is turned off in an infalling galaxy, and how
metals are mixed into the group or cluster ICM \citep[see
e.g.][]{Larson1980, Balogh2000, Domainko2006}. The hydrodynamical
interaction processes between galaxies and groups/clusters may thus
influence important observational effects, such as the density
morphology relation \citep{Dressler1980, vanderWel2010, Boselli2006},
or the existence of a tight red cluster galaxy sequence
\citep{Baldry2004, Cortese2009}.  Another interesting aspect besides
the stripping is the generation of turbulence by galaxies infalling
into clusters.  Numerical simulations \citep{Dolag2005, Vazza2006,
  Iapichino2008, Dolag2009} indicate that turbulence can be generated
in the wake of an infalling galaxy.  This interesting phenomenon may
play an important role in determining the dynamics of the cluster gas
and could in principle be detected through X-ray observations of the
broadening of sharp metal lines \citep{Sunyaev2003}.

Modelling these processes reliably goes back to
\citet{GunnGott} and is a considerable challenge, as direct
hydrodynamic simulations of the relevant processes are very difficult
to model due to the large dynamic range, and the uncertainties
associated with the modelling of star formation and feedback
processes.  Furthermore, it is not clear whether the hydrodynamical
techniques presently in use are strongly affected by systematic errors
in this regime \citep [e.g.][]{Agertz}.  Indeed, previous studies of
galaxy-cluster interactions with different numerical schemes disagreed
strongly about how fast a galaxy loses its gas to the cluster. For
example, a grid-based simulation of \citet{Quilis2000} predicted a
much higher stripping rate than SPH-based work by \citet{Abadi1999}.

In this paper, we concentrate on the numerical aspects of the
stripping of a galaxy's gas as it interacts with the ICM.
\citet{Agertz} has recently shown that the popular smoothed particle
hydrodynamics (SPH) technique has problems to properly account for
fluid instabilities, prompting significant concerns about a possible
unphysical suppression of stripping processes \citep[see
also][]{Price2008, Wadsley2008,Read2010,Cha2010,Valcke2010,Junk2010,Abel2011}.
In a recent study \citep{Hess2010} we have therefore proposed a new
`Voronoi particle hydrodynamics' (VPH) method that improves on the
widely used SPH technique \citep{Lucy1977, Gingold_Monaghan,
  Larson1978} in several respects. By employing a Voronoi tessellation
for the local density estimate, a consistent decomposition of the
simulation volume is achieved, contact discontinuities can be resolved
much more sharply, and a `surface tension' effect across them is
avoided. Our preliminary tests of VPH based on the `blob-test' of
\citet{Agertz} already suggested that stripping is more efficient in
VPH compared with SPH. Here we shall investigate this in more detail
using more realistic set-ups that mimic galaxy evolution processes. In
addition to the two particle-based Lagrangian schemes for
hydrodynamics, SPH and VPH, we will also carry out comparison
simulations with the moving-mesh code {\small AREPO}
\citep{Springel2010}. Whereas {\small AREPO} uses a Voronoi
tessellation as well, this code employs an entirely different
methodology for fluid dynamics, based on a finite volume Godunov
scheme that calculates hydrodynamical fluxes with a Riemann solver
across mesh boundaries. The comparison of this diverse set of three
numerical methods is useful to understand and quantify the systematic
uncertainties of the different methods.
 
In our tests simulations, we ideally want to investigate as realistic
conditions as possible, including a full treatment of gravity and dark
matter, as well as star formation and cooling. This in principle calls
for simulations in which a well-resolved galaxy model is dropped into
an equally well represented cluster of galaxies. Because the
substantial computational cost of such a set-up severely limits the
resolution that can be achieved, we base part of our study on a number
of more idealised simulations, for example by placing galaxy models
into a `wind tunnel' that mimics the impinging flow of gas onto a
galaxy in orbit in a cluster. Finally, we also carry out cosmological
simulations of the formation of the `Santa Barbara cluster'
\citep{Frenk}, primarily to study how well VPH performs in
non-radiative cosmological simulations of cluster formation compared
to the other techniques.  Already in past studies, the Santa Barbara
cluster comparison project has led to important insights into how
numerical effects impact the thermodynamic structure of simulated
clusters, and for this test problem, there is already a considerable
body of results in the literature \citep[e.g.][]{Wadsley2004,gadget2,Thacker2006}.
  
This paper is structured as follows. In Section~\ref{SecBasics}, we
introduce the numerical methods we will use in our numerical
comparison study. In Section~\ref{IsolatedGal}, we check how well the
different numerical schemes represent the evolution of an isolated
galaxy, and whether there are already significant differences at this
level. We then consider in Section~\ref{WindTunnel} wind-tunnel
experiments in which we expose the isolated galaxy models to a
supersonic wind, allowing a detailed examination of the stripping
process. In Section~\ref{Infall}, we follow up on these experiments by
studying the behaviour of a galaxy model that falls into an isolated
spherical cluster, comparing the VPH results with those obtained with
SPH and {\small AREPO}.  Finally, in Section~\ref{Cosmo}, we
investigate the performance of VPH in cosmological simulations of
cluster formation, using re-simulations of rich clusters identified in
the Millennium Simulation as well as the Santa Barbara cluster initial
conditions. We give a discussion of our results in the context of a
vortex test problem in Section~\ref{SecConclusions}, combined with a
summary of our conclusions.

\section{Methodology} \label{SecBasics}

We begin by introducing the different codes and computational methods
we use in this study. In the interests of brevity, we shall only describe the
most important characteristics of the codes; full details can be found
elsewhere \citep{gadget2, Hess2010, Springel2010}.

\subsection{Smoothed particle hydrodynamics}

Smoothed particle hydrodynamics (SPH) is a particle based approach to
fluid dynamics that has seen widespread use in astronomy since it was
first introduced more than three decades ago \citep{Lucy1977,Gingold_Monaghan}.
SPH discretises the mass of the fluid in terms of particles, whose
dynamics is governed by a fluid Lagrangian of the form
\begin{equation}
	L = \sum_i \left[ \: \frac{1}{2}\: m_i \: \vec{v}_i^2 - m_i \:
          u_i(\rho_i, s_i) \: \right] .
\label{lagr}
\end{equation}
Here $u_i$ describes the thermal energy per unit mass, which for an
ideal gas depends only on density $\rho_i$ and specific entropy
$s_i$. An adaptive, spherically symmetric smoothing kernel is employed
to estimate the density based on the spatial distribution of an
approximately fixed number of nearest neighbours. (As a standard value
within this work we set the number of neighbours to $N=48$.) The
Lagrangian in Eq.~(\ref{lagr}) then uniquely determines the equations
of motion in a form that simultaneously conserves energy and entropy
\citep{springel_hernquist_02,Price2007,Rosswog2009,Springel2010b}.
However, an artificial viscosity needs to be added to allow for
dissipative shock capturing and to produce a damping of post-shock
oscillations. If not explicitly denoted otherwise we use the viscosity
parametrisation of the {\small GADGET} code with a viscosity strength
of $\alpha=1$, and we apply the viscosity limiter in the presence of
shear introduced by \citet{Balsara1995}.

Among the primary advantages of SPH are its very good conservation
properties, the manifest Galilean invariance at the discretised level
of the equations, and the absence of preferred spatial directions.
However, it has recently become clear that contact discontinuities, in
particular, are challenging for the method. For example, they are
associated with a spurious surface tension effect in SPH
\citep{Springel2010b}, and fluid instabilities such as the
Kelvin-Helmholtz instability are suppressed across them
\citep{Agertz}.

\subsection{Voronoi particle hydrodynamics}

Because of the accuracy problems of SPH in certain situations, we have
recently developed another particle-based hydrodynamic method that
differs from SPH in important ways, while still sharing a number of
similarities. In this `Voronoi Particle Hydrodynamics' (VPH) method
\citep{Hess2010}, the fluid is also discretised in terms of mass
elements $m_i$, and the same fluid Lagrangian as in Eq.~(\ref{lagr})
is used. Also identically to SPH we assume a polytropic equation of
state $P=A\,\rho^\gamma$ with an adiabatic coefficient $\gamma$ in all
our tests. However, the kernel estimation technique of SPH is replaced
by a very different approach to calculate the gas density.  This is
done by constructing a Voronoi tessellation for the point set, in
which to each of the particles the volume of all the space which is
closer to this point than to any other point is assigned. The density
can then be obtained as the particle mass divided by the volume of the
associated Voronoi cell. We note that such a tessellation technique
allows a close to optimum exploitation of the density information
contained in the particle distribution \citep{Pelupessy}, and in
particular, very sharp discontinuities can be resolved over the
distance of a single cell.

With the density computed, and given specified entropies $s_i$ for every
particle, we can use the ideal gas Lagrangian to derive unique
equations of motion for VPH. The key quantity that is needed in doing
this is the derivative of the particle volume with respect to the
coordinates of the surrounding particles, but thanks to the
mathematical properties of the Voronoi tessellation, the corresponding
geometric factors can be computed relatively easily \citep{Serrano,Hess2010}
and have a clear geometric meaning. For example, the primary force
between two VPH particles is their pressure difference times the area
of their common tessellation face. Because the dynamics in VPH is
derived from the Lagrangian given in Eqn.~(\ref{lagr}), energy,
momentum and entropy are conserved exactly, just like in SPH. However,
there is no surface tension effect, and the VPH particle volumina add
up to the total simulated volume exactly, something that is not the
case in SPH in general. One can hence hope that VPH provides an
interesting improvement over SPH. Indeed, in \citet{Hess2010} we have
shown that this is the case for several test problems, such as the
growth of Kelvin-Helmholtz instabilities at contact discontinuities
with a large density jump. It is one of the primary goals of this
paper to see whether these improvements are also relevant in
cosmological structure formation problems.

In VPH, we need to rely on an artificial viscosity to treat shocks,
similarly to SPH.  We here follow the standard SPH approach by
\citet{Gingold_Monaghan} and \cite{Balsara1995} in introducing an extra
friction force that reduces the kinetic energy and transforms it into
heat in regions of rapid compression. Due to its similarities to its
SPH counterpart we usually use a value of $\alpha=1$ for our
artificial viscosity in VPH as well, but we also vary $\alpha$
explicitly to demonstrate its implications. In addition, in VPH we can
also add some weak, non-dissipative ``shape forces'' that are designed
to maintain a reasonably regular mesh geometry, which tends to improve
the overall accuracy of the VPH dynamics. As discussed extensively in
\citet{Hess2010}, there are different possibilities to introduce such
extra forces. We derive them by adding extra terms to the Lagrangian
which penalise highly distorted mesh cells. For the shape correction
scheme we usually choose $\beta_0=0.2$ and $\beta_1=0.01$. However,
the optimum choice for the strength of these terms is unfortunately
not (yet) clear. For this reason, we have typically carried out all
our tests with several different settings for the strength of the
artificial viscosity and the shape forces, finding in most cases that
our results are quite insensitive to the detailed choices over a broad
range of parameter settings.

Both the SPH and VPH simulations we study here have have been carried
out with {\small GADGET-3}, an improved version of the publicly
available cosmological code {\small GADGET-2} \cite[last described
  in][]{gadget2}. The code is parallelized for distributed memory
machines and uses a hierarchical tree algorithm for calculation the
self-gravity of the gas. A collision-less stellar or dark matter
component can be optionally included as well. 

\subsection{Hydrodynamical moving-mesh code}

Finally, the third numerical technique for hydrodynamics we examine is
implemented in the moving-mesh code {\small AREPO}
\citep{Springel2010}. It uses an entirely different approach in which
the {\em volume} is discretised.  The hydrodynamical equations are
solved in terms of a Godunov scheme where at each mesh interface a
Riemann problem is calculated and the resulting fluxes are used for an
exchange of conserved fluid quantities between the cells. For higher
spatial accuracy, this is combined with a piece-wise linear
reconstruction of the gas distribution and a second-order time
integration scheme, yielding overall second-order integration accuracy
in smooth parts of the flow.

The particular mesh used by {\small AREPO} is given by the Voronoi
tessellation of a set of mesh-generating points. The use of this
particular type of mesh allows the method to be formulated such that
the mesh can move along with the gas, continuously adjusting its
topology to the fluid motion. The method exploits the property of a
Voronoi tessellation that the mesh changes continuously when the
generating points are moved, without showing problematic mesh-tangling
effects. If the points are moved with the local flow velocity (which
is the standard way to operate the code), the method then has a
similar Lagrangian character as SPH/VPH. However, unlike in SPH and
VPH, the masses of {\small AREPO}'s resolution elements (the cells)
are not required to remain strictly constant.  In fact, if the
mesh-generating points are being fixed on a static Eulerian grid, then
{\small AREPO} is equivalent to the MUSCL-Hancock scheme, a popular
method for second-order Eulerian gas dynamics on structured
grids. {\small AREPO} should hence be viewed as being closely related
to ordinary Eulerian mesh-based hydrodynamics, except that its ability
to move the mesh along with the flow considerably reduces advection
errors.

\section{Isolated galaxies and their evolution} \label{IsolatedGal}

Before we analyse the interaction of individual galaxies within a
cluster environment, we here study galaxy models in isolation to see
how well the different numerical schemes treat them and whether our
initial conditions are sufficiently stable. This is also intended to
identify possible subtle systematic differences between the numerical
schemes that may be difficult to detect in more complicated set-ups.
We hence use identical initial conditions for the three codes such
that differences introduced by varying start-up procedure are
minimised. We we will see however that even for identical particle
positions, masses, and temperatures, there can be variations in
density (and hence specific entropy since we start with a prescribed
temperature) directly after the start of a simulation, as a result of
the different density estimation methods. We are in fact especially
interested in the question whether these small differences have any
bearing on the further evolution of isolated galaxy models.

To set-up compound galaxies in isolation, we construct dark matter
halo models and stellar disks in dynamical equilibrium through an
approximate solution of the Jeans equations
\citep{Hernquist1993,Springel1999,Springel2005b}. The primary
assumption made is that the local velocity dispersion can be
approximated reasonably well with a triaxial Gaussian. For the models
considered here, a dark matter halo with a Hernquist profile is used,
but alternatively a truncated NFW profile could be used as well, with
no difference to our conclusions. The baryonic matter is represented
with a central stellar bulge, an exponential stellar disk, and an
exponential gaseous disk. In addition, we adopt a gas distribution of
very low density in the halo, taken to be in hydrostatic equilibrium
in the total potential such that this gas is close to the virial
temperature. This component helps to avoid problems in VPH and 
{\small AREPO}
due to ``empty space'' -- unlike SPH, these codes can not easily deal
with vacuum boundary conditions as they tessellate space. By having
the low density background gas in the halo, which we extend to a box
region around the galaxy, we can avoid this problem.
 
Following the formalism and nomenclature of \citet{Mo1998} and
\citet{Springel2005b}, we have chosen a virial velocity of $V_{200}=
207\,{\rm km\, s^{-1}}$, yielding a total mass of $M_{\mathrm{total}}=
V_{200}^3/ (10 \,G H_0) = 2.1 \times 10^{12}\, h^{-1} {\rm M}_{\odot}$ for
the system, of which a fraction
$M_{\mathrm{disk}}/M_{\mathrm{total}}=0.041$ is assumed to be in the
disk, a fraction $M_{\mathrm{bulge}}/M_{\mathrm{total}}=0.01367$ of
the mass is in the bulge and the rest is distributed in a halo with a
concentration of $c=9.0$ and a spin parameter of $\lambda =
0.033$. The Hubble constant is set to $h = H_0 / (100\, {\rm
  km\,s^{-1} Mpc^{-1}}) = 0.7$.  Since we want to concentrate on the
gas dynamics, we make the disk very gas-rich by giving it a gas
fraction of $50\%$.  The gas component in the halo is given a mass
fraction $M_{\mathrm{gas,halo}}/M_{\mathrm{total}}=0.0025$,
considerably less than the amount of gas in the disk.  This additional
gas in the halo is represented with particles/cells of mass equal to
the gas particles in the disk, and its spatial distribution is clipped
outside a box of side-length $2000\,h^{-1}{\rm kpc}$ centred on the
galaxy, allowing us to impose periodic boundary conditions for the gas
there.

\begin{figure}
\begin{center}
\resizebox{9cm}{!}{\includegraphics{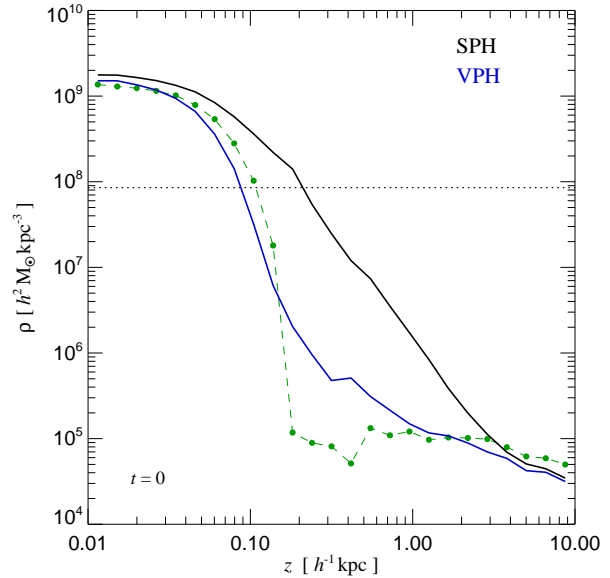}}
\caption{Vertical gas density profile of our isolated galaxy model at
  the initial time $t=0$ when simulated with different numerical
  techniques. The green dashed lines gives the {\em actual} initial
  density profile obtained by averaging the mass distribution over the
  radial range $1.5\,h^{-1}{\rm kpc} < R < 12\,h^{-1}{\rm kpc}$.  The
    black lines correspond to the mean density estimated for the
    particles falling in the corresponding region in our SPH
    simulation. The blue line is the corresponding measurement for
    VPH. Note that the actual particle distribution is {\em identical}
    for SPH and VPH at time $t=0$, so the difference merely reflects
    different systematics in the density estimation techniques
    employed by the two codes. We also note that the result for
    {\small AREPO} is the same as the one for VPH, since they both use
    the same initial conditions and base their density estimate on the
    Voronoi tessellation.}
\label{SPH_zDens}
\end{center}
\end{figure}

\begin{figure}
\begin{center}
\resizebox{9cm}{!}{\includegraphics{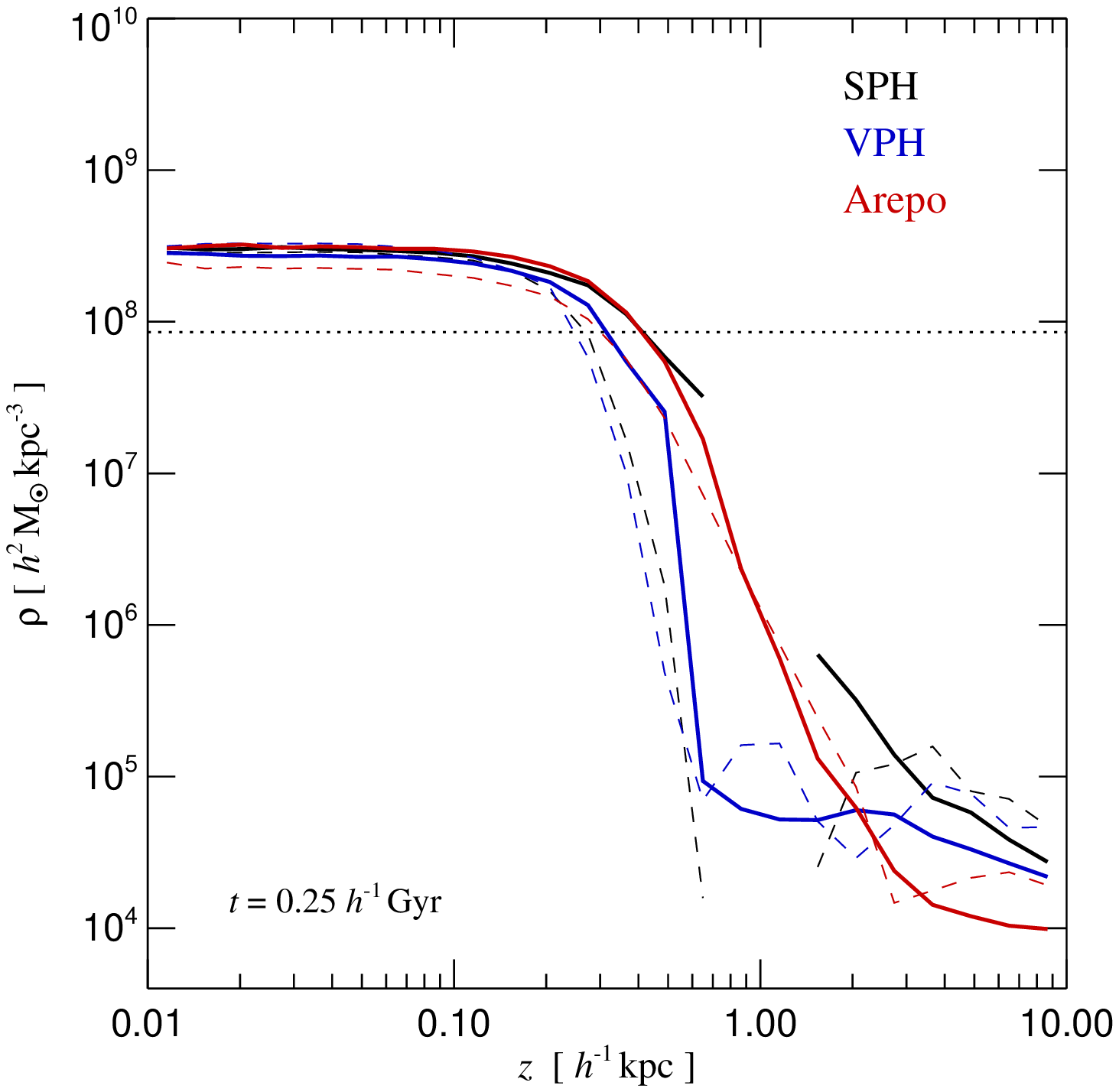}}\\
\resizebox{9cm}{!}{\includegraphics{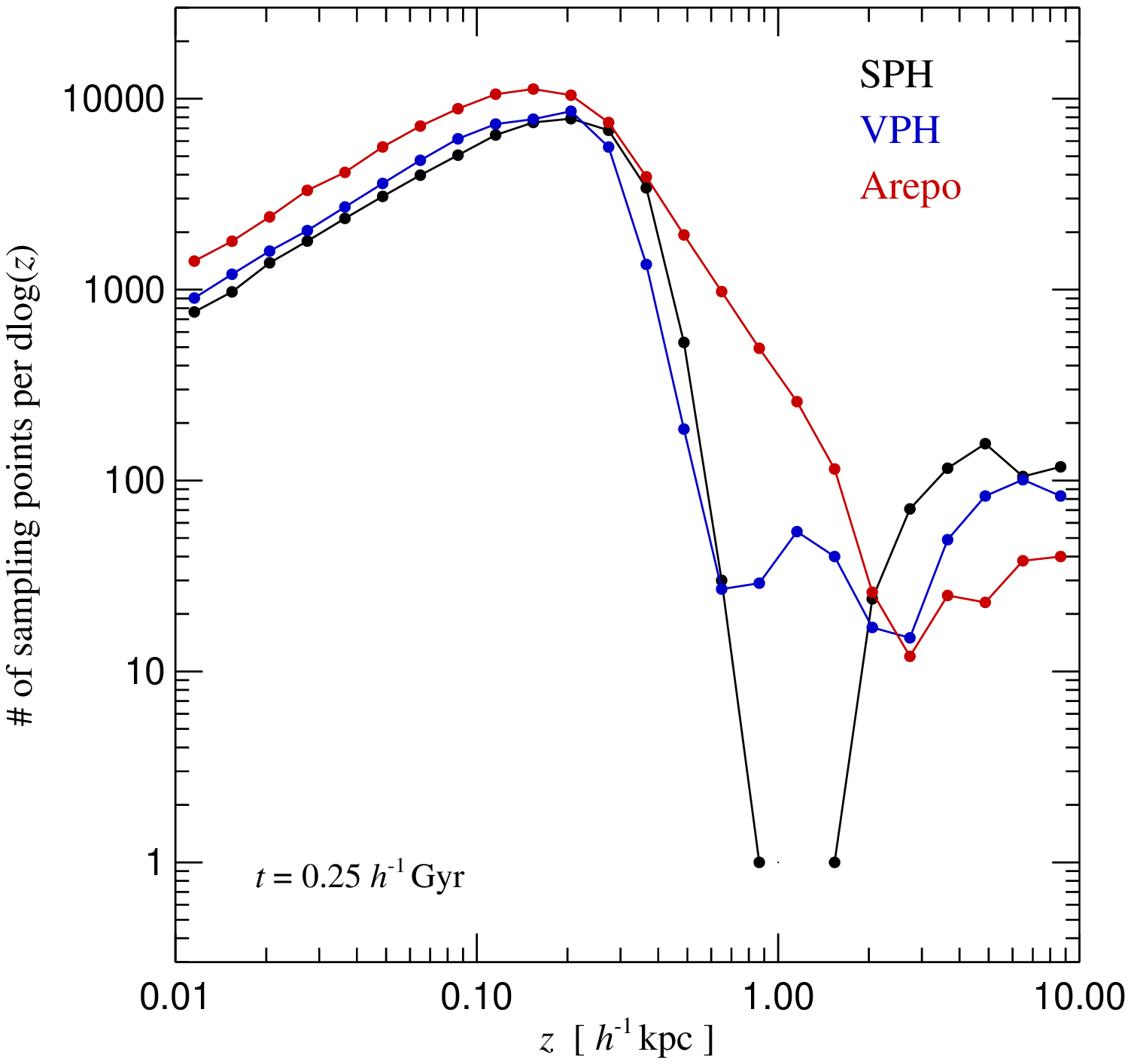}}\\
\caption{Vertical gas density profiles and sampling point densities 
at an
evolved time ($t=0.25\,h^{-1}\mathrm{Gyr}$) for  an isolated galaxy model,
simulated either with SPH (black), VPH (blue) or {\small AREPO}
(red). The top panel gives the gas density profile obtained by
averaging the density estimates obtained for each simulation
particle/cell as solid lines, while the dashed lines are the actual
mean mass profile by summing up the masses and dividing by the bin
volumes. We see that for VPH and  {\small AREPO} these measures agree
reasonably well with each other, while in SPH some larger systematics
at the strong contact discontinuity between cold/dense disk and
hot/tenuous corona appear. In particular, there is in fact a small
region with
essentially no tracer particles in SPH, which is more explicitly seen
in the bottom panel, which simply gives the total number of tracer
particles (fluid particles or mesh-generating points, respectively)
found in the corresponding region.}
\label{isolated_z_profiles}
\end{center}
\end{figure}

We include radiative cooling (for a primordial composition), and model
star formation and supernova feedback with the simple multi-phase
model for the ISM of \citet{SpringelHernquist2003}.  For the supernova
feedback model we adopt an elevated effective temperature of $T_{\rm
  SN}=10^9\rm{K}$ and an evaporation efficiency of $A_0=10^5$. This is
done in order to shift the density threshold for the onset of
star-formation to a 10 times higher values than in the default model
of \citet{SpringelHernquist2003}, while keeping the star-formation
timescale at $t^0_{\star} = 2.1\,{\rm Gyr}$ and maintaining a good
agreement with the \citet{Kennicutt1998} relation.  With this change,
the star formation threshold corresponds to a hydrogen number density
of $n_{\rm H} = 1.3\,{\rm cm}^{-3}$. Such an elevated star formation
density threshold has recently been advocated to facilitate the
formation of more realistic spiral galaxies \citep{Guedes2011}, and it
will sharpen any potential issues brought about by the density
contrast between disk and gaseous corona, which is what we are
specifically interested in in this work. If not denoted otherwise we
employ this model throughout this paper whenever star formation and/or
cooling is required.

In our default intermediate resolution (hereafter called `R4'), the
gas in the disk is represented by $80000$ particles and an equal
number of collision-less particles for the stellar disk. The bulge is
represented by $40000$ particles, and the dark matter halo by $120000$
heavier particles that are given a larger gravitational softening
length of $0.5 \, h^{-1} \rm{kpc}$ compared to $0.25 \, h^{-1}
\rm{kpc}$ for the softening length of the baryonic particles. The gas
in the halo out to the boundaries of the simulation box is represented
with an additional $9756$ particles. Besides this default resolution
we have also simulated realisations of the same galaxy model both at
lower and higher resolutions, in order to get a good sense of
numerical resolution effects. To this end we lowered the resolution in
three steps by factors of 2 in all particle components, and we
increased it in three steps by factors of 2 as well. We hence arrived
at 7 different resolutions R1-R7, with $10000$, $20000$, $40000$,
$80000$, $160000$, $320000$, and $640000$ resolution elements in the
gaseous disk, respectively (see Table~\ref{isoGal_paramTable}). The
  gravitational softening was varied in proportion to the cube root of
  the mass resolution, $\epsilon \propto m^{1/3}$, in these
  simulations. All our simulations with SPH, VPH and the {\small
    AREPO} code were started from the same identical initial
  conditions files in order to yield a direct comparison that also
  allows an identification of different start-up systematics. For the
  simulations with {\small AREPO}, we have activated on-the-fly
  refinements and derefinements similar to \citet{Vogelsberger2011} in
  order to keep the mass resolution always close to the initial value.

\begin{table*}
\begin{tabular}{ l | r | r |  r | r | r |  r  }
\hline  
run name & gas particles & disk + bulge particles &
softening gas/stars  & DM particles & softening DM  \\
  \hline   
R1 	& $10000$	& $10000 +
5000$ & $0.5 \; h^{-1} \rm{kpc}$	& $15000$ 	& $1.0
\; h^{-1} \rm{kpc}$\\
R2 	& $20000$	& $20000 +
10000$ & $0.4 \; h^{-1} \rm{kpc}$	& $30000$ 	& $0.8
\; h^{-1} \rm{kpc}$\\
R3 	& $40000$	& $40000 +
20000$ & $0.31 \; h^{-1} \rm{kpc}$	& $60000$ 	& $0.6
\; h^{-1} \rm{kpc}$\\
R4 	& $80000$	& $80000 +
40000$ & $0.25 \; h^{-1} \rm{kpc}$	& $120000$ 	& $0.5
\; h^{-1} \rm{kpc}$\\ 
R5 	& $160000$	& $160000 +
80000$ & $0.2 \; h^{-1} \rm{kpc}$	& $240000$ 	& $0.4
\; h^{-1} \rm{kpc}$\\ 
R6 	& $320000$	& $320000 +
160000$ & $0.16 \; h^{-1} \rm{kpc}$	& $480000$ 	& $0.32
\; h^{-1} \rm{kpc}$\\
R7 	& $640000$	& $640000 +
320000$ & $0.125 \; h^{-1} \rm{kpc}$	& $960000$ 	& $0.25
\; h^{-1} \rm{kpc}$ \\
  \hline  
\end{tabular}
\caption{List of numerical parameters for the simulations shown in the
  resolution study of Figure~\ref{Iso_sfr}.
  The columns give the simulation name, 
  the initial number of gas particles, the initial number of star
  particles in disk and bulge, the gravitational softening
  lengths for the gas particles, the number of dark matter particles,
  and finally the gravitational softening length of the collisionless
  dark matter particles in the halo.}
\label{isoGal_paramTable}
\end{table*}

\subsection{Gas density maps and structure of the disk}

\begin{figure}
\begin{center}
\resizebox{7.5cm}{!}{\includegraphics{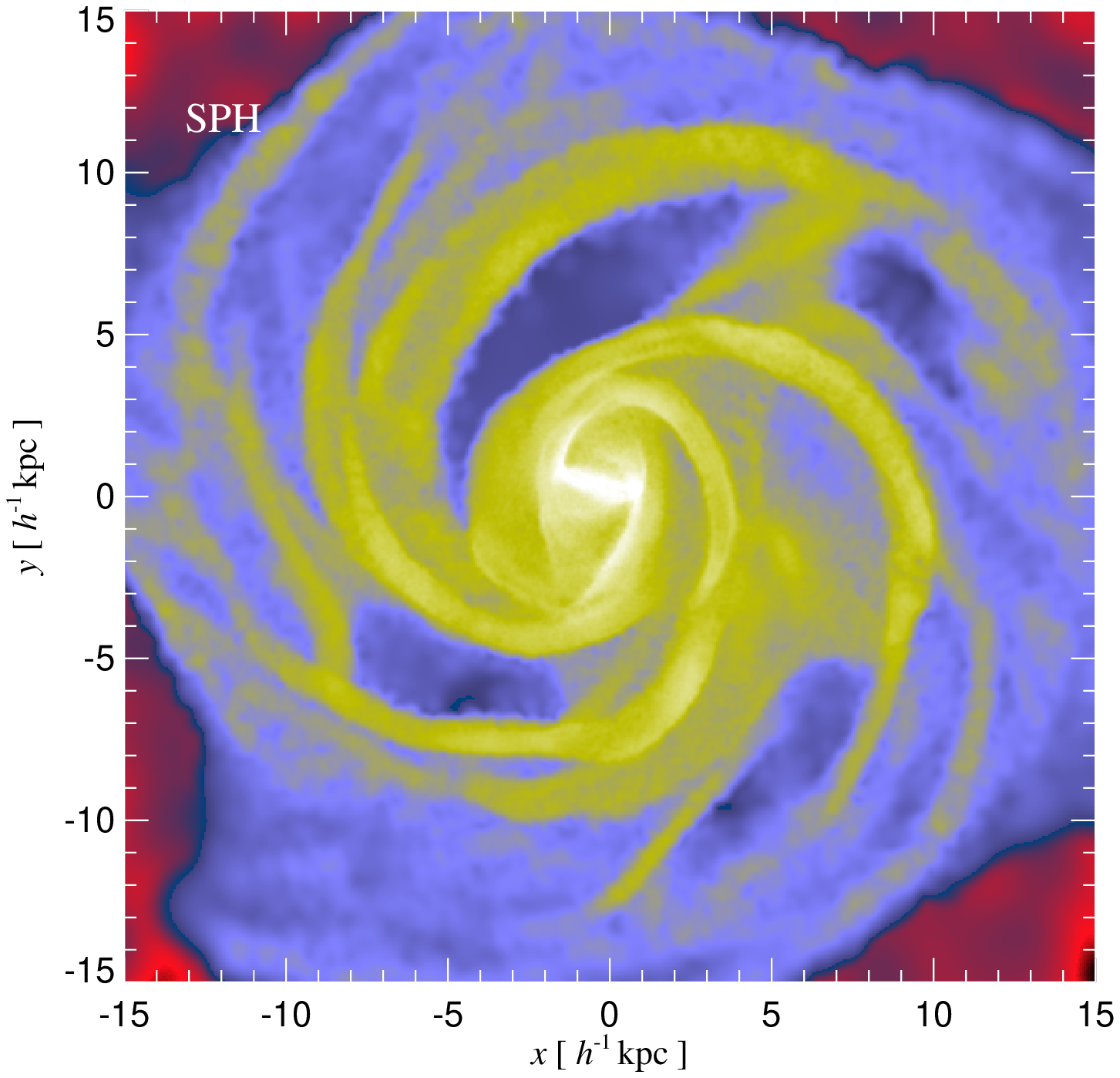}}\vspace*{-0.85cm}\\
\resizebox{7.5cm}{!}{\includegraphics{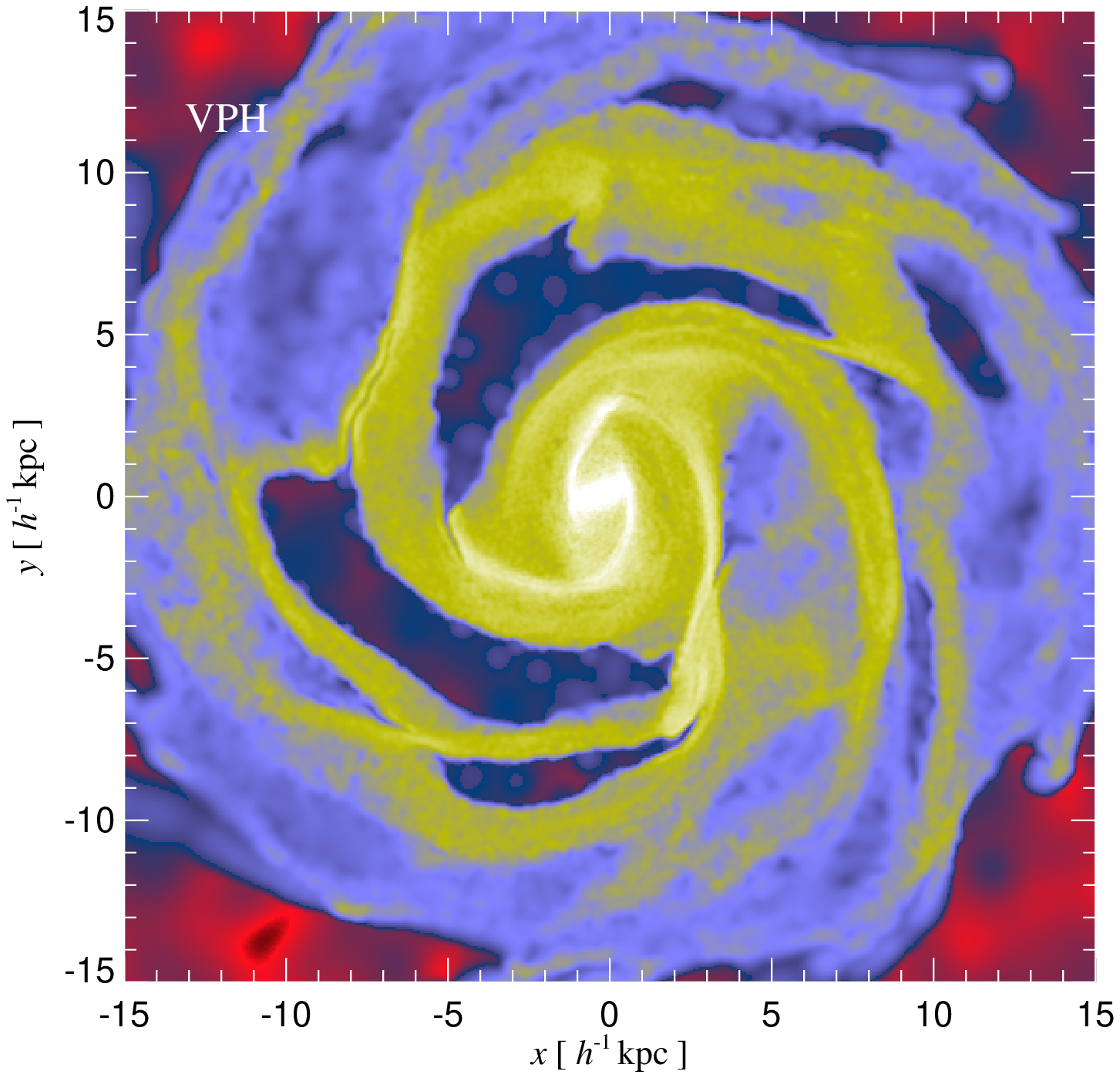}}\vspace*{-0.85cm}\\
\resizebox{7.5cm}{!}{\includegraphics{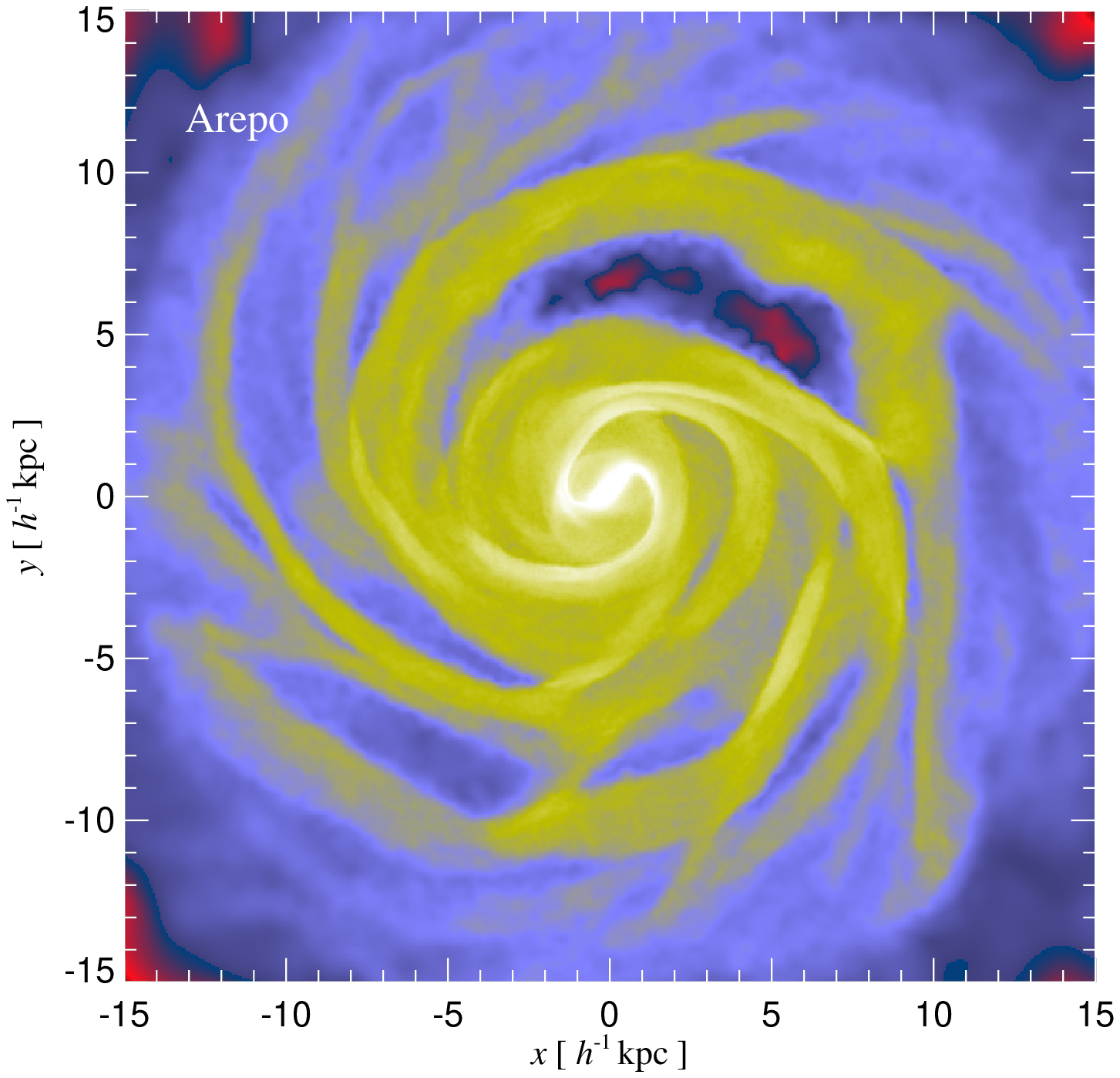}}\vspace*{-0.5cm}\\
\resizebox{7.5cm}{!}{\includegraphics{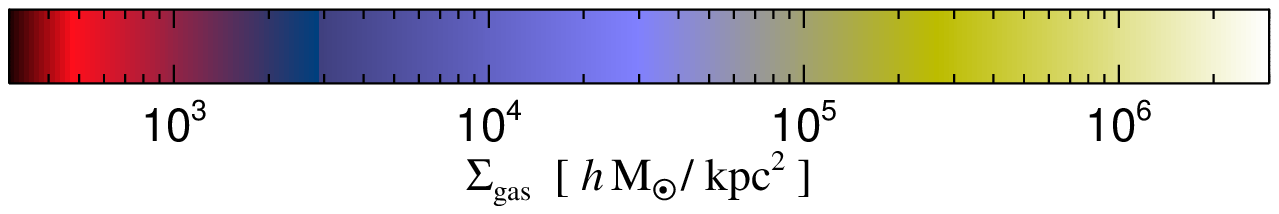}}\\
\caption{Projected gas density maps at $t= 0.5\,h^{-1} \mathrm{Gyr}$
  of our isolated galaxy model simulated with SPH (top), VPH
  (middle) and {\small AREPO} (bottom). Each map measures $30\,h^{-1}{\rm
    kpc}$ on a side, employs a logarithmic colormap, and was
  constructed with an identical adaptive binning method.}
\label{density_maps_isolated}
\end{center}
\end{figure}

\begin{figure}
\begin{center}
\resizebox{7.5cm}{!}{\includegraphics{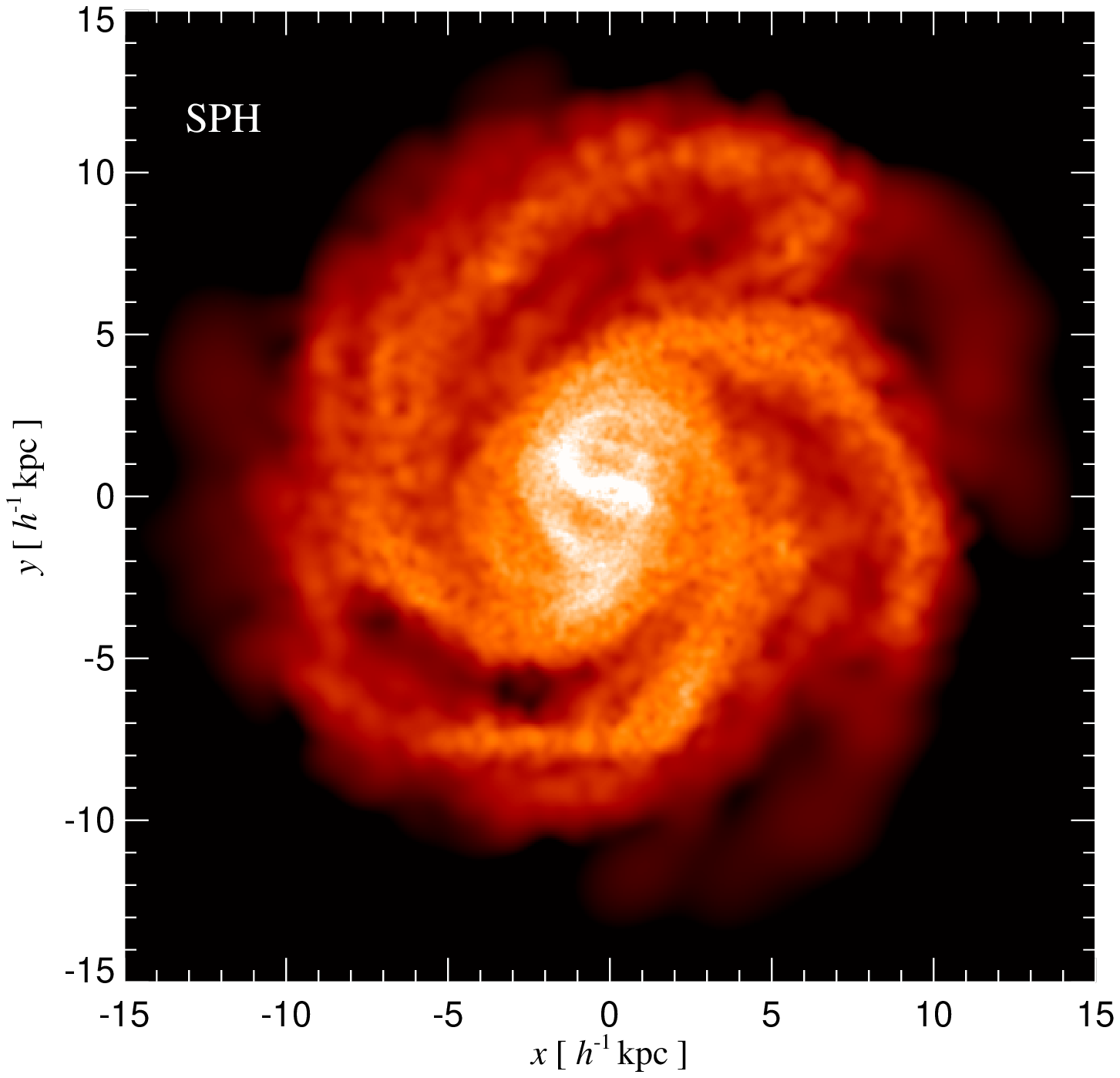}}\vspace*{-0.85cm}\\
\resizebox{7.5cm}{!}{\includegraphics{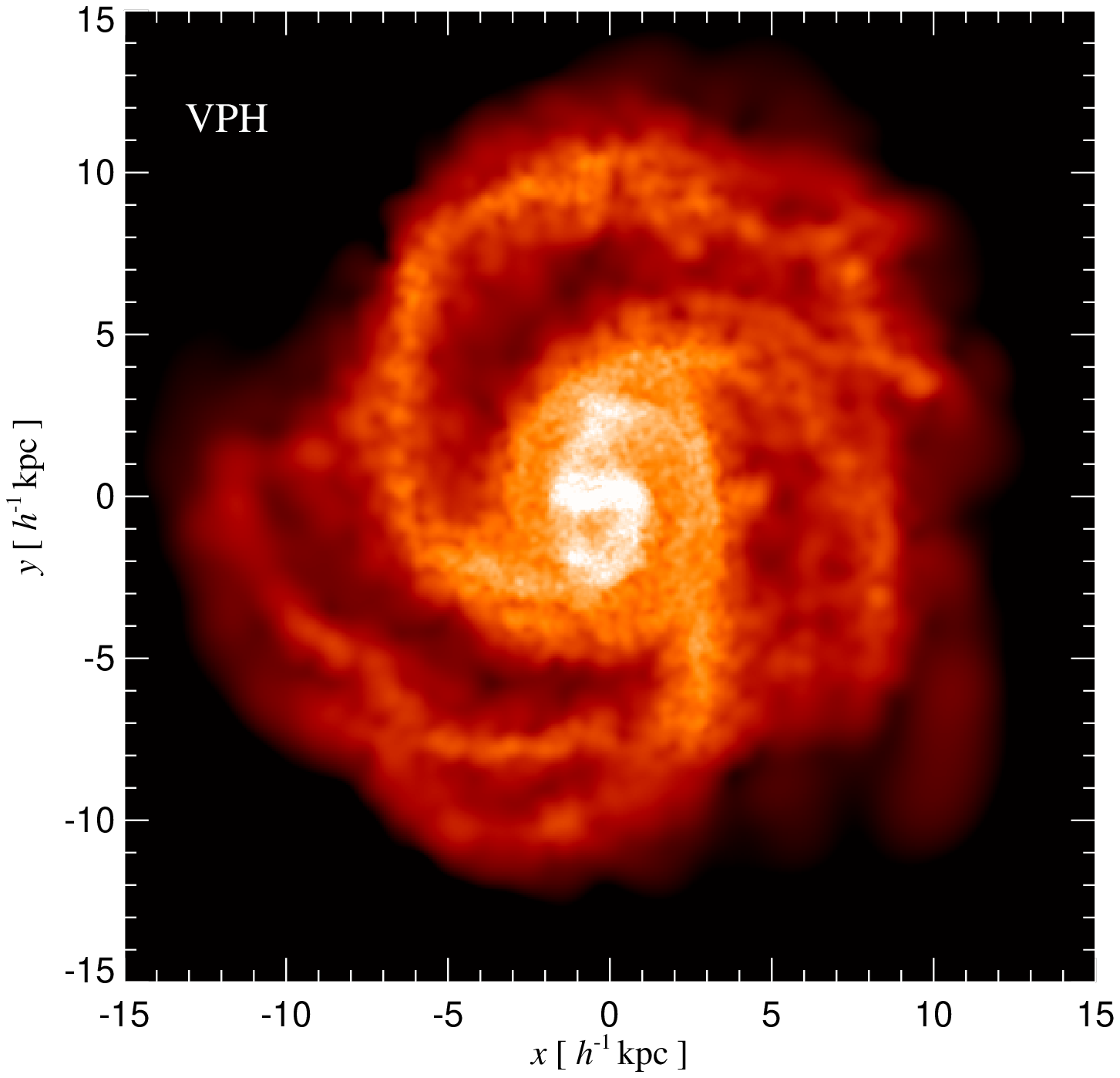}}\vspace*{-0.85cm}\\
\resizebox{7.5cm}{!}{\includegraphics{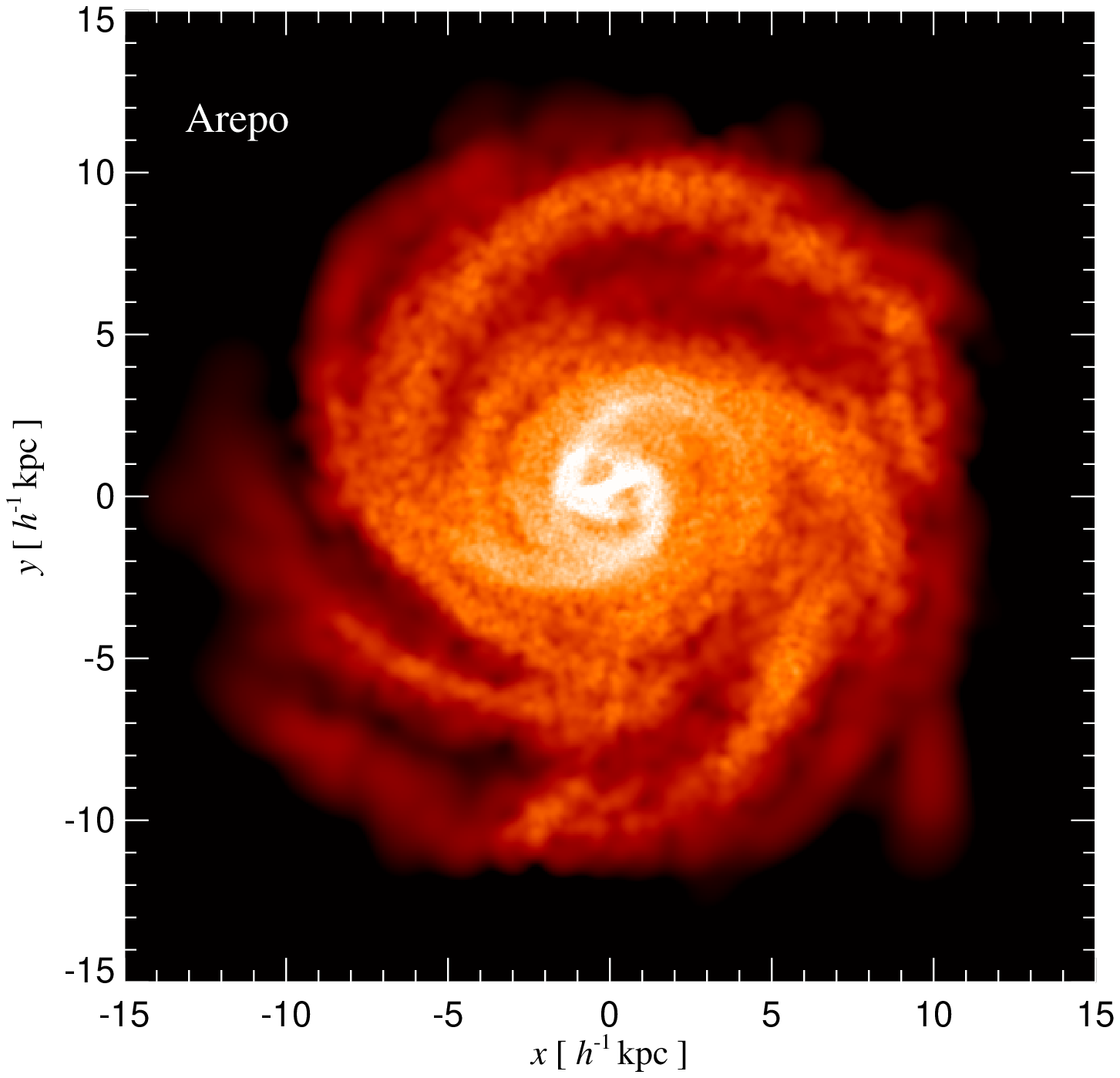}}\vspace*{-0.5cm}\\
\resizebox{7.5cm}{!}{\includegraphics{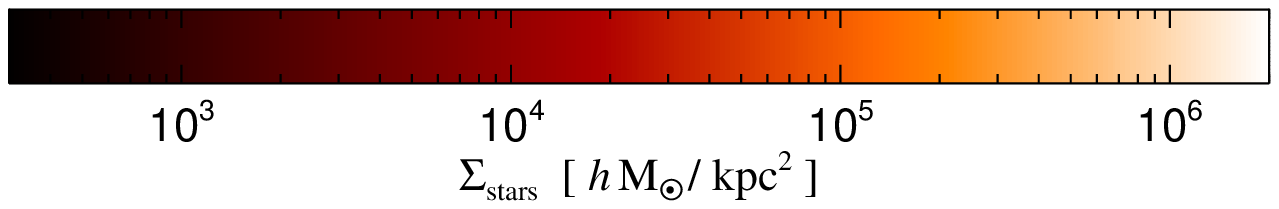}}\\
\caption{Projected mass density of newly formed stars (i.e.~omitting
  stars already present in the stellar disk of the initial conditions)
  in a galaxy model evolved to time $t=0.5\,h^{-1} \mathrm{Gyr}$ with
  three different numerical schemes, SPH, VPH, and {\small AREPO},
  from top to bottom. The maps measure $30\,h^{-1}\rm{ kpc}$ on a
  side. }
\label{Iso_LogDens_stars}
\end{center}
\end{figure}

As we have remarked above, identical particle positions and masses do
not necessarily imply equal densities in our different schemes. To
examine this point, we consider in Figure~\ref{SPH_zDens} each
method's estimate of the initial gas density profile in the
$z$-direction, perpendicular to the disc. The solid curves represent
averages of the densities estimated for the individual particles/cells
as a function of distance from the disk plane, while the dashed line
is the {\em actual} average mass density distribution obtained by
binning the particles directly.  Since both VPH and AREPO use a
Voronoi tessellation for the density estimate, they agree on the
densities at the initial time, but the kernel estimates of SPH (based
on $N_{\rm ngb}=48$ smoothing neighbours) show an interesting
difference. In particular, the particles close to the dense disk gas
(for $z\sim 0.2-2.0\,h^{-1}{\rm kpc}$) show an elevated density
estimate because they ``see'' some of the dense particles within their
smoothing radius. This effect is not unexpected given that sharp
discontinuities will always be smoothed out to some extent in SPH, by
construction. The density estimate of VPH has a somewhat better
resolving power, but here a different systematic effect becomes
noticeable. Particles in the surface layer of the dense disk sometimes
extend their volume quite far into the region above the disk until
they ``see'' a particle of the background corona, thereby
underestimating the density in this region.  Finally, there is an
interesting systematic difference between SPH and VPH at the very
centre of the disk, where SPH shows a small positive bias in the
density estimate. This comes about in part due to the Poisson sampling
present in the initial conditions, which in SPH causes larger positive
excursions in the density estimates. For a system in pressure
equilibrium, this effect is weaker, but it not necessarily vanishes
because also here the SPH density may be biased and the sum of the
volumes associated with each particle is not guaranteed to add up to
the total volume.

In Figure~\ref{isolated_z_profiles}, we analyse the vertical structure
of the gas after the isolated galaxy has been evolved for a time of
$0.25\,h^{-1}{\rm Gyr}$. In the top panel of the figure, we again
compare measurements of the mean density estimated for particles/cells
as a function a distance $z$ above the disk mid-plane with the actual
mean mass profile present in the simulation. We give results for all
three numerical methods (solid lines), and since the actual gas
distribution can have evolved differently in the three methods, this
is given individually as well (dashed lines). We see that the two
measures of the density stratification track each other well both in
VPH and AREPO. However, the moving-mesh code shows higher density just
above the disk compared with VPH. If anything, SPH shows still higher
densities in this region, although its actual mass distribution is
very similar to that of VPH. This situation arises due to a
significant difference SPH exhibits in the estimated density versus
the real mass distribution in the regime of the `shoulder' of the
disk's density distribution. Also, there is actually a `gap' in the
run of SPH's density estimates, arising simply because we find no SPH
tracer particles for measuring the density there. This gap becomes
more explicit in the bottom panel of Figure~\ref{isolated_z_profiles},
where we show counts of fluid tracer particles in logarithmic bins in
the same region above the disk. The sampling gap at $z\sim 1\,
h^{-1}{\rm kpc}$ in SPH is evident in this figure, but it does not
occur in VPH. We note that \citet{Agertz} pointed out that such a gap
is seen generically at strong density jumps in SPH, and that it may be
a primary cause for the suppression of fluid instabilities across such
contact discontinuities. Recently, a number of authors
\citep{Price2008,Read2011} have suggested that SPH should be outfitted
with additional dissipation schemes that smooth out such structures to
improve on this behaviour.

\begin{figure}
\begin{center}
\hspace{-5mm}
\resizebox{9cm}{!}{\includegraphics{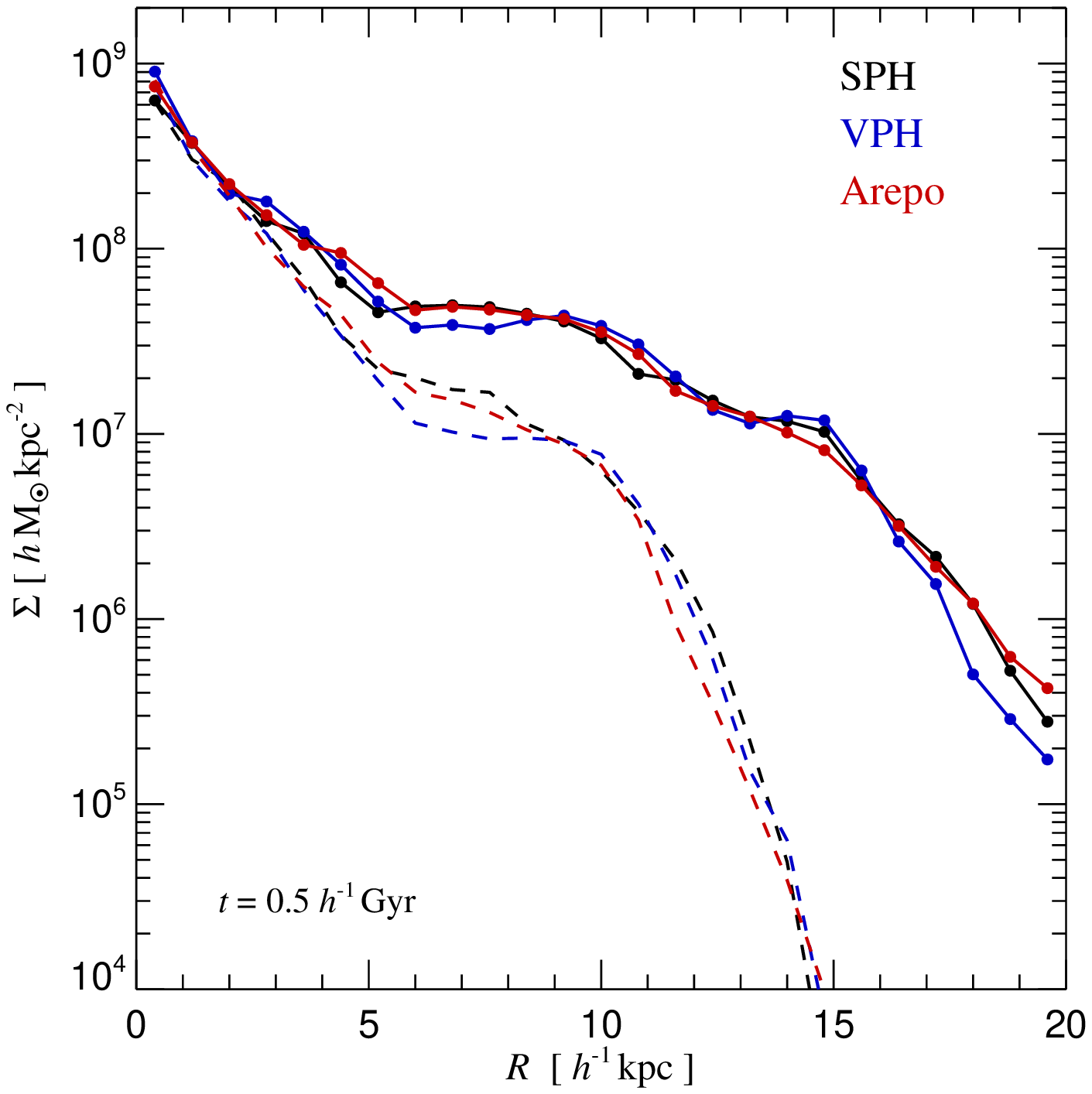}}
\caption{Surface density profiles of gas (solid lines) and newly
  formed stars (dashed lines) at time $t=0.5\,h^{-1}\rm{Gyr}$ for our
  isolated galaxy model when simulated with either SPH (black), VPH
  (blue) or {\small AREPO} (red). }
\label{isolated_Sigma_profiles}
\end{center}
\end{figure}

\begin{figure}
\begin{center}
\resizebox{8.5cm}{!}{\includegraphics{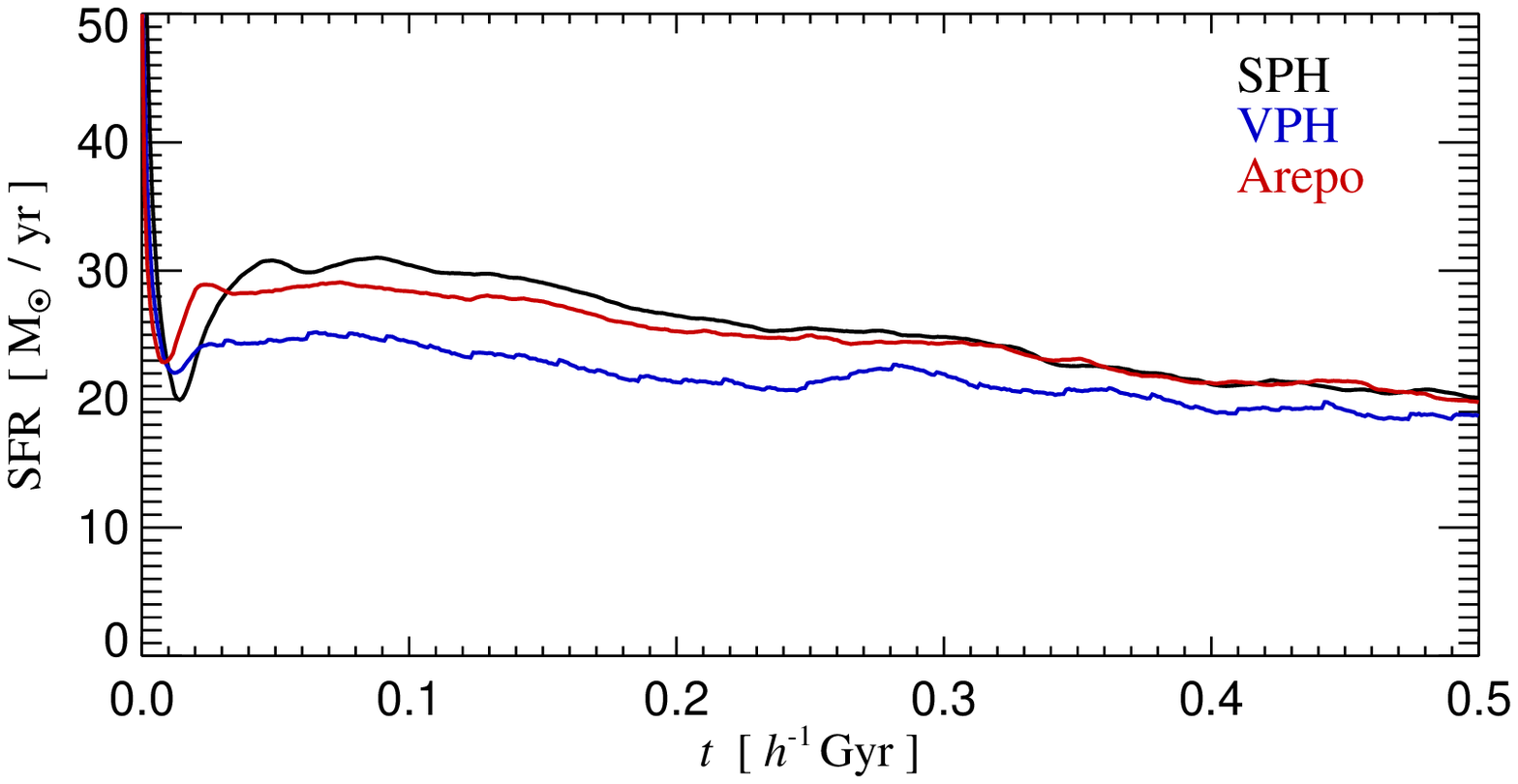}}\\
\resizebox{8.5cm}{!}{\includegraphics{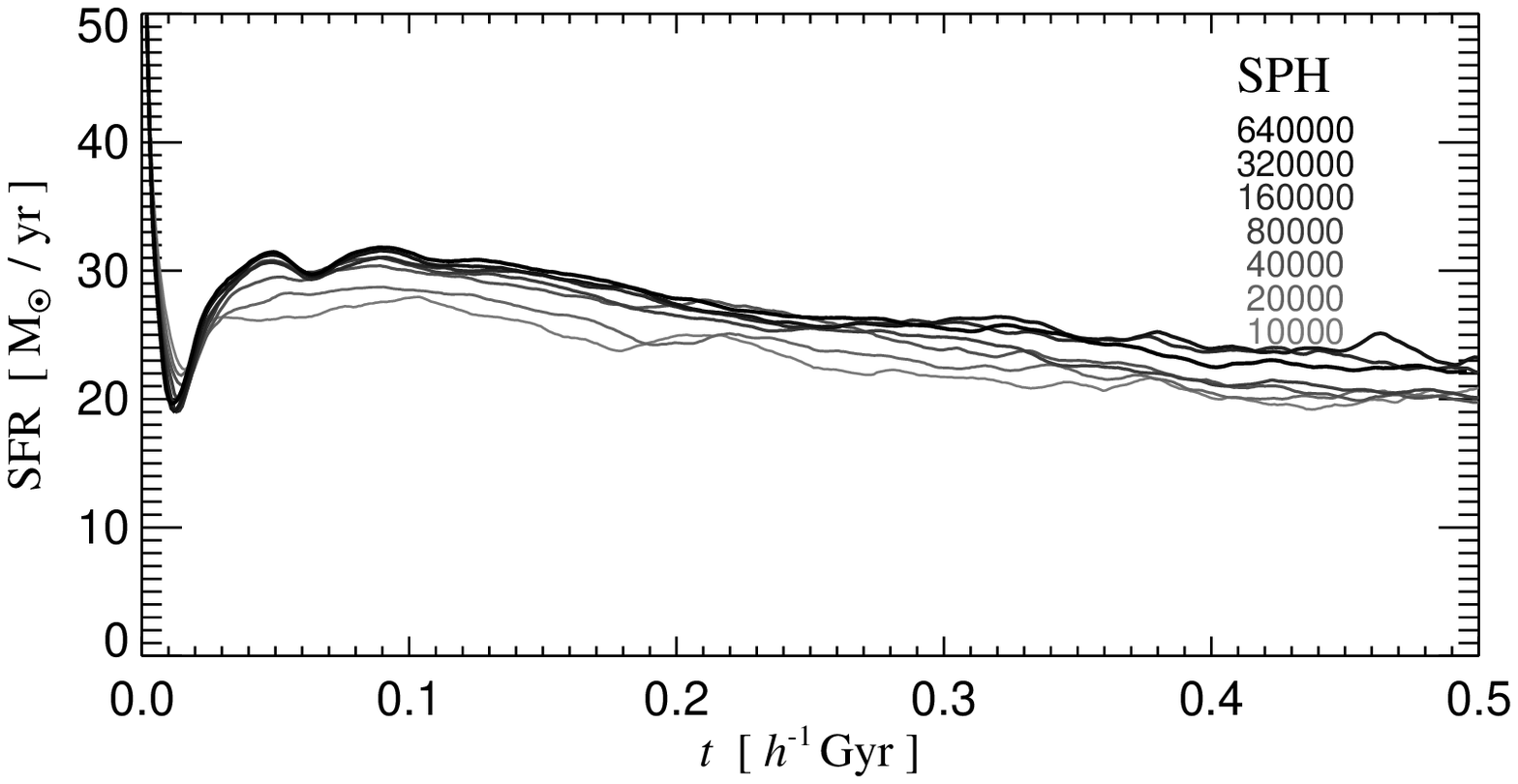}}\\
\resizebox{8.5cm}{!}{\includegraphics{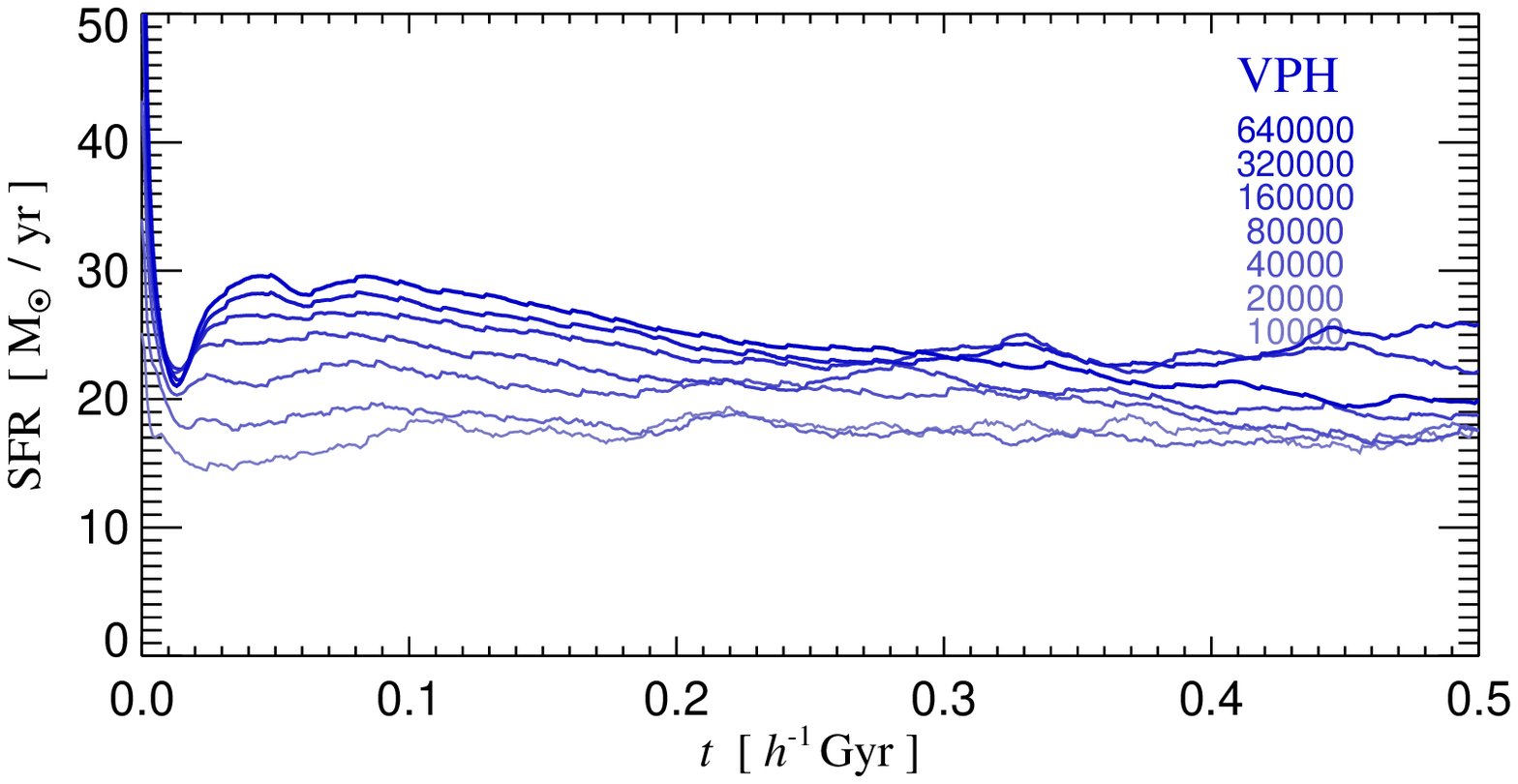}}\\
\resizebox{8.5cm}{!}{\includegraphics{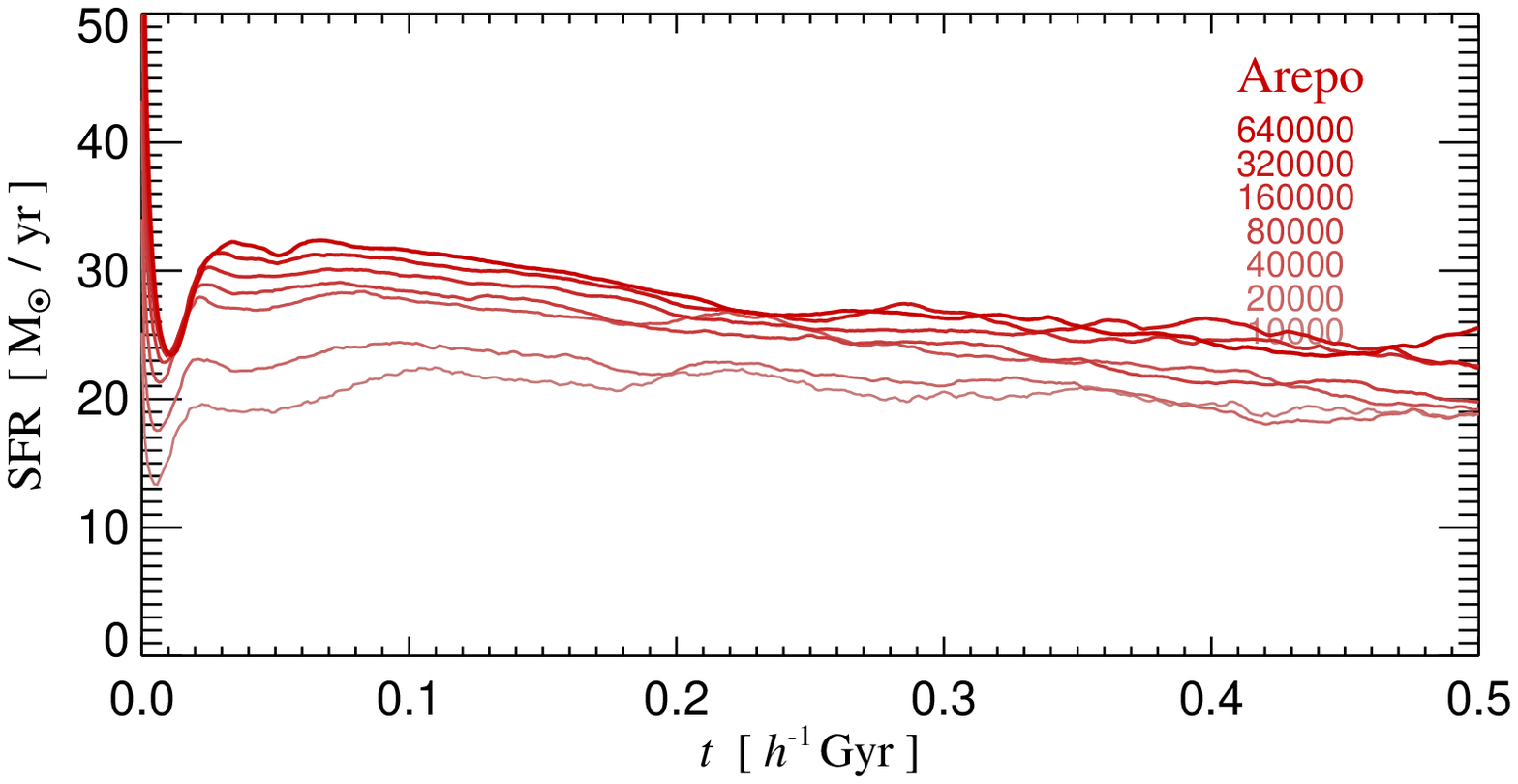}}\\
\caption{Comparison of the time evolution of the star formation rate
  of an isolated galaxy model, simulated with SPH (black), VPH (blue),
  and {\small AREPO} (red). The top panel compares the three different
  simulation techniques directly with each other, at the
  comparatively high resolution of $160000$ gas particles/cells in
  the initial disk. The other three panels show resolution studies
  for each of the codes individually, where the resolution in the gas
  was varied between 10000 and 640000 resolution elements. The number
  of dark matter and stellar bulge/disk particles has been varied in
  proportion in each of the runs of the series. We note that the lower
  resolution runs in VPH and  {\small AREPO} are somewhat more
  strongly affected
  by resolution than for SPH, but in general all the methods converge
  and give consistent results with each other at high resolution.}
\label{Iso_sfr}
\end{center}
\end{figure}

\begin{figure}
\begin{center}
\resizebox{8cm}{!}{\includegraphics{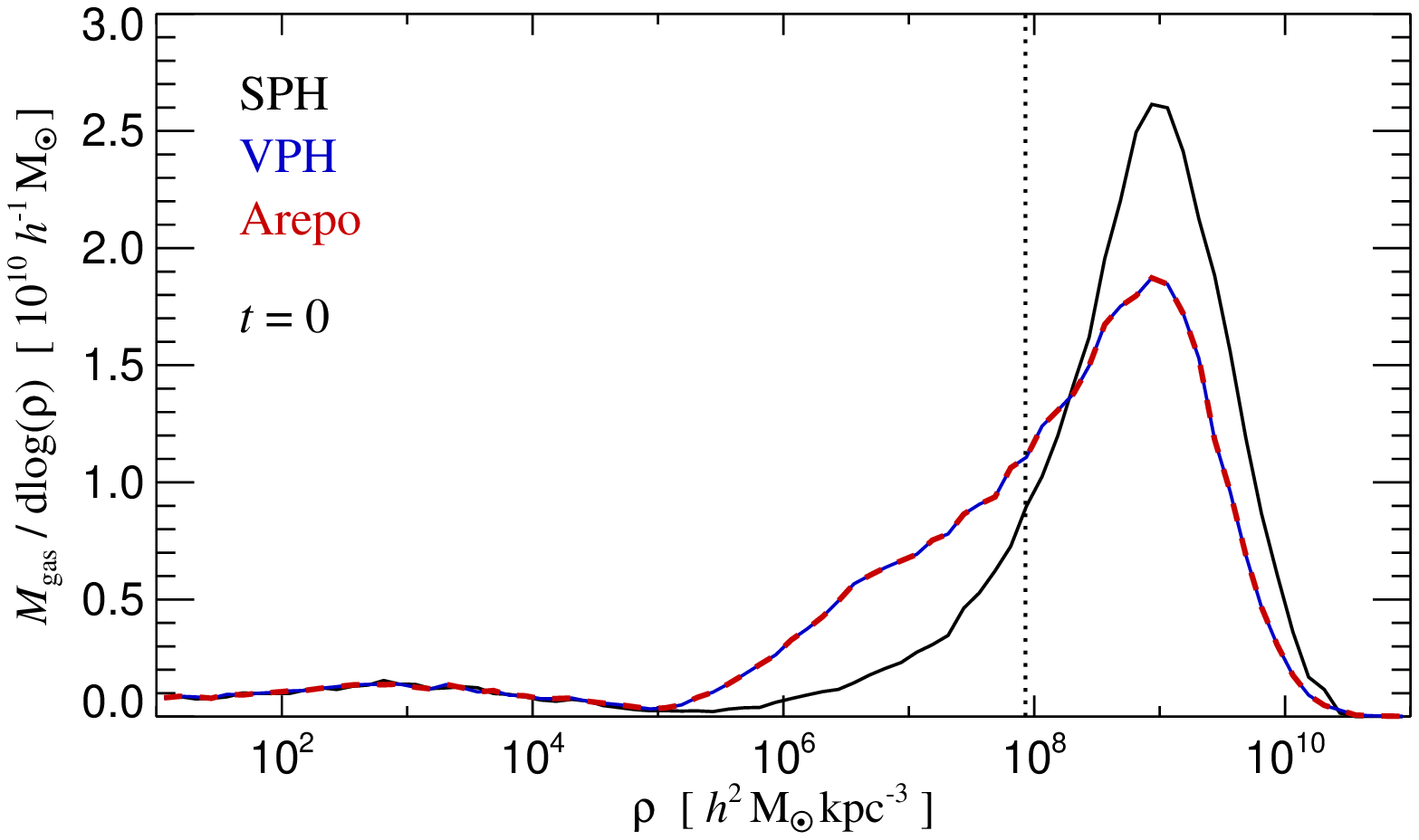}}\\
\resizebox{8cm}{!}{\includegraphics{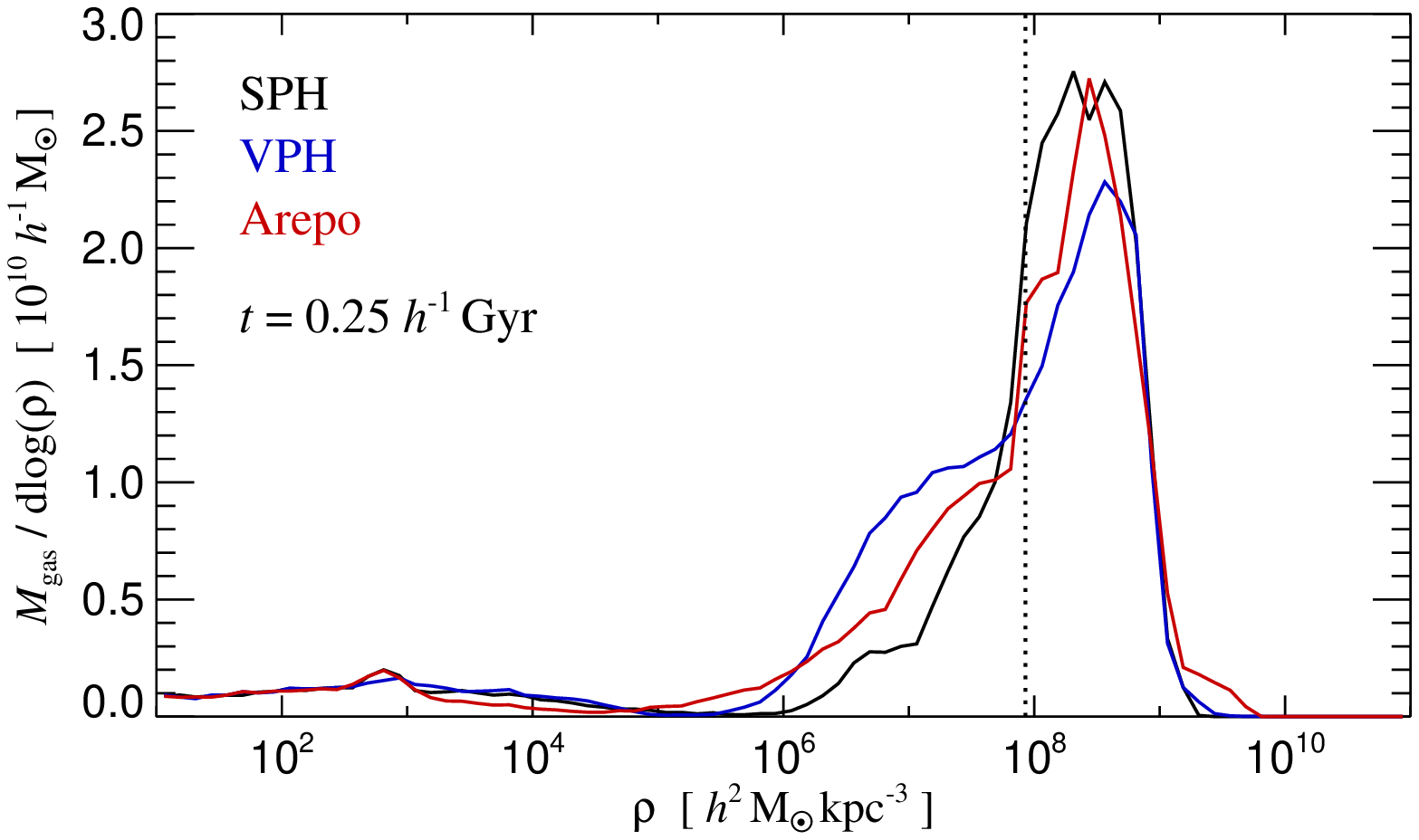}}\\
\caption{Distribution of gas mass as a function of density {\em
    estimated} by our different numerical techniques (SPH in black,
  VPH  in blue, and AREPO  in red) in a simulation of an 
isolated disk galaxy. In the top panel,
we show the distribution of gas mass in the initial conditions at
$t=0$, where the particle/cell distribution is identical for all three
runs. Nevertheless, SPH estimates higher densities for some of the gas
in or close to the surface of the gas disk, such that more gas mass
ends up above the density threshold for star formation (vertical
dotted line). At the later time of 
$t=0.25 \,h^{-1} \rm{Gyr}$ shown in the bottom panel, the difference
has become considerably smaller, but there is
still an important systematic offset between VPH and SPH. 
\label{Iso_denshist}}
\end{center}
\end{figure}

In Figure~\ref{density_maps_isolated}, we show face-on maps of the
galaxy's projected gas density distribution, after a period of
$t=0.5\,h^{-1} \mathrm{Gyr}$ of evolution. It is reassuring but probably not
too surprising that the overall morphology is extremely similar, as
the gravity from the dark matter halo and the stellar disk are primary
driving forces of the disc dynamics.  However, upon close inspection,
one can identify some interesting systematic differences between the
techniques that manifest themselves in the gas density maps.  Compared
with SPH, the higher effective spatial resolution of VPH produces
crispier but marginally noisier looking features such as spiral
arms. Perhaps the most prominent difference is however the lower
density of the gas found in VPH compared to SPH in some regions
between spiral arms.  The gas distribution of the mesh-based {\small
  AREPO} looks a bit sharper and less noisy than both particle-based
schemes.

Nevertheless, there is reassuring similarity of the gas surface
density distribution between the different techniques, an impression
that is corroborated by maps of the projected stellar mass density in
the disks, which we show in Figure~\ref{Iso_LogDens_stars}. Here we
show only the newly formed stars in the simulations, so that any
structure in this component directly reflects the underlying gas
dynamics.  We find good agreement for these collisionless particles,
which can also be interpreted as evidence that the gravitational
dynamics of collisionless particles is followed with equal quality in
all three codes. This verifies that differences in the evolution of
our galaxy models are expected to arise only from the different
treatment of the hydrodynamics, and not from differences in the way
the collisionless dynamics of stars and dark matter is followed.

Despite the systematics we identified in the density estimates and in
the projected gas distributions, we overall find that all three codes
evolve the galaxy model in a similar and stable fashion.  This is
further confirmed by radial surface density profiles of the gas and
the newly formed stars shown in Figure~\ref{isolated_Sigma_profiles}.
All three methods are clearly able to retain the initial structure of
the galaxy model to a similar degree, yielding only very minor
differences after some time.

\subsection{Star formation rate evolution} \label{sfr_iso}

Despite the good agreement between the different methods noted above,
some minor but interesting discrepancies are revealed when looking in
detail at the time evolution of the star formation rate at different
resolutions, as done in Figure~\ref{Iso_sfr}. In the top panel, we
compare the results for SPH, VPH and {\small AREPO} at our default
resolution of R4, which lies in the middle of our extended set of
models used in our resolution study. There is in general good
agreement between the runs at late times, where SPH and {\small AREPO}
agree extremely well and VPH lies at most a few per cent
lower. However, at earlier times some large differences are
noticeable. First of all, at $t=0$, the star formation rate for SPH is
about $\sim 20$ per cent higher than that of VPH and {\small AREPO}, a
difference entirely caused by different density estimation methods
(see below). But after an initial transient, when the disk settles
from the approximate equilibrium realised in the initial conditions to
a proper self-consistent equilibrium, somewhat larger differences are
present for an extended period of time.  VPH tends to show a slightly smaller star
formation rate than SPH, with {\small AREPO} lying somewhere in the
middle.

Some clues to the origin of these differences are provided by the
lower three panels of Figure~\ref{Iso_sfr}, which give the evolution
of the SFR for all seven numerical resolutions we considered for our
isolated galaxy model, and for each numerical method.  We see that all
three techniques show some residual drift in the SFR at low
resolution, but they tend to converge reasonably well towards higher
resolution. Given the fact that the mass resolution is varied by a
factor of 64 here, a scatter in the star formation rate of order
$10-20$ per cent can probably be considered
satisfactory. Nevertheless, there are clearly interesting differences
in the strength of the resolution-dependence of the numerical
techniques. In particular, VPH shows a somewhat stronger reduction of
the SFR at the lowest resolution compared to SPH, whereas the latter
is
remarkably resilient to resolution changes. Perhaps
counter-intuitively, this arises despite the relatively strong biases
in SPH's density estimate at the interface between  disk and
corona. 

Another look at this issue is provided by Figure~\ref{Iso_denshist},
in which we show how much gas mass is estimated by the codes to lie at
a certain density value. In the top panel of Fig.~\ref{Iso_denshist},
this is given for the initial time. The fact that SPH systematically
estimates more gas to lie at a density value above the star formation
threshold than the Voronoi-tessellation schemes (despite an {\em
  identical} point distribution in this case) explains the offset in
the SFRs at $t=0$. However, also at later times, as seen in the bottom
panel of Fig.~\ref{Iso_denshist}, such a systematic difference
persists.  The variations in the star formation rates calculated by
the different codes appear hence to be primarily driven by the way the
sharp edge of the star-forming disk is represented. These differences
are however quite minor overall and can be largely ignored for our
subsequent investigations.

\section{Galaxy in a wind tunnel} \label{WindTunnel}

Simulating the interaction of a galaxy with the sparse intra-cluster
gas of a galaxy cluster at high-resolution is computationally
challenging
\citep[e.g.][]{Abadi1999,Tittley2001, Schulz2001,Roediger2006,
Roediger2007,Roediger2008},
especially with the particle-based methods SPH and VPH, which do not
easily lend themselves to adaptive refinement techniques. In fact,
they basically require particles of equal or very similar mass to work
well \citep{Ott2003}. Hence, if one simply wants to let a galaxy model
fall into a massive cluster, the latter needs to be represented with
very high particle number (because it is so massive), such that one
ends up spending only a tiny fraction of the computational effort onto
the galaxy and the surrounding gas, where the actual interaction takes
place. A further complication is that some of the stripped gas mass
may be distributed over a large region across the cluster,
corroborating the need to simultaneously account for the whole cluster
at high resolution.

\begin{figure*}
\begin{center}
\includegraphics[height=0.305\textheight]
{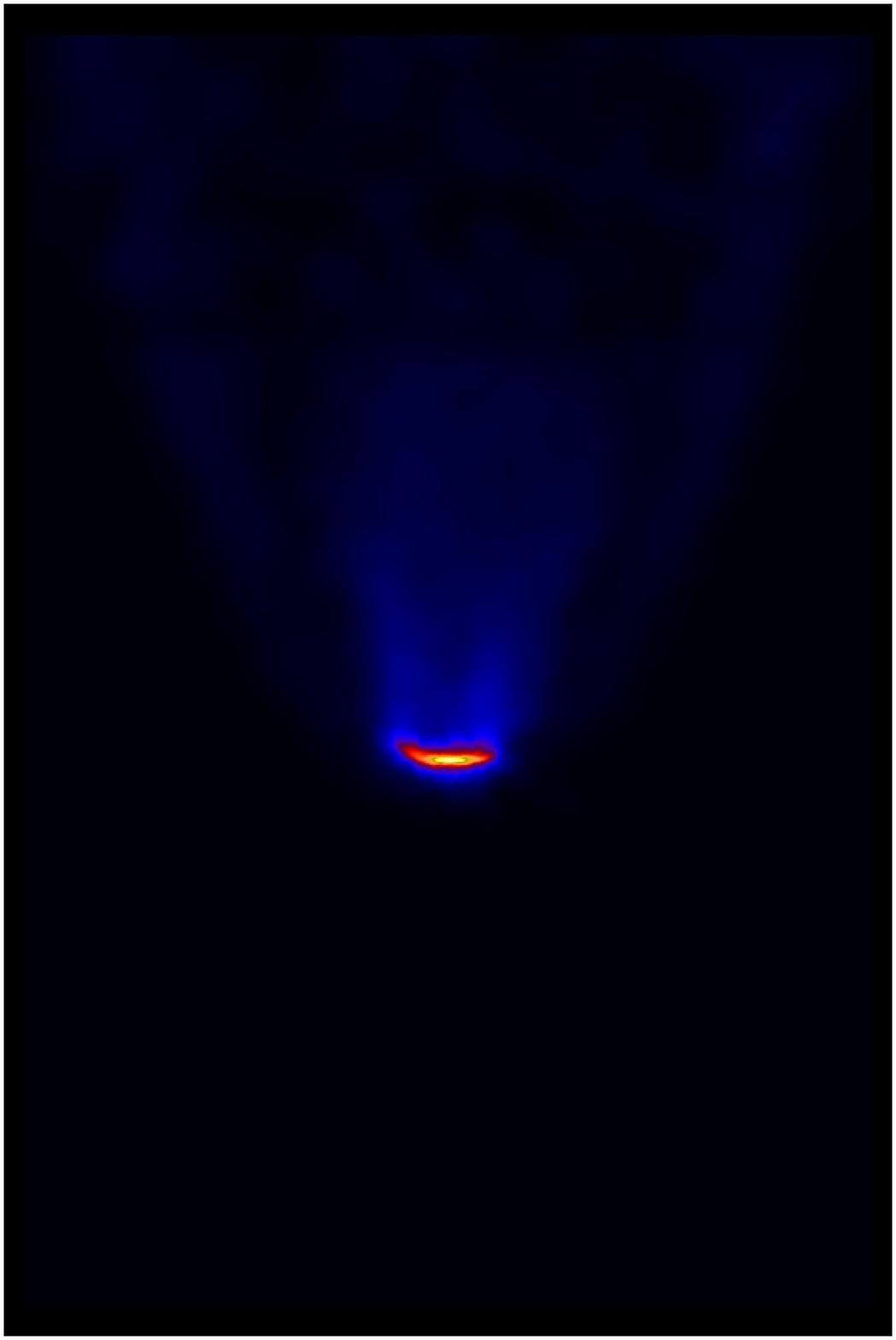}
\includegraphics[height=0.305\textheight]
{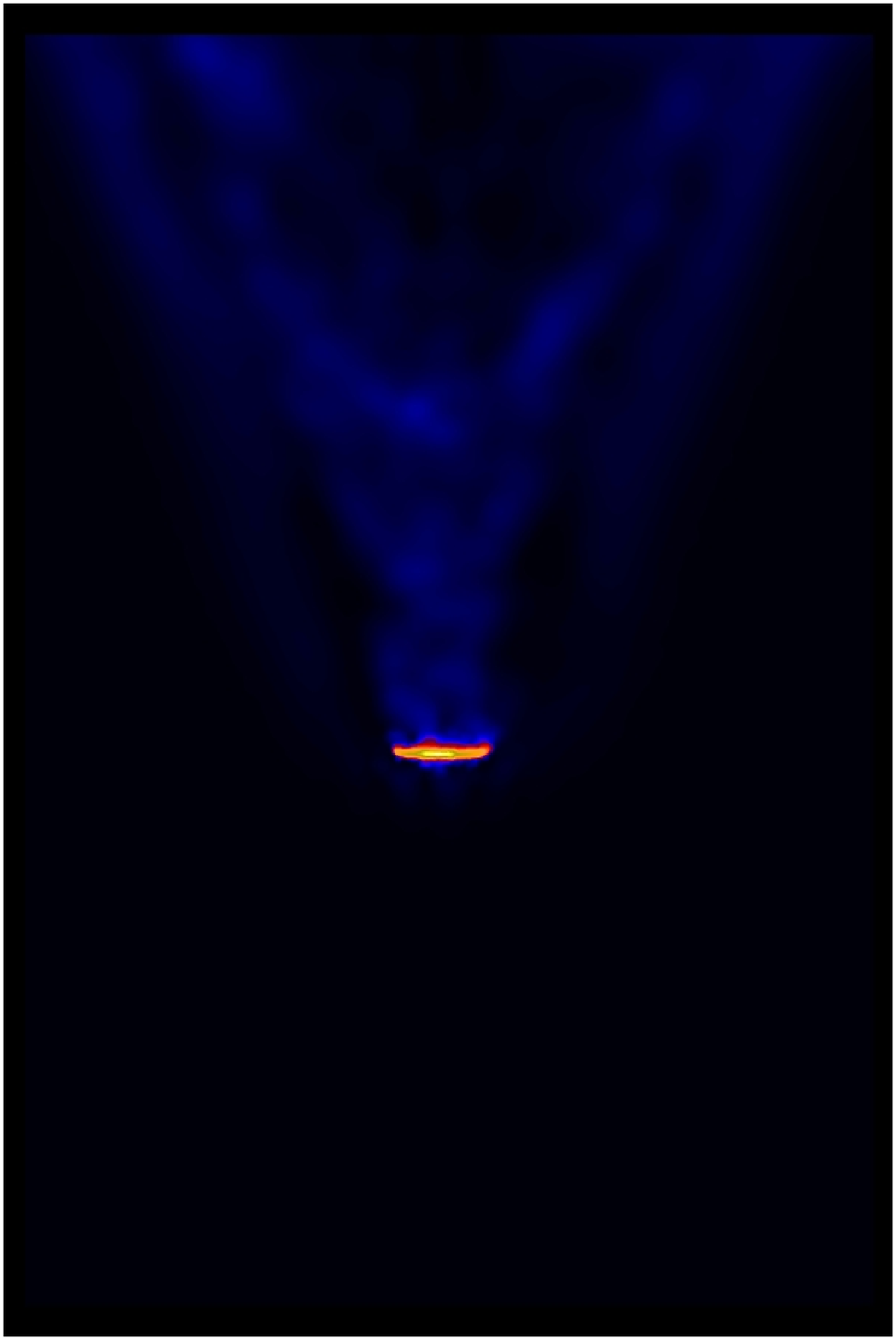}
\includegraphics[height=0.305\textheight]
{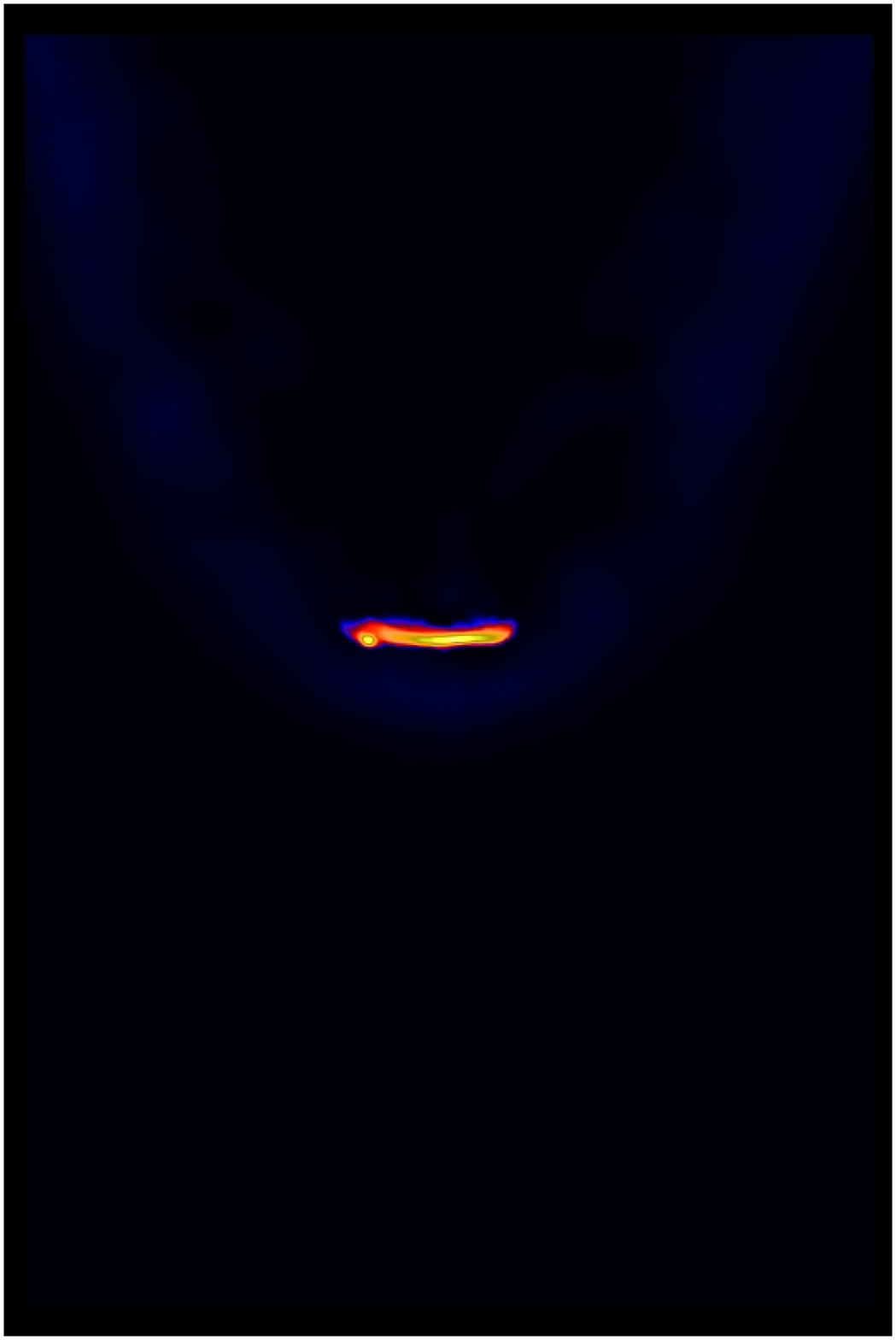}
\hspace{-3mm}
\includegraphics[height=0.305\textheight]
{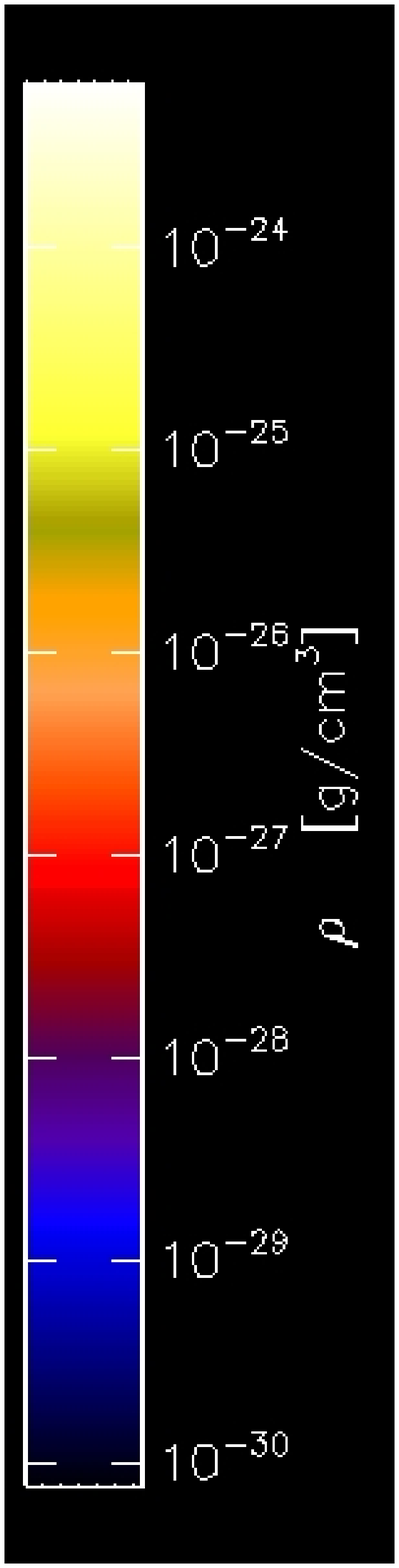}\\
\vspace{-3mm}
\includegraphics[height=0.305\textheight]
{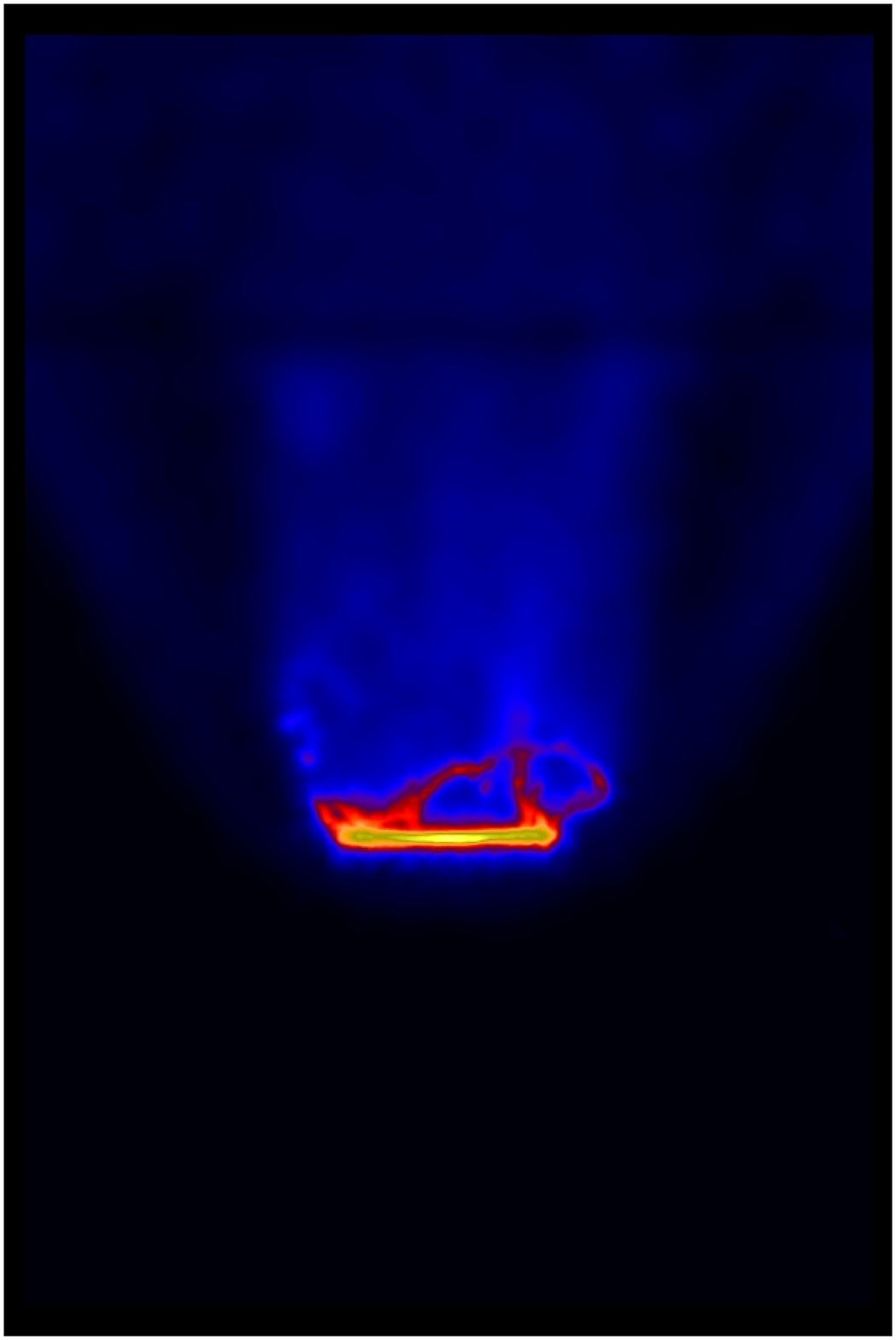}
\includegraphics[height=0.305\textheight]
{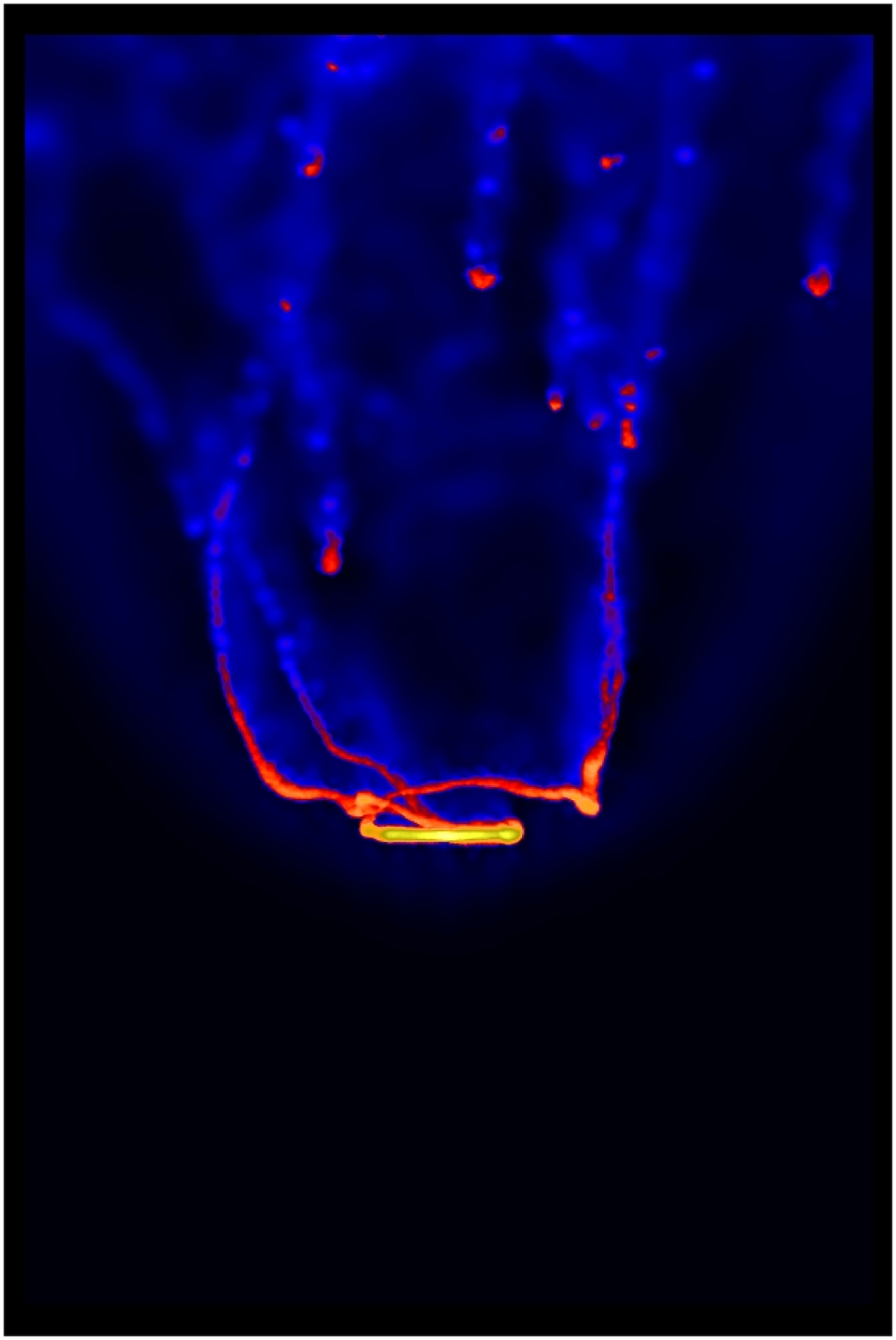}
\includegraphics[height=0.305\textheight]
{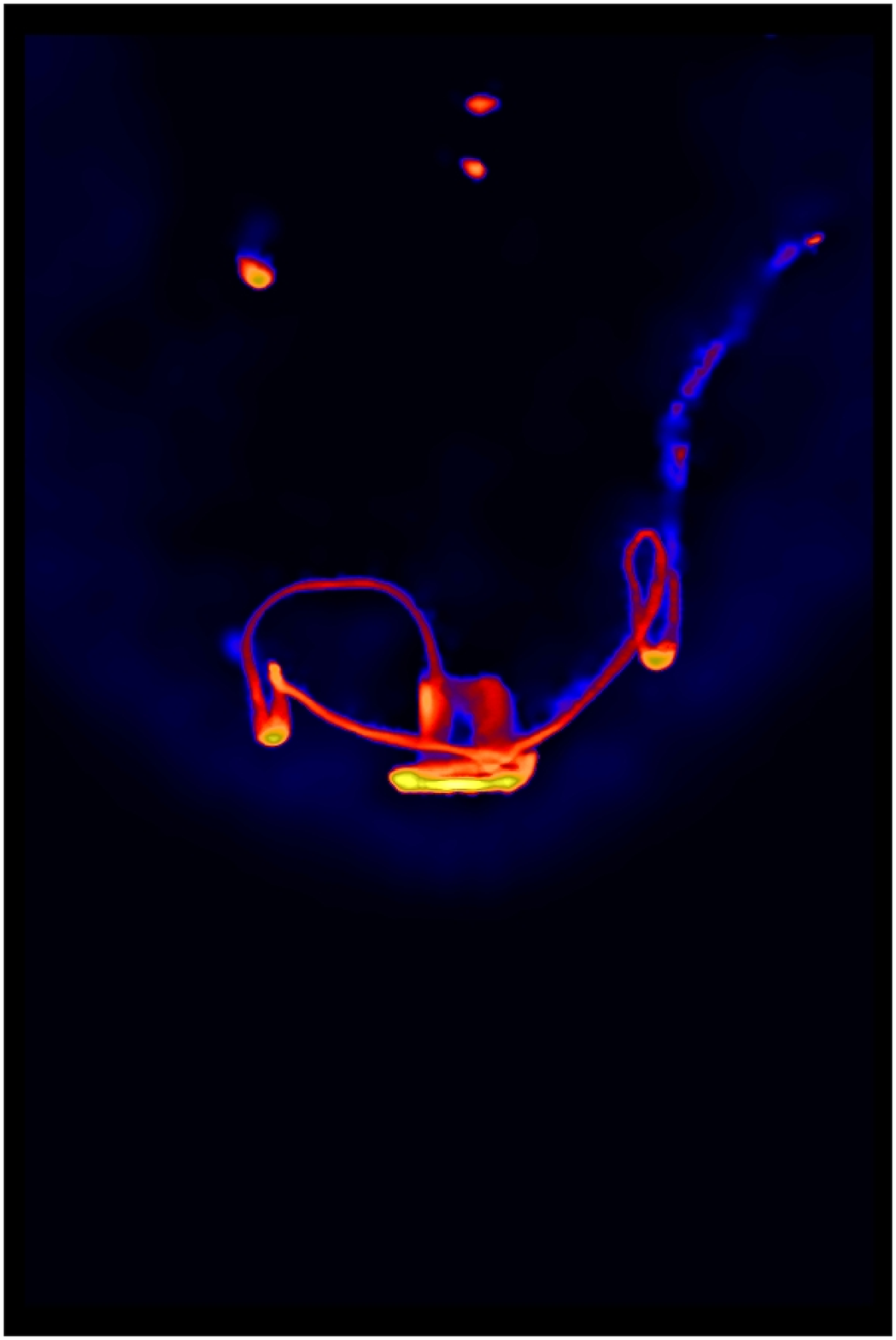}
\hspace{-3mm}
\includegraphics[height=0.305\textheight]
{plots/WindTunnel/maps/Dens_colorbar.eps}\\
\vspace{-3mm}
\includegraphics[height=0.305\textheight]
{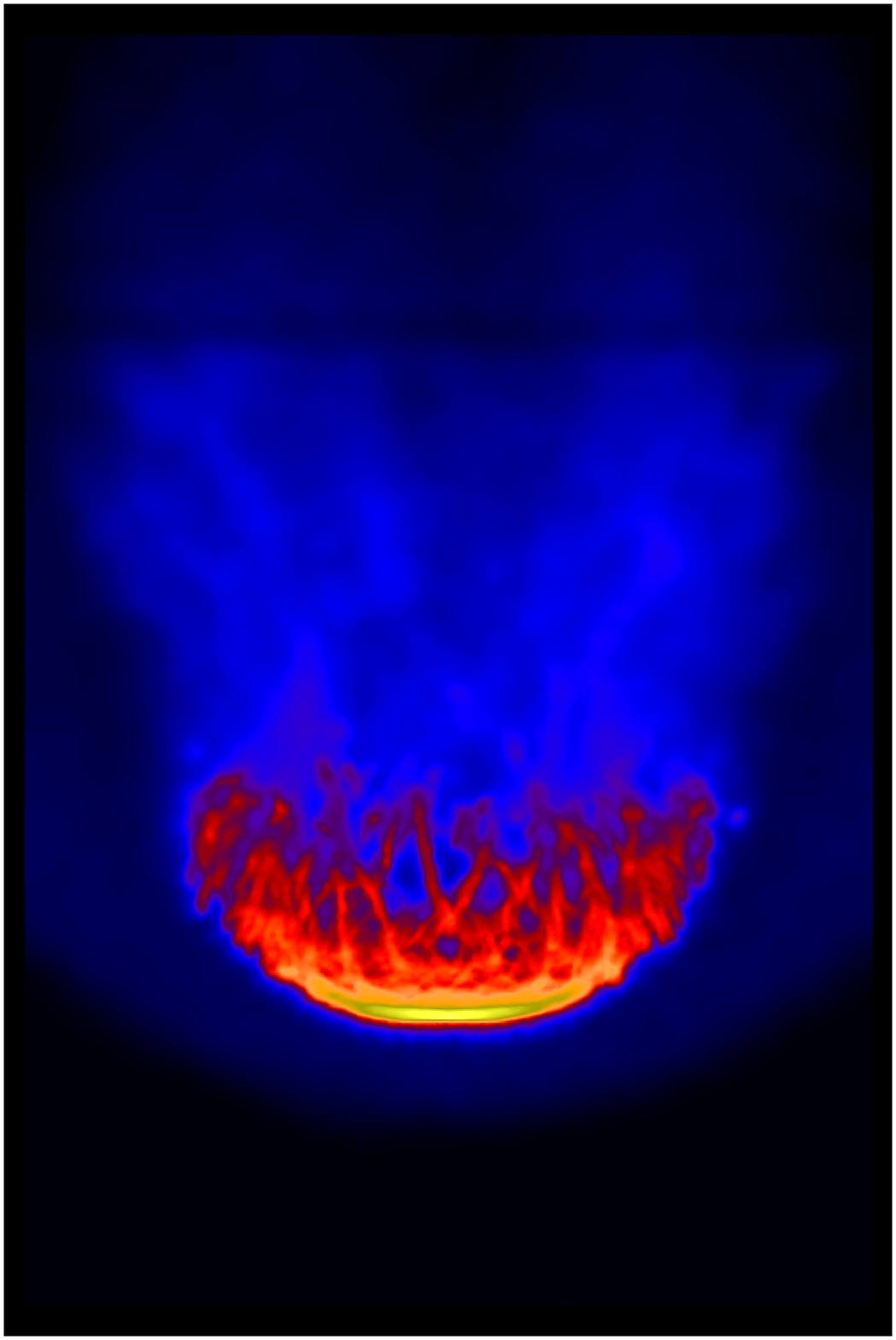}
\includegraphics[height=0.305\textheight]
{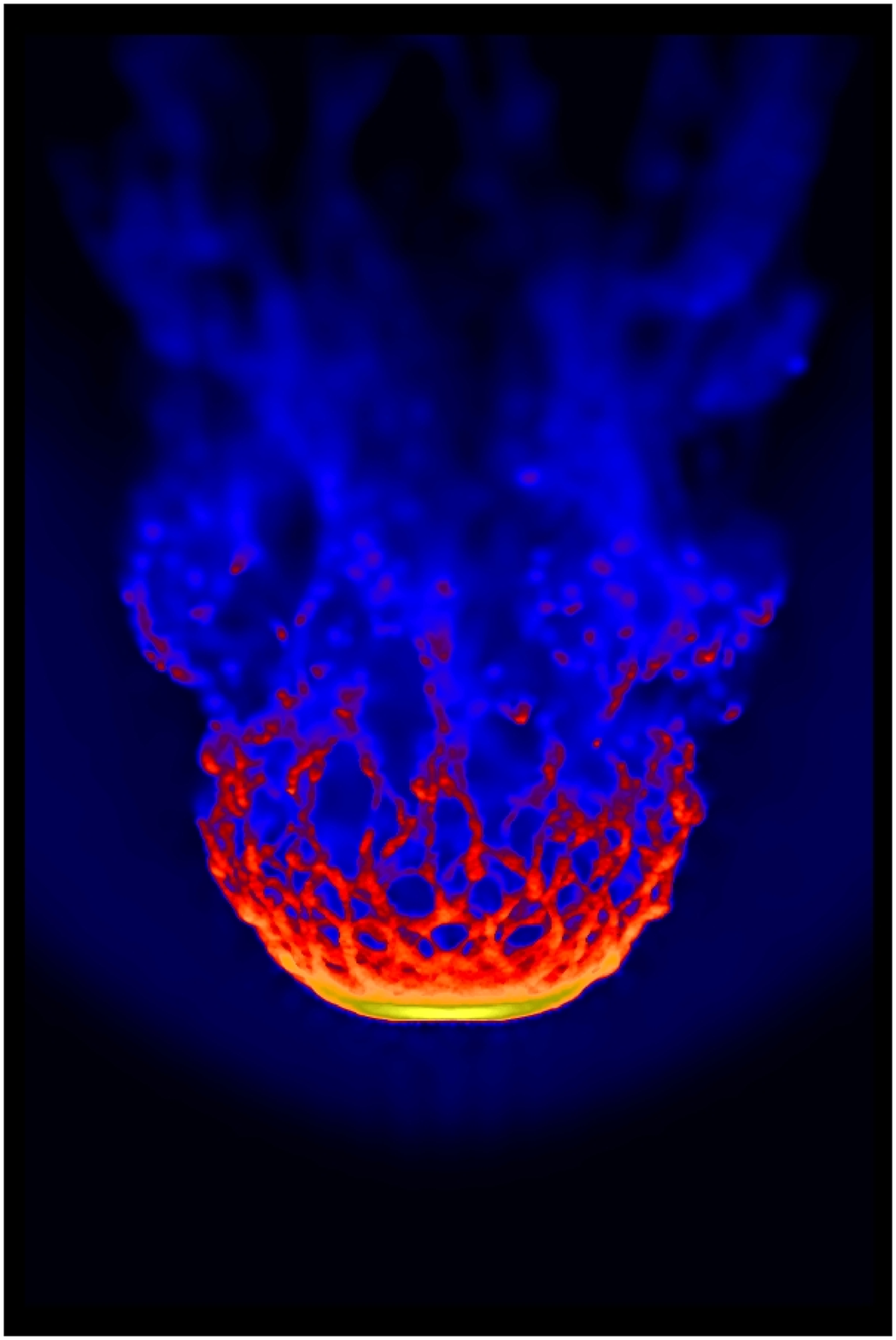}
\includegraphics[height=0.305\textheight]
{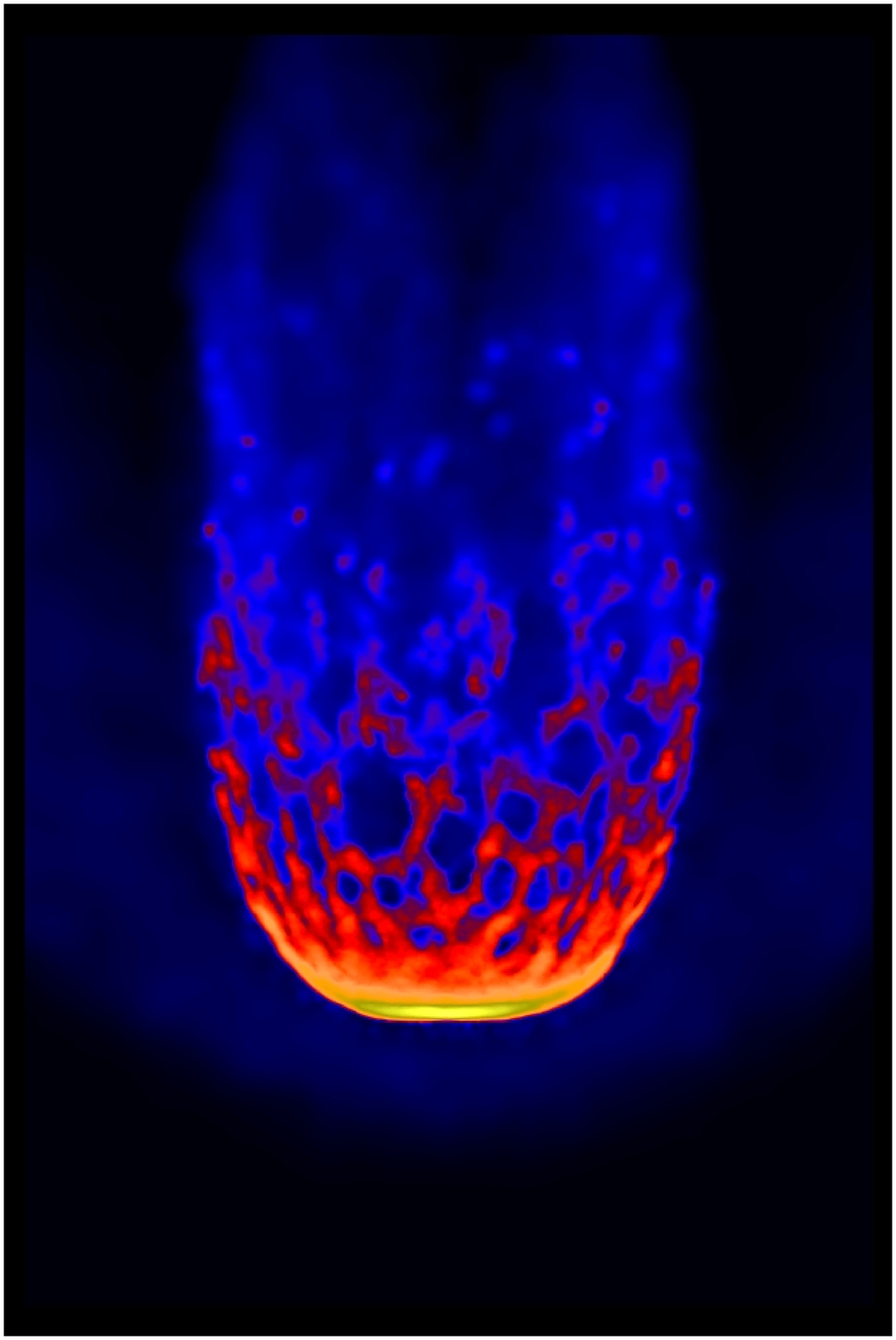}
\hspace{-3mm}
\includegraphics[height=0.305\textheight]
{plots/WindTunnel/maps/Dens_colorbar.eps}
\caption{Projected gas density maps of a galaxy (seen edge-on) that
  experiences a supersonic wind from below. The different panels show
  the logarithm of the density field at different times, from
  $T=0.04\,h^{-1} \mathrm{Gyr}$ (bottom row), $T=0.4\, h^{-1}\mathrm{Gyr}$
  (middle row) to $T=0.7\, h^{-1}\mathrm{Gyr}$ (top row), and for different
  numerical techniques, {\small AREPO} (left column), VPH (middle
  column) and SPH (right column). All maps are generated with the same
  adaptive smoothing algorithm and have an extension of $30\,h^{-1} {\rm
    kpc} \times 45\,h^{-1}{\rm kpc}$. }
\label{WT_maps}
\end{center}
\end{figure*}

We will nevertheless consider such simulations later on in
Section~\ref{Infall} to check the validity of our approach, but here
we first investigate an alternative, more idealised set-up, which
makes it much easier to reach an adequate resolution. In this
approach, we effectively put a galaxy model into a `wind tunnel',
i.e.~a rectangular box in which we let gas stream onto the galaxy with
prescribed density, velocity and temperature, matched to what we
expect during a cluster passage. The much smaller volume that needs to
be simulated around the galaxy in this situation drastically lowers
the computational expense, and even more importantly, the
simplification brought about by such a controlled set-up makes it much
easier to distinguish between different numerical effects.

\subsection{Initial conditions}

For definiteness, we put our model galaxy into a rectangular box of
side-length $80\,h^{-1} \mathrm{kpc}$. At one side of the box, gas is
constantly injected, which we realise in the case of SPH and VPH by
creating new particles in a suitable fashion, mimicking an infinitely
extended grid of particles that moves into the simulation domain. On
the opposite side of the box, we effectively implement outflow
boundary conditions by removing particles once they start moving out of
the box. Since we here focus on supersonic winds, the removal of
particles at the outflow side does not lead to the propagation of
perturbations upstream to the inflowing gas. In the case of the
{\small AREPO} code, the inflow and outflow regions are realised in an
equivalent, yet technically different fashion. We here use the
on-the-fly refinement and derefinement features of the {\small AREPO}
code to produce new mesh-generating points in the inflow region, and
to remove them near the outflow side. The net result is identical to
the SPH and VPH cases, i.e.~the galaxy at the centre of the box is hit
by a wind of particles/cells with prescribed density and
velocity. This impinging wind has the same gas mass resolution as our
target galaxy model.

For all the boundaries of the box that form the enclosing sides of our
wind tunnel, we adopt periodic boundary conditions for the gas as far
as the hydrodynamics is concerned. However, periodic boundary
conditions are not imposed for self-gravity, which we fully include
(note that our galaxy is held together by gravity).  Since the mass of
the gas in the wind is quite small, most of it is unbound to the
galaxy, and our restricted treatment of self-gravity to the region of
the box is still a good approximation. We note that a number of
similar wind tunnel experiments have previously been carried out
\citep[e.g.][]{Agertz, Iapichino2008}, but without a treatment of self
gravity. While our setup is not capable of accounting for
the tidal forces that arise when an extended galaxy travels through
the gravitational potential of a cluster, these forces can be expected
to be sub-dominant compared to the hydrodynamical forces in the outer
parts of a cluster, which is the region we are most interested in.
However, gravitational forces between dark matter, stars and gas
inside the galaxy are very important, and our approach is the first
that allows a detailed study of the response of the gravitationally
coupled disk-halo system to the ram-pressure of the impinging wind. In
particular, this allows a realistic measurement of the acceleration of
the whole galaxy due to the ram-pressure force exerted by the wind.

In our default set-up, we have chosen a constant density of $\rho=1.90
\times 10^{27}\,{\rm g\, cm^{-3}}$ and a velocity of $v=3000\,{\rm km
  \, s^{-1}}$ for the wind, letting it flow onto the galaxy in a
face-on orientation. A density and velocity of this magnitude would be
typical close to apocentre for a galaxy that has fallen into a large
cluster. Of course, in reality the strength of the wind will be a
function of the orbital phase and of parameters such as cluster size
and angular momentum of the galaxy orbit. Changing these values should
however not change our results at a qualitative level. For the galaxy
model studied in our simulations, we have adopted structural
properties identical to those described in
Section~\ref{IsolatedGal}. We place the galaxy suddenly into the
supersonic wind, refraining from any attempt to impose a more gradual
start-up in which the wind would somehow be slowly turned on.

\begin{figure}
\begin{center}
\hspace{-8mm}
\includegraphics[width=0.5\textwidth]
{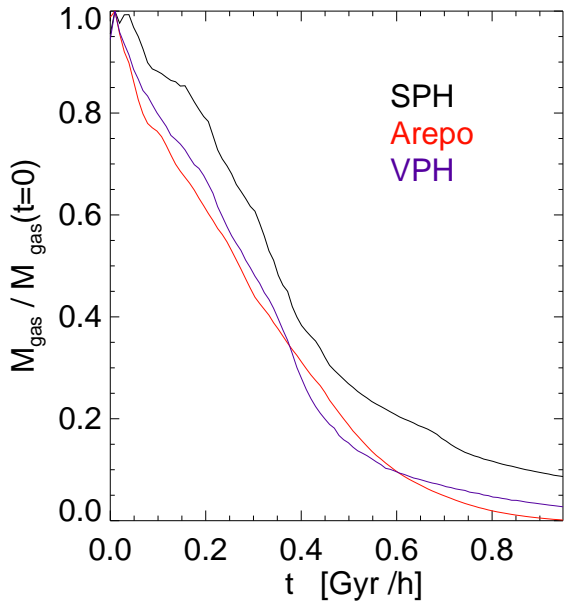}
\vspace{-5mm}
\caption{Loss of star-forming ISM gas as a function of time due to
  stripping through a supersonic wind.  The gas remaining in the ISM
  of the galaxy is here measured according to the criteria of
  Eq.~(\ref{partofcloud}). The three
  solid lines show the measurements for our simulations with SPH, VPH
  and {\small AREPO}, as labelled. We find that the stripping is quite
  similar in VPH and {\small AREPO}, but significantly less efficient
  in SPH.   }
\label{WTgasloss}
\end{center}
\end{figure}

\begin{figure}
\begin{center}
\includegraphics[width=0.11\textheight]{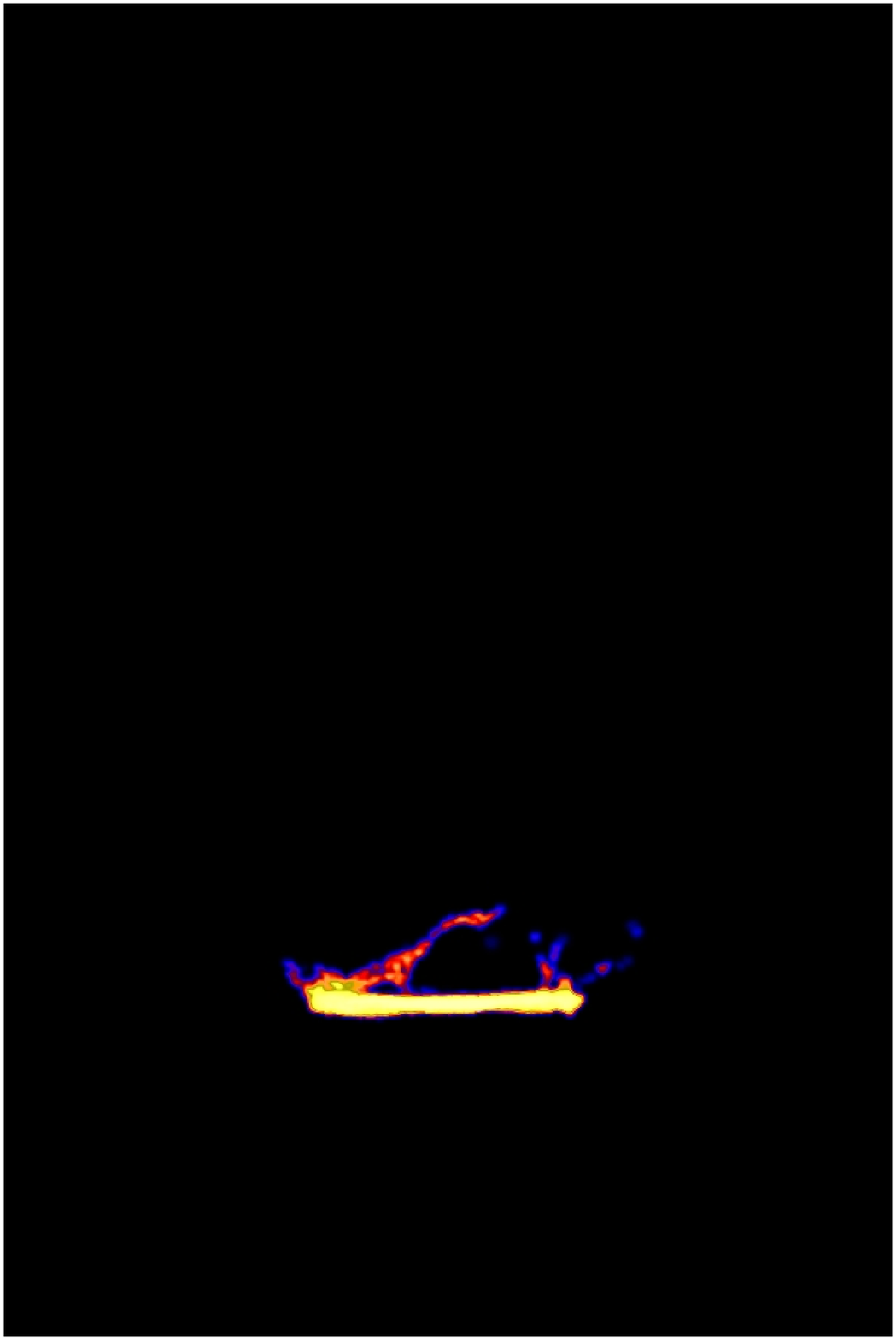}%
\hspace{1mm}%
\includegraphics[width=0.11\textheight]{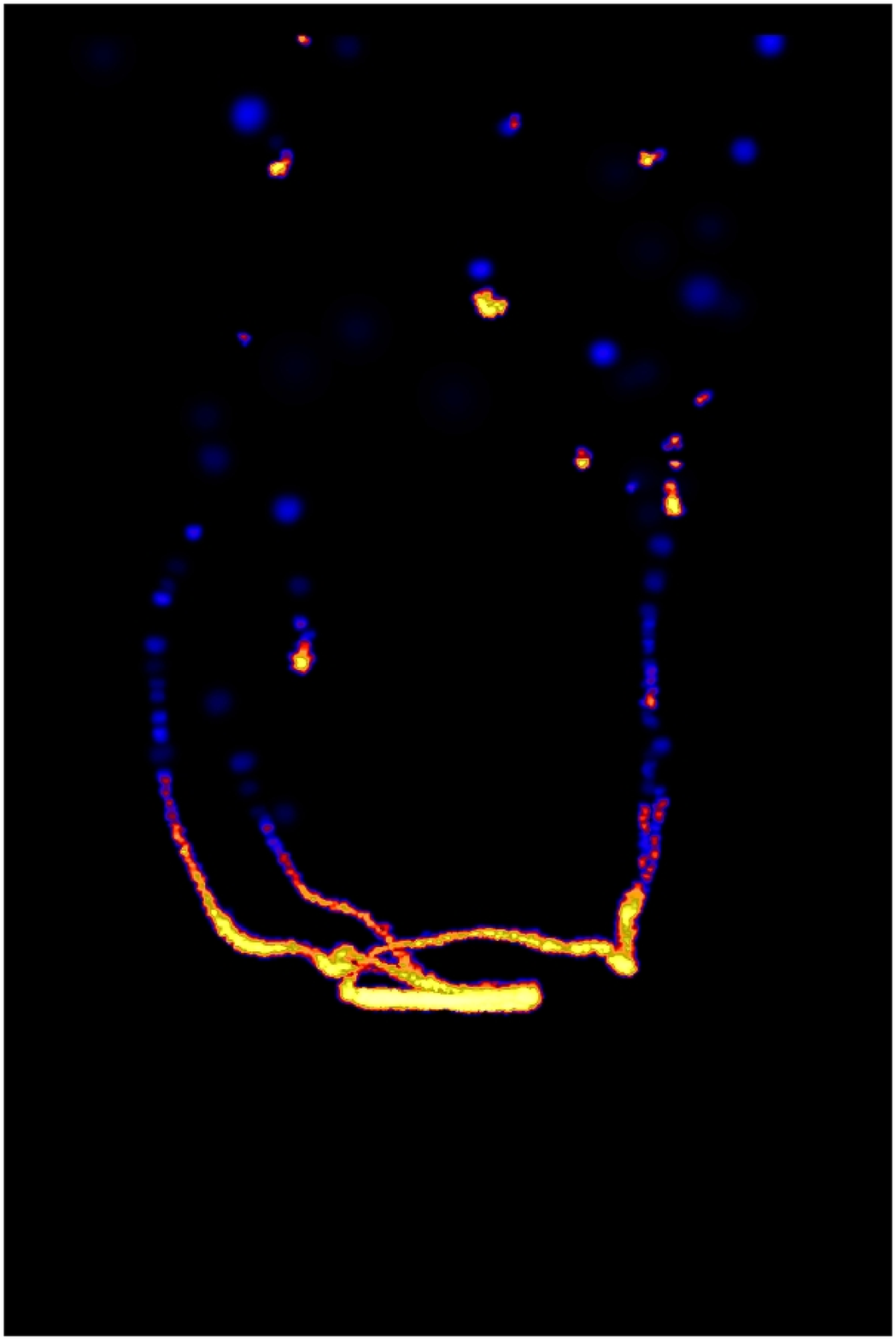}%
\hspace{1mm}%
\includegraphics[width=0.11\textheight]{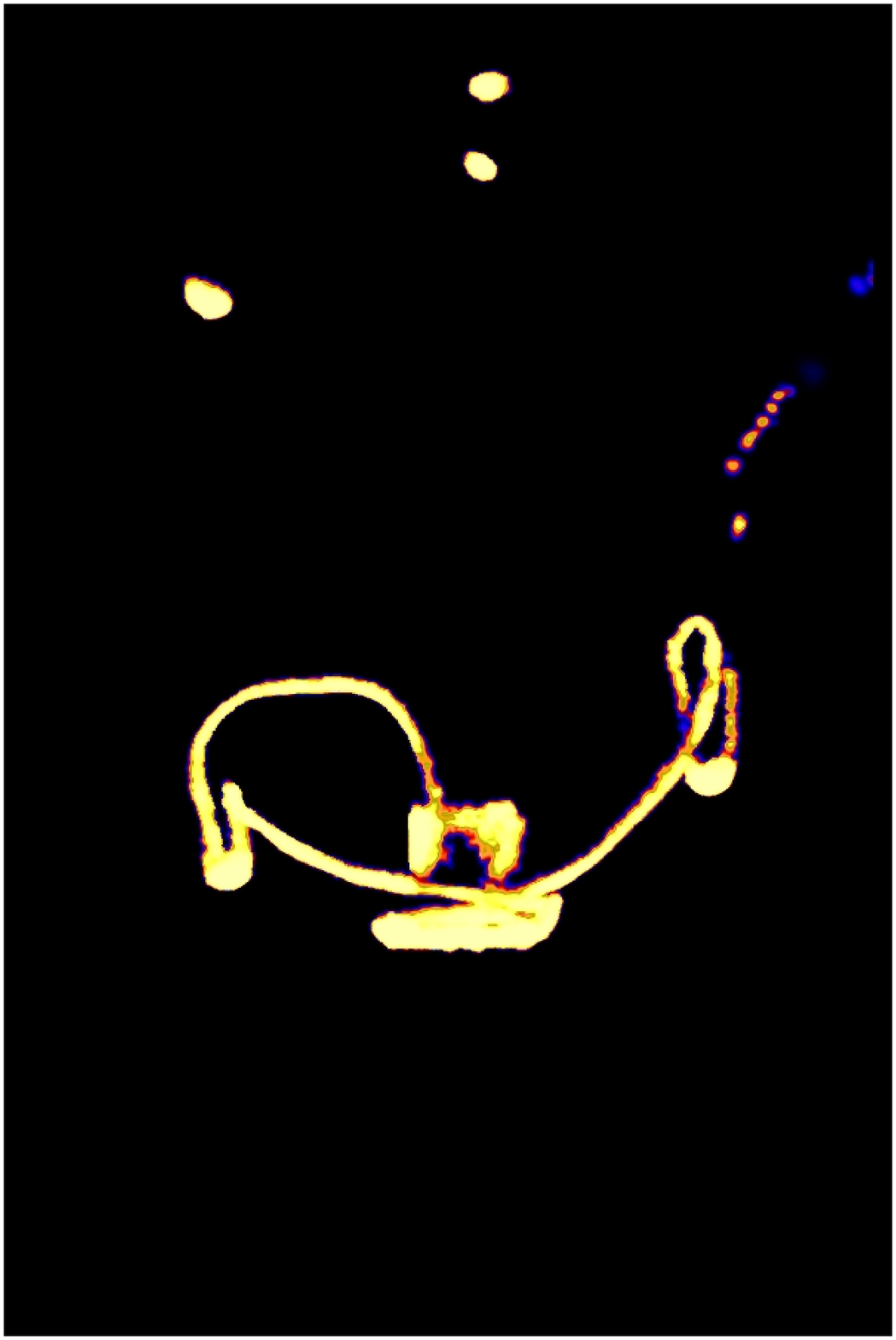}
\caption{Maps of the projected star formation rate density in a galaxy
  model (seen edge-on) that encounters an IGM wind in the upwards
  direction. We show results at $T=0.4\,h^{-1} \mathrm{Gyr}$ for our wind
  tunnel experiments, carried out with {\small AREPO} (left), VPH
  (middle) and SPH (right). Each panel extends over $36\,h^{-1}{\rm kpc}
  \times 48\,h^{-1}{\rm kpc}$.
\label{WT_StarFmaps}}
\end{center}
\end{figure}

\subsection{Stripping}

Hydrodynamically, the galaxy represents an obstacle in the supersonic
wind, causing a bow shock in the upstream region ahead of the
galaxy. Inside the bow shock and ahead of the front side of the
galaxy, the compressed wind is slowed down and exerts significant
ram-pressure onto the galaxy \citep{GunnGott}, whereas around the
sides, a region of strong shear flow is produced as the wind streams
around the galaxy. In this region, the formation of Kelvin-Helmholtz
instabilities is expected, which accelerate the stripping of gas from
the galaxy's disk and mix it with the downstream flow. 

\citet{Agertz} has previously investigated the simpler case of a
spherical, non-self-gravitating gaseous obstacle (the `blob-test') and
found substantial discrepancies between different hydrodynamical
methods for the rate of gas stripping and the time until eventual
complete disruption of the gas blob.  Here, we are particularly
interested to see whether such differences also occur between SPH, VPH
and {\small AREPO} when the more realistic disk-wind setup we are
investigating is considered.

In Figure~\ref{WT_maps}, we show projected gas density maps of the
galaxy in the wind tunnel in an edge-on orientation, with time
evolving from bottom to top. While the morphology of the stripping
process looks similar at early times in all three codes considered,
significant differences develop over time. Both of the particle-based
methods loose small clumps of gas and exhibit string-like features of
quite dense gas. This effect is particularly strong in SPH. In
contrast, the {\small AREPO} disk stays comparatively coherent, with
all of the stripped gas moving to lower densities quickly. Hence there
remains no dense debris in the downstream flow.  At the latest time
shown in the figure (top row), the residual gas disks of {\small
  AREPO} and VPH are visually significantly smaller than in SPH,
suggesting that more gas has been stripped.

\begin{figure}
\begin{center}
\hspace{-8mm}
\includegraphics[width=0.5\textwidth]
{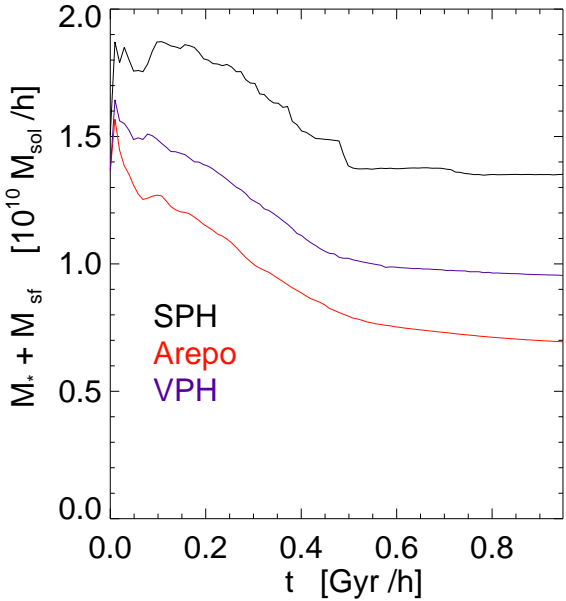}
\vspace{-5mm}
\caption{Time evolution of the sum of the mass of star-forming gas and
  the mass of all stars formed in the whole simulation box since the
beginning of the simulations
  of a galaxy exposed to an impinging IGM wind in a wind tunnel. We
  show results for different simulation techniques, including an
  SPH-based simulation (black solid), VPH (blue solid), and {\small
    AREPO} (red).  }
\label{WTsform_plus}
\end{center}
\end{figure}

\begin{figure}
\begin{center}
\includegraphics[height=0.20\textheight]
{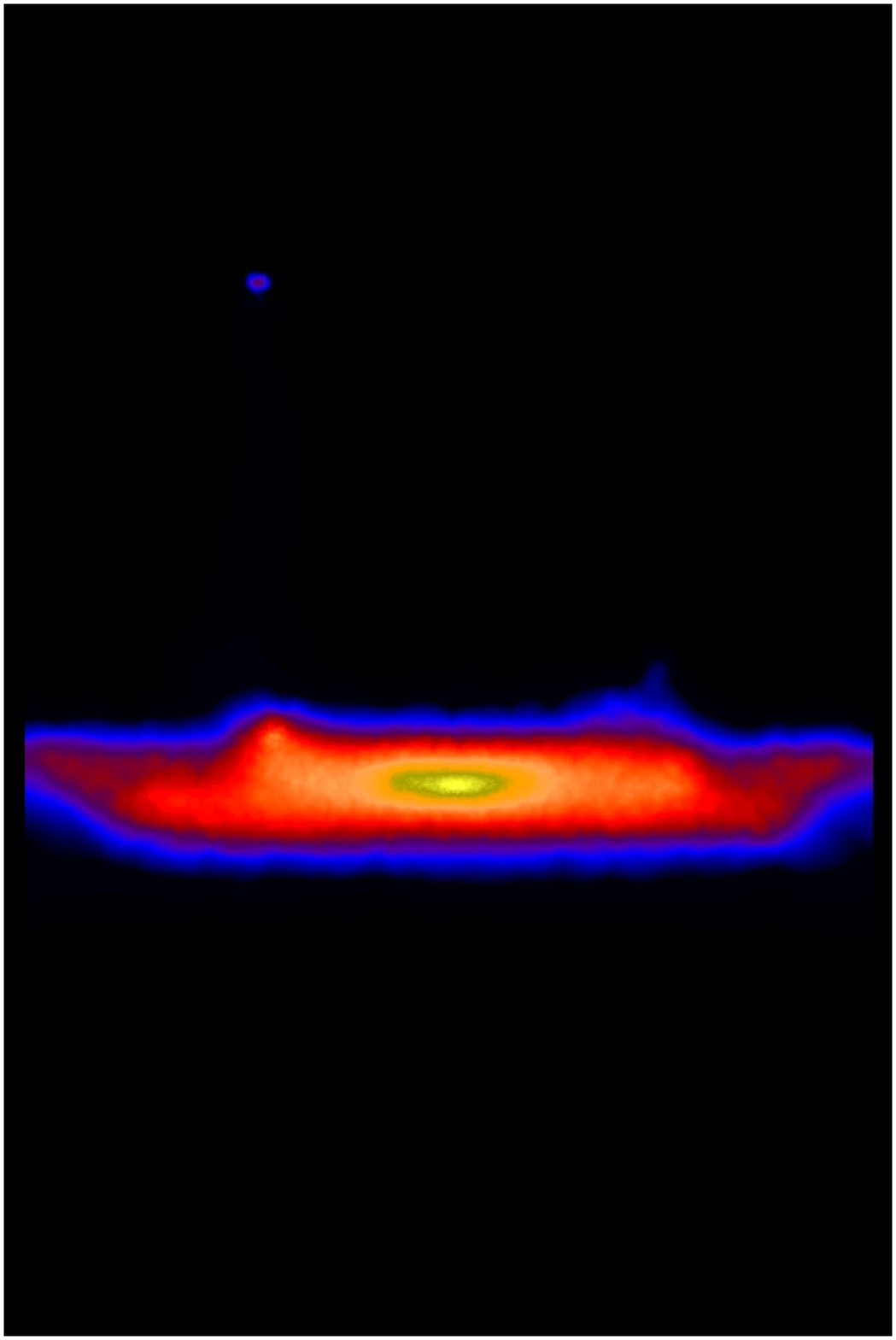}%
\hspace{1mm}%
\includegraphics[height=0.20\textheight]
{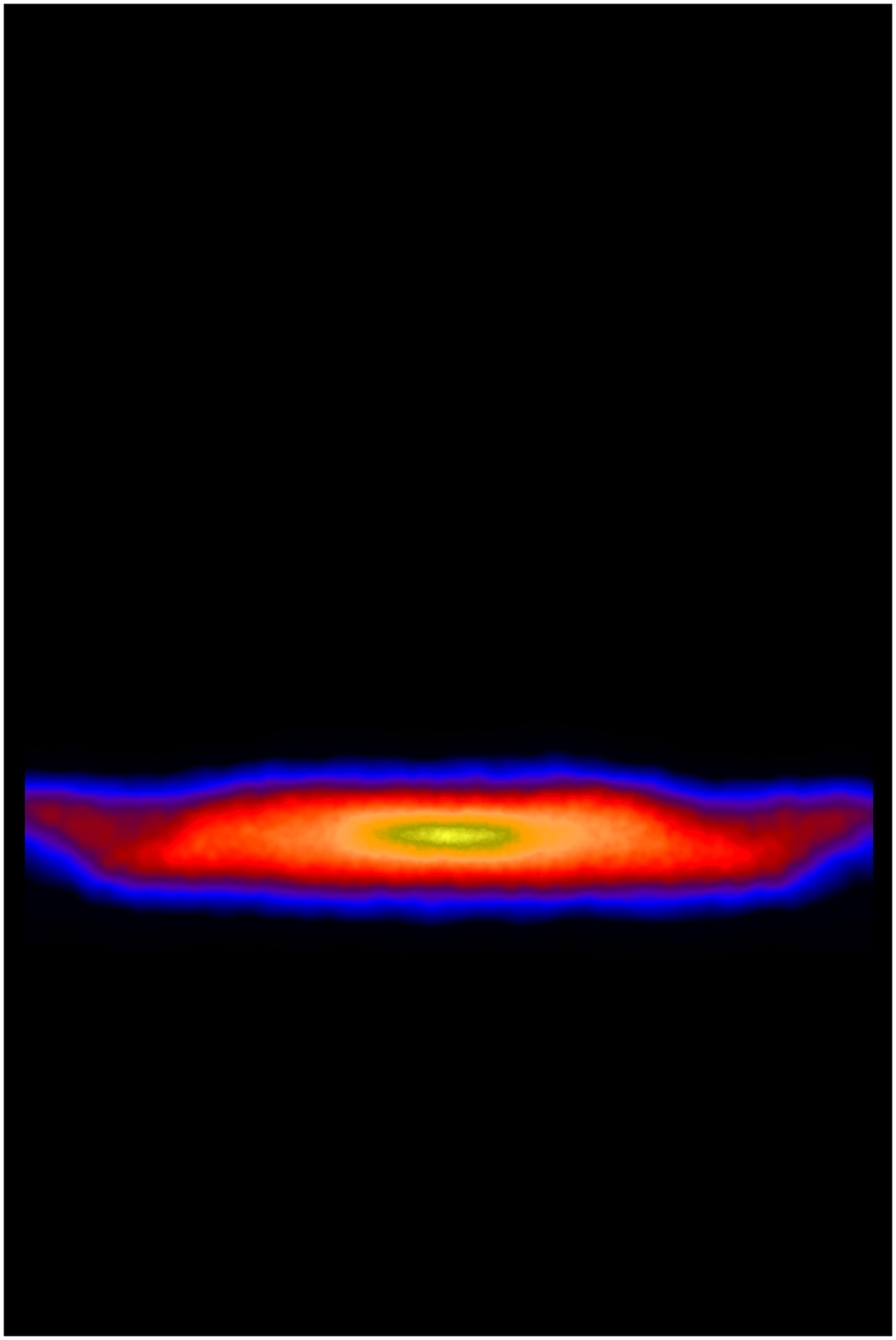}%
\hspace{1mm}%
\includegraphics[height=0.20\textheight]
{plots/WindTunnel/maps/Dens_colorbar.eps}
\caption{Stellar density maps of model galaxies (seen edge on) exposed
  to a supersonic IGM wind, at $T=0.4\,h^{-1}\mathrm{Gyr}$. The two panels
  show results for SPH (left) and VPH (right), respectively.
  Interestingly, the SPH galaxy is accelerated more strongly in the
  downstream direction, as a result of a larger effective ram pressure force due
  to slower stripping of its gaseous disk.  The SPH simulation also
  shows signatures of the wind imprinted onto the morphology of the
  stellar disk, a behaviour that is not seen in VPH or {\small
    AREPO}. The two maps show the same spatial region of the wind
  tunnel and have an extension of $30\,h^{-1}{\rm kpc} \times 45\,h^{-1}{\rm
    kpc}$.  }
\label{WT_StellarDiskmaps}
\end{center}
\end{figure}

This impression is born out by a quantitative measurement of the gas
mass that is still in the ISM phase of the galaxy model.
We consider a gas particle (or cell) to be part of the ISM of the galaxy when
it is sufficiently dense and cold. Specifically, we require the
 conditions
\begin{equation}
	\rho > 2000 \, \rho_{\mathrm{wind}}  	\quad \mbox{and} \quad 
	T < 0.5  \, T_{\mathrm{wind}}
\label{partofcloud} 
\end{equation}
to be met, where $\rho_{\rm wind}$ and $T_{\rm wind}$ are the density
and temperature of the wind.  Furthermore, we impose an additional
condition on the maximum allowed separation $r$ of a particle from the
centre-of-mass of all gas that fulfils the criterion of
Eq.~(\ref{partofcloud}). Only for $ r < 8 \, \mathrm{kpc}$ the gas
particle is considered to be part of the ISM of the primary
galaxy. This condition is introduced in order to discard cold clumps
of gas that have been stripped out of the galaxy but have still stayed
cold and dense. Indeed, it turns out that this secondary criterion is
quite important in our SPH simulations, where the galactic disk tends
to shed dense clumps under the action of the ram pressure. These
fragments can still fulfil Eqn.~(\ref{partofcloud}) despite being
stripped, but as they separate from the main galaxy, they eventually
violate the distance condition and are then counted as stripped gas.

In Figure~\ref{WTgasloss} we show the ratio of the remaining ISM gas
mass to the initial mass a function of time, comparing wind tunnel
simulations carried out with the three different numerical
techniques. Clearly, the mass loss in SPH is smallest overall, with
about 10\% of the initial mass remaining after $1\, {\rm Gyr}$,
whereas at this time the ISM of the VPH and {\small AREPO} runs is
almost stripped completely. Note that this measure of the mass loss
counts most of the dense gas blobs that stay coherent in SPH as lost
gas. If we were to measure the amount of ``non-dispersed'' gas
instead, the difference would be considerably larger because the gas
that is stripped in SPH from the disk is largely not dispersed,
unlike in the other two approaches (see below). The relative
similarity of the overall shape of the gas mass evolution in
Figure~\ref{WTgasloss} is therefore not simply a time delay in SPH
relative to the other methods, even though at intermediate times the
mass loss rates are similar.

It is interesting to note that the stripped dense gas in the SPH
simulation, and to a lesser extent in the VPH run, can continue to
form stars at some level. This is particularly evident in
Figure~\ref{WT_StarFmaps}, which shows maps of the star formation rate
density in an edge-on projection, at time $t = 0.4\,h^{-1}{\rm Gyr}$.
This highlights that in the case of the SPH simulations the stripped
gas clumps are not only very concentrated, they also continue to form
stars.  This phenomenon is still present in VPH, albeit at a much
weaker level and with decreasing star formation rate over distance. In
contrast, {\small AREPO} does not show such a behaviour.

\begin{figure*}
\begin{center}
\hspace{-8mm}
\includegraphics[width=0.5\textwidth]
{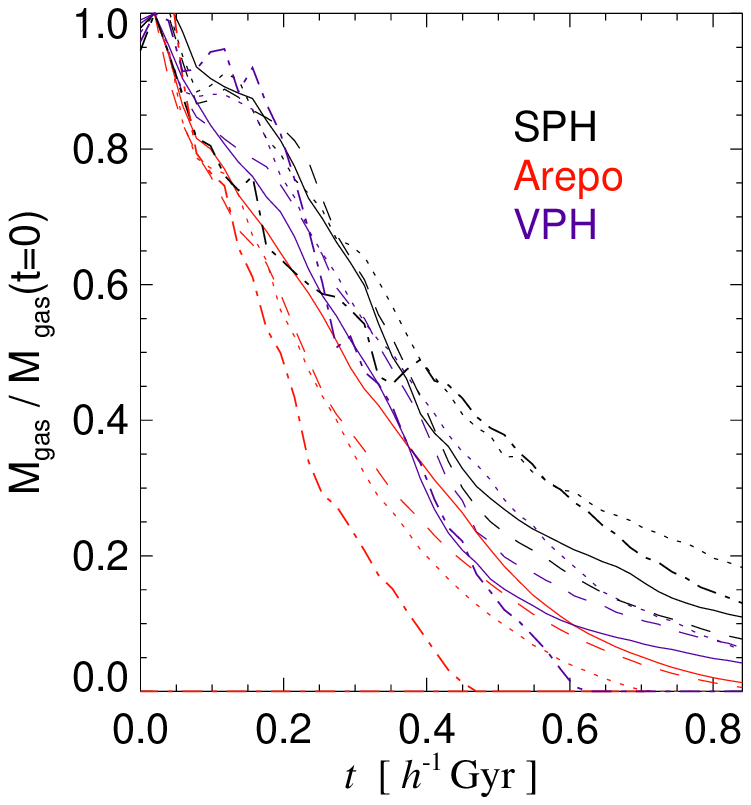}
\includegraphics[width=0.5\textwidth]
{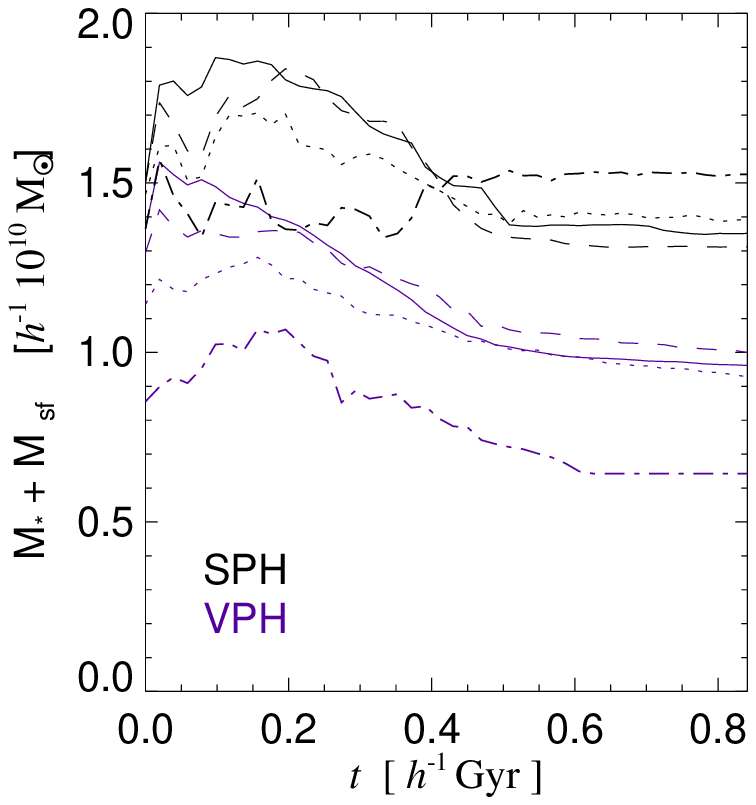}
\vspace{-2mm}
\caption{Resolution study of the mass loss of a galaxy in a wind
  tunnel. The left panel measures the gas remaining in the ISM
  (defined through Eq.~\ref{partofcloud}) as a function of time, for
  simulations carried out with VPH, SPH, and {\small AREPO}, as
  labelled. In each case, four resolutions are shown, corresponding
  to $4000$ gas particles (very low resolution, dotted-dashed),to
  $20000$ gas particles (low resolution, dotted), $80000$ gas
  particles
  (intermediate resolution, dashed), and $320000$ particles (high
  resolution, solid) in the initial galaxy model.  The panel on the
  right compares the sum of the remaining star forming gas plus the
  newly formed stellar mass in the whole simulation box as a function
of time in VPH and SPH,
  using the same simulation set. Here a quite good convergence can be
  observed for the different particle-based techniques, but SPH is
  offset relative to VPH and has a significantly smaller amount of
  stripped gas.}
\label{WTgasloss_RES}
\end{center}
\end{figure*}

In SPH, there are at least two effects that can help to explain the
enhanced star formation in the stripped clumps. First, it is known
that spurious surface tension forces in SPH exist that will slightly
compress the gas of the clumps and keep them coherent.  Second, there
is the issue of a potential artificial suppression of gas mixing in
the turbulent wake behind the galaxy.  Both SPH and VPH, by
construction, prevent that the low entropy gas of the ICM is mixed at
the particle level with the high entropy gas of the impacting
wind. This imposed Lagrangian behaviour ignores any small-scale mixing
processes that could happen on scales below the spatial resolution
limit.  In contrast, the mesh-based approach of {\small AREPO}
implicitly allows for such processes. Here, the mixing manifests
itself through mass exchanges between the cells, giving rise to a
much more diffuse wake behind the galaxy (left column in
Figure~\ref{WT_maps}) where no gas remains that is still dense enough
to support star formation.  In the mesh-based approach, one may have
in fact an opposite problem compared to the particle schemes. Here the
need to advect the fluid over mesh interfaces can easily lead to
excessive numerical mixing. However, the moving-mesh technique of
{\small AREPO} should reduce such errors significantly compared to
more traditional Eulerian methods.

Another view on the different behaviour with respect to
stripping is provided by Figure~\ref{WTsform_plus}, which shows the
time evolution of the sum of the star forming gas mass (which is just
the total gas mass above the star formation threshold) plus the
stellar mass formed since the beginning of the simulation.
Interestingly, this quantity is nearly constant in time in the case of
SPH, showing that essentially none of the initially dense and star
forming gas is dragged to low enough density to stop star formation.
This is consistent with our earlier findings, but also makes it
quantitatively evident that essentially no mixing of this low entropy
gas with higher entropy material occurs. In contrast, the sum of star
forming gas mass plus stellar mass declines significantly in {\small
  AREPO}, by more than a factor of 2 over the timescale of
$1\,h^{-1}\mathrm{Gyr}$ that is shown here. The results for VPH
are intermediate,
showing some evidence for gas moving to higher entropy or lower
densities during the stripping process and thus reducing the amount of
total gas available for star formation.

Interestingly, the discrepancies in the stripping rates also lead to
different accelerations of the whole galaxies, because they experience
unequal ram pressure forces as a result of different effective areas
of the remaining gas disks.  Since the gas in the disk is
gravitationally coupled to the stellar disk and the dark matter halo,
it is important to note that the wind not only strips the gas
component of the galaxy, it also accelerates the {\em whole}
galaxy. In Figure~\ref{WT_StellarDiskmaps}, we show maps of the
projected stellar densities of the stellar disks in SPH and VPH at an
identical time, corresponding to $T=0.4\,h^{-1} \mathrm{Gyr}$.
Clearly, the {\em whole} SPH galaxy has been pushed more into the
downstream direction, demonstrating that it has experienced a higher
effective ram pressure, as a consequence of the reduced stripping rate
of its disk. In real galaxy clusters, this will in turn lead to a
larger dynamical friction force and a somewhat faster decay of the
orbit of the galaxy. An additional effect is that in the SPH case we
see small features emanating in the downstream direction from the
stellar disk. Figure~\ref{WT_StellarDiskmaps} shows two small humps in
the stellar disk, as well as a faint stellar blob at some distance
further downstream. These are in fact stars of the original stellar
disk that have been gravitationally pulled out of the disk by dense
gas. Both effects are absent in the equivalent VPH and {\small AREPO}
calculations, emphasising again how much more (artificial)
``coherence'' the dense gas phase exhibits in the case of SPH, where
comparatively massive gas clumps are removed from the disk that stay
coherent afterwards.

\subsection{Dependence on resolution and artificial viscosity}

\begin{table}
\begin{tabular}{ l | c | r |  r  }
\hline  
Code & resolutions & soft. baryons & softening DM \\
  \hline   
VPH 	& 4k, 20k, 80k, 320k	& $0.05 \; h^{-1} \rm{kpc}$	& $0.1
\; h^{-1} \rm{kpc}$\\
SPH 	& 4k, 20k, 80k, 320k	& $0.05 \; h^{-1} \rm{kpc}$	& $0.1
\; h^{-1} \rm{kpc}$\\
{\small AREPO}	& 4k, 20k, 80k, 320k	& $0.05 \; h^{-1} \rm{kpc}$
& $0.1 \; h^{-1}
\rm{kpc}$\\
  \hline  
\end{tabular}	
\caption{Gravitational softening lengths for simulations in
section~\ref{WindTunnel}. The column ``resolutions'' denotes the
initial number of gas particles in the disk where ``k'' denotes one
thousand. The following columns show the gravitational softening
lengths for baryons, that is for gas and
stellar particles. The column on the right gives these values
for the collisionless DM halo particles.}
\label{WT_SoftTable}
\end{table}

In order to examine the quantitative robustness of our wind tunnel
results with respect to numerical resolution, we repeated our
simulations with SPH, VPH and {\small AREPO} using both a lower and a
higher resolution than in our default runs discussed thus far, where
the galaxy is resolved with $80000$ gas particles in the initial gas
disk.  With respect to our default resolution, our low resolution
simulations have 4 times fewer particles in each component, whereas
the high resolution runs have 4 times mores particles, as denoted in
Table~\ref{WT_SoftTable}.  We here use gravitational softening lengths
independent of resolution.

\begin{figure}
\begin{center}
\hspace{-8mm}
\includegraphics[width=0.5\textwidth]
{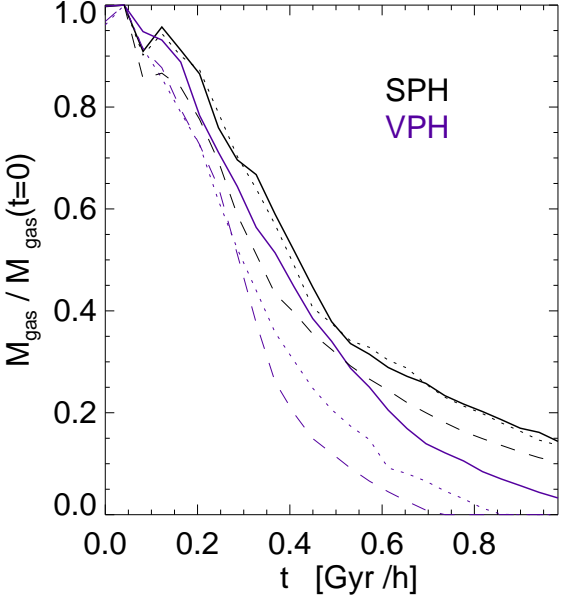}
\vspace{-5mm}
\caption{Dependence of the wind tunnel stripping efficiency in VPH (blue
lines) and SPH (black lines) on the artificial viscosity. We
  show the loss of star-forming gas mass (defined as in
  Eq.~\ref{partofcloud}) of our default galaxy model as a function of
  time, for high artificial viscosity $\alpha=1$ (solid lines),
  for intermediate strength $\alpha=0.5$ (dotted lines) and for
  low artificial viscosity $\alpha=0.25$ (dashed lines).}
\label{WT_AV}
\end{center}
\end{figure}

\begin{figure}
\begin{center}
\hspace{-8mm}
\includegraphics[width=0.5\textwidth]
{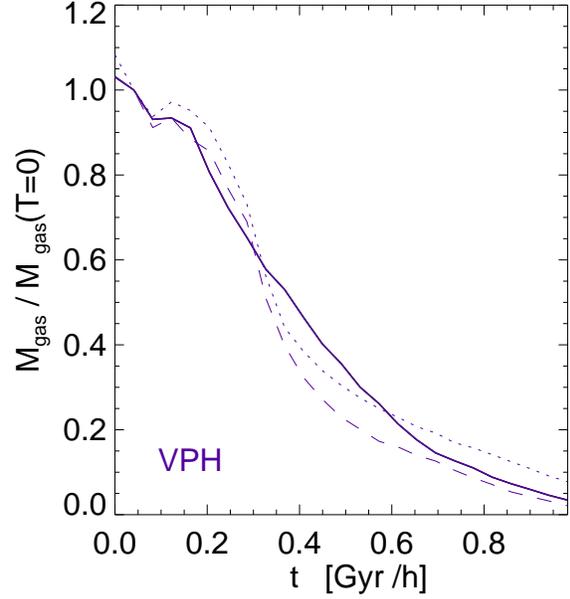}
\vspace{-5mm}
\caption{Dependence of the stripping efficiency in VPH on the strength
  of mesh shape correction forces, which consist of small,
  non-dissipative forces that tend to steer mesh cells towards a
  `rounder' shape in case they have become highly asymmetric. We show
  the loss of star-forming gas mass (defined as in
  Eq.~\ref{partofcloud}) of our default galaxy model as a function of
  time, for our standard choice of $\beta_0=0.2, \beta_1=0.01$ (solid
  line) for parametrisation the shape correction forces. We also show
  results for two different settings, for $\beta_0=0.6, \beta_1=0.03$
  (dashed line) and for the somewhat extreme choice of $\beta_0=1.8,
  \beta_1=0.09$ (dotted line), respectively.  }
\label{WT_SHP}
\end{center}
\end{figure}

In Figure~\ref{WTgasloss_RES}, we show results of our resolution study
for the gas stripping out of the ISM as a function of time, and for
the sum of remaining star forming gas and newly formed stellar mass.
We find that the general trends remain very similar at all three
resolution levels, i.e., gas is stripped the fastest in {\small AREPO}
and slowest in SPH, with VPH taking on an intermediate
role. Interestingly, in the two particle codes, the stripping proceeds
more slowly when the resolution is poorer, whereas in {\small AREPO}
the opposite trend is observed, here the stripping tends to accelerate
slightly for better resolution. We hence find that the results become
more similar for better resolution. In particular, in the high
resolution case, VPH is in fact quite close to {\small AREPO}.  When
the sum of the star forming gas mass and the newly formed stars is
considered, we see that this quantity is very robust for SPH and VPH
as a function of resolution, with the offset probably simply
reflecting different systematics related to the density estimation at
the interface between gas disk and surrounding medium, as in
Figure~\ref{Iso_denshist}.  On the other hand, the relative constancy
of this quantity at late times reflects the inherent inability of
these particle schemes to mix low entropy gas with gas of much higher
entropy.

\begin{figure*}
\begin{center}
\includegraphics[width=0.4\textwidth]
{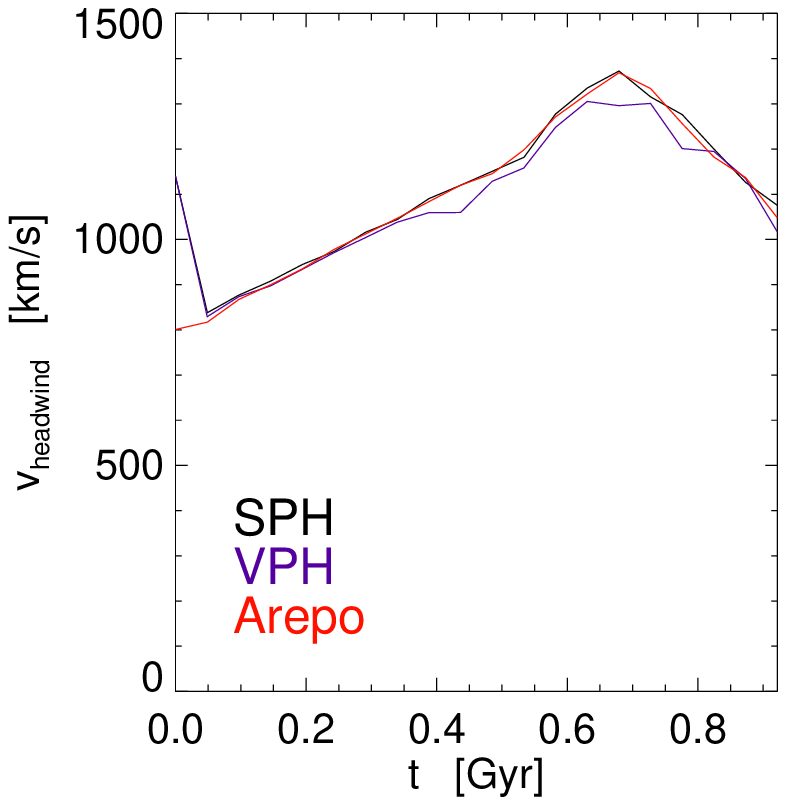}
\includegraphics[width=0.4\textwidth]
{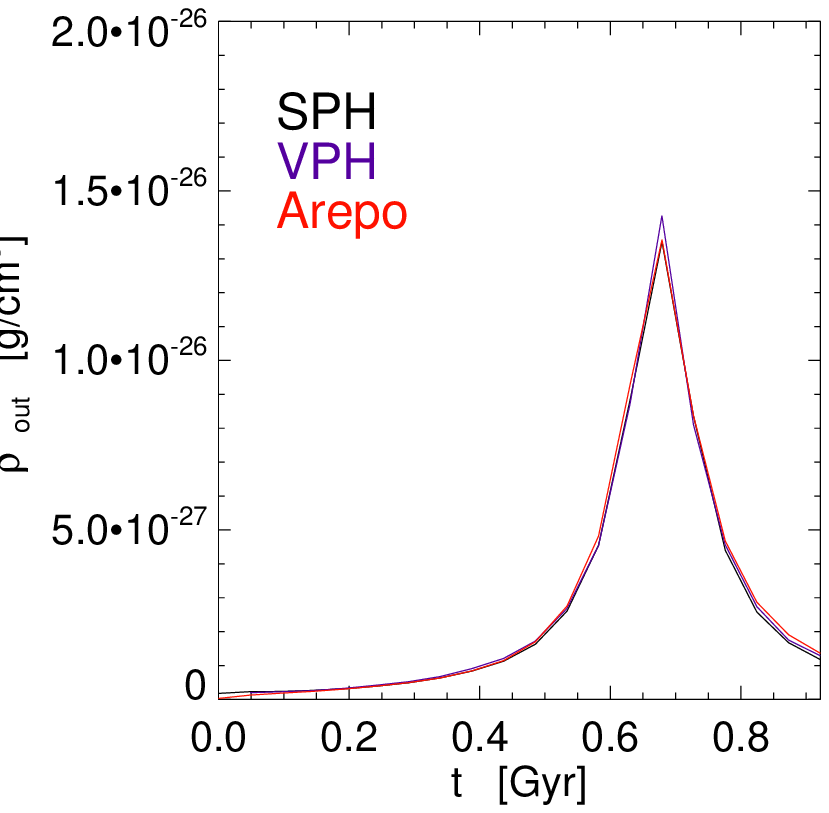}\\
\vspace{-4mm}
\includegraphics[width=0.4\textwidth]
{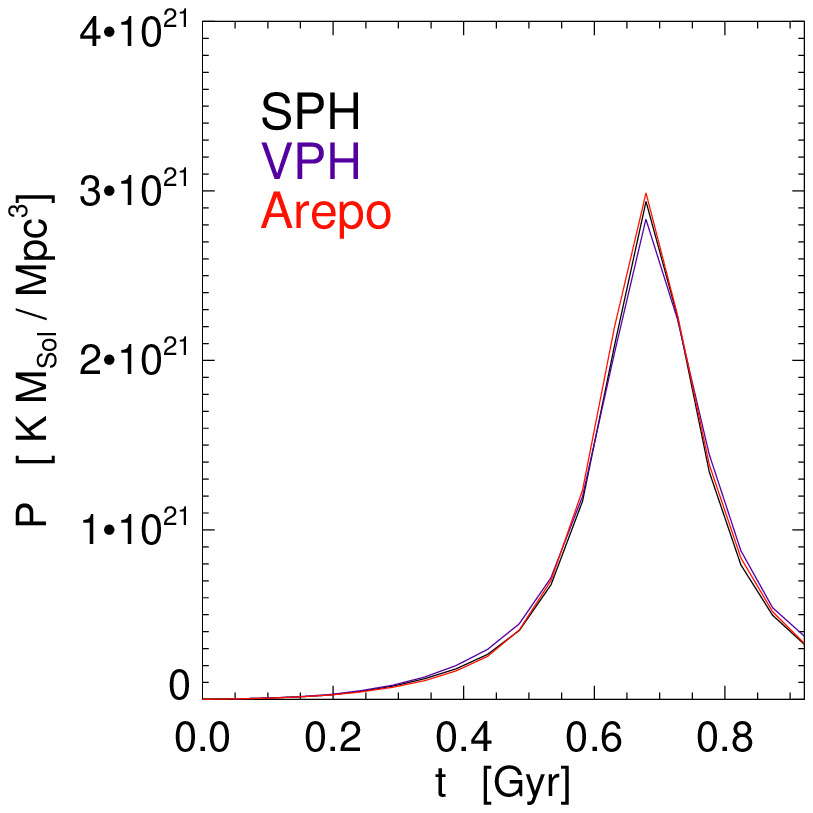}%
\includegraphics[width=0.4\textwidth]
{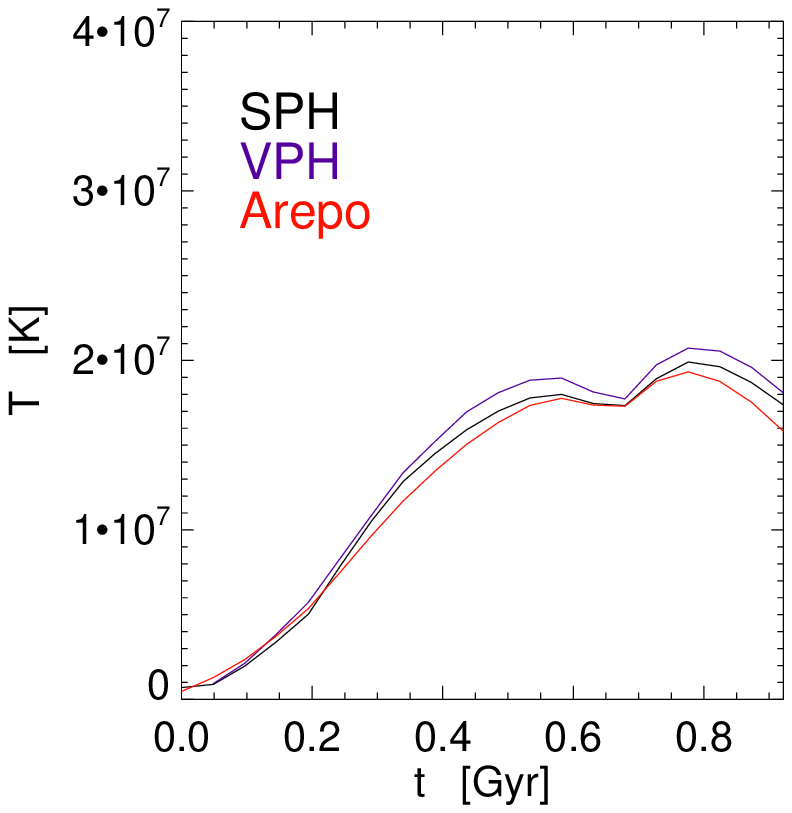}
\vspace{-4mm}
\caption{Properties of the `headwind' encountered by the in-falling
  galaxy.  The velocity difference is calculated between the
  centre-of-mass of the gas in the galaxy (defined according to
  Eqn.~\ref{crit_1stpart}) and the environment, which is taken as a
  sector of a spherical shell with an opening angle of $30^{\rm o}$ in
  the direction of the velocity vector and at a distance of $[45,
    99]\,h^{-1}\rm{kpc}$ with respect to the galaxy centre. All simulations
  have been conducted with cooling and star formation (the cited
masses and times carry a factor of $h^{-1}$ in their units). In each
panel,
  we show results for simulations with VPH (blue), SPH (black), and
  {\small AREPO} (red), comparing from top left to bottom right the
  relative velocity, density, pressure and temperature of the wind
  encountered by the galaxy as a function of time.  }
\label{GalStr_headwind}
\end{center}
\end{figure*}

\begin{figure*}
\begin{center}
\hspace{-12mm}
\includegraphics[width=0.30\textwidth]
{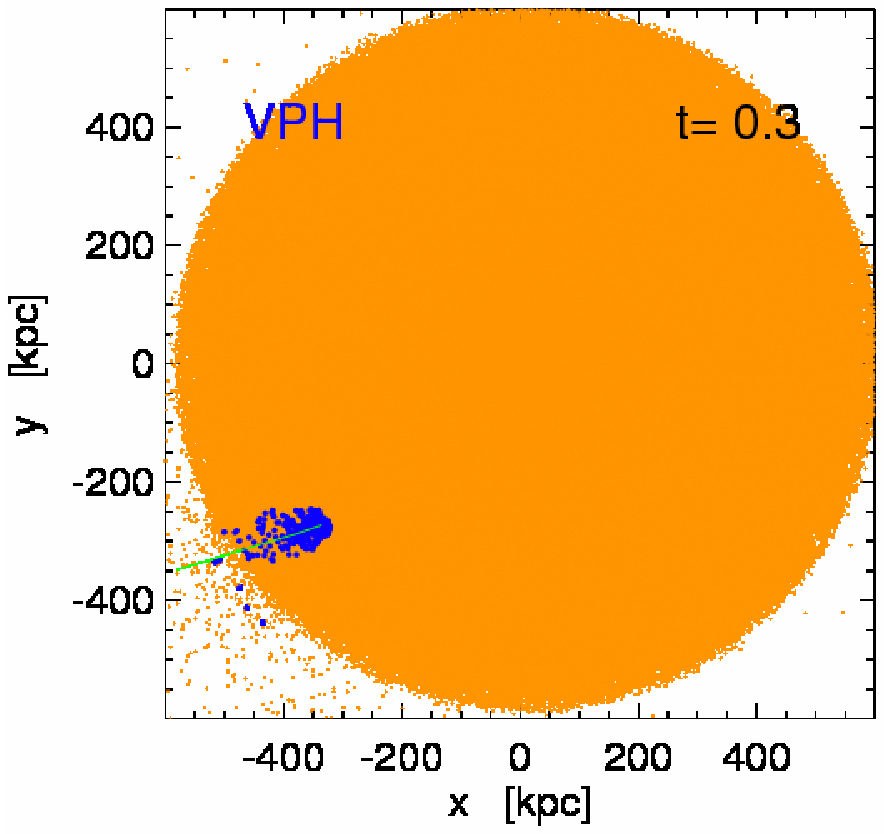}
\includegraphics[width=0.30\textwidth]
{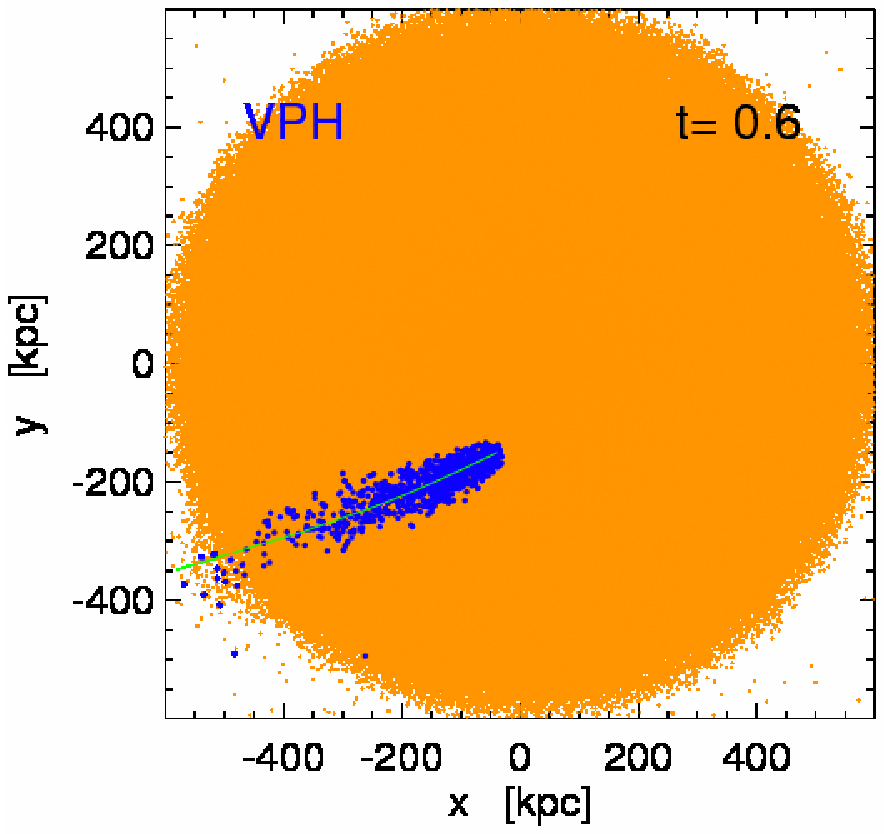}
\includegraphics[width=0.30\textwidth]
{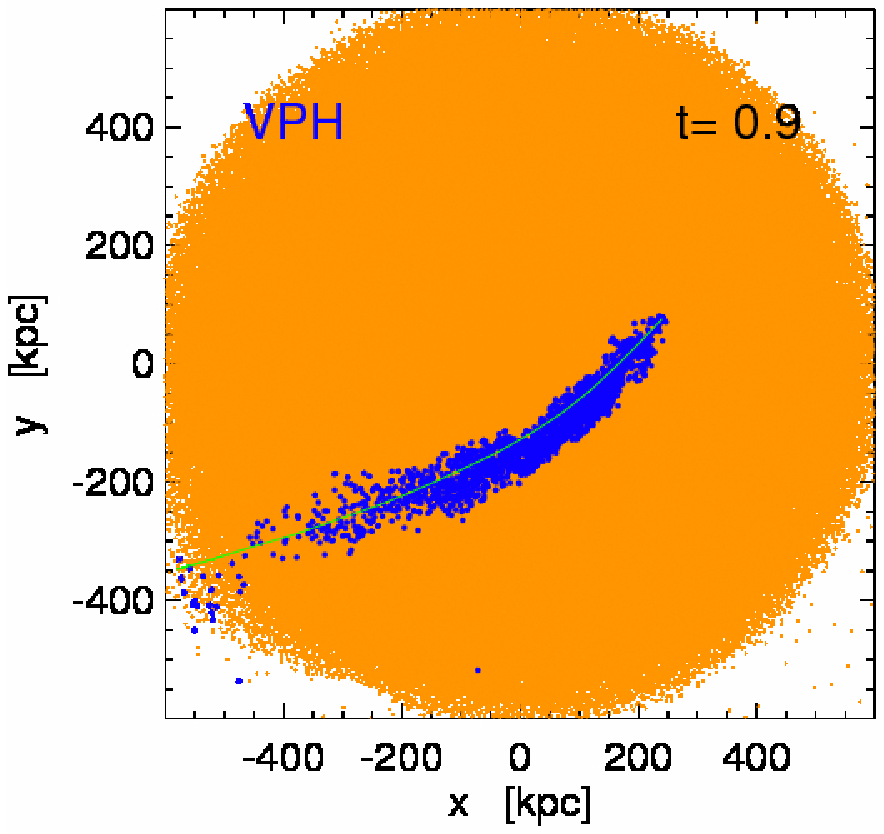}\\
\hspace{-12mm}
\vspace{-8mm}
\includegraphics[width=0.30\textwidth]
{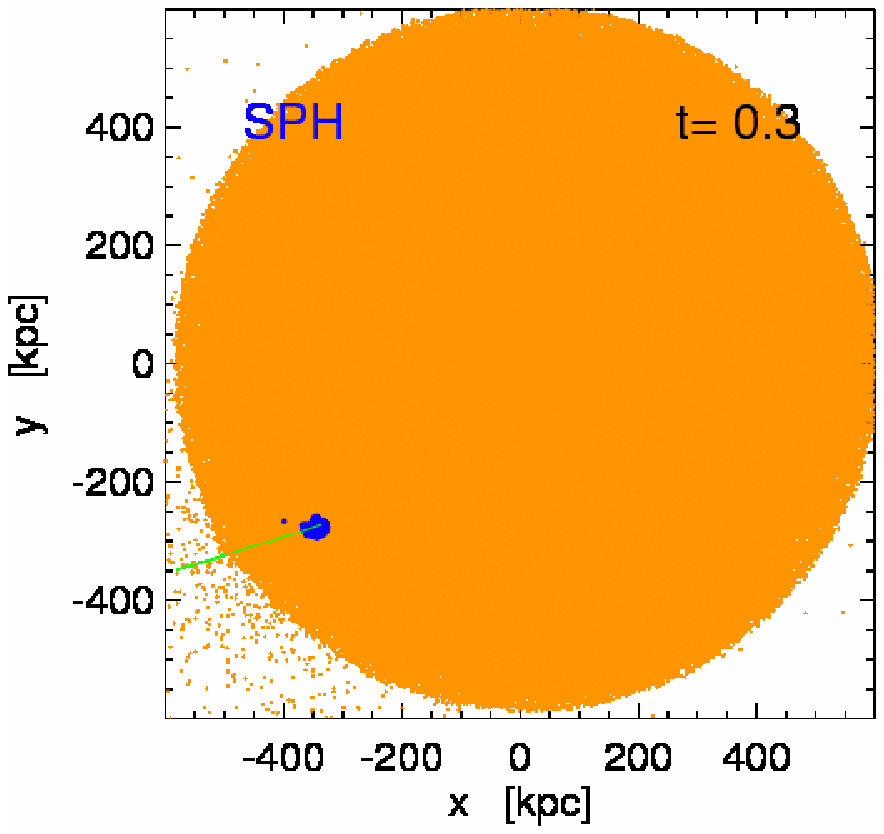}
\includegraphics[width=0.30\textwidth]
{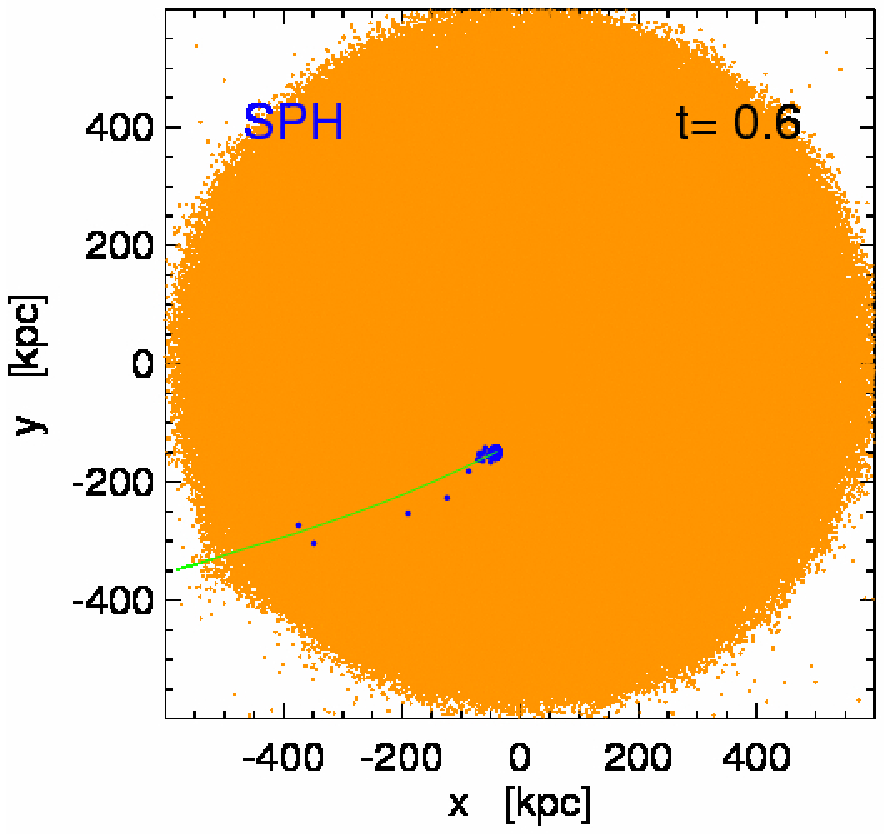}
\includegraphics[width=0.30\textwidth]
{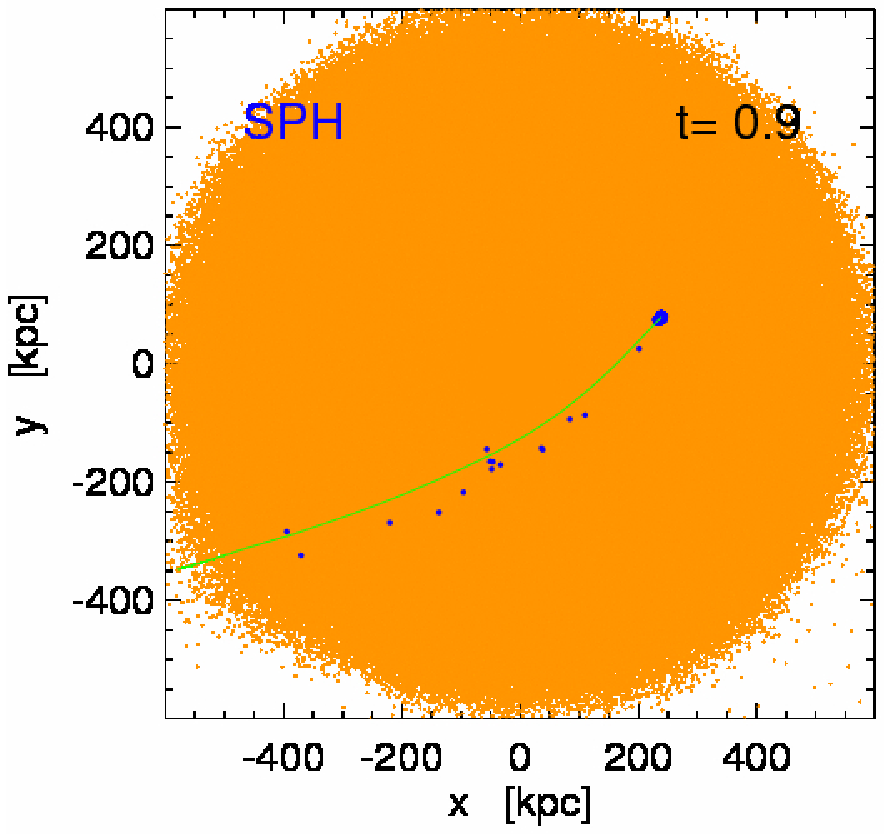}
 \vspace{6mm}
 \caption{Visual comparison of the spatial distribution of stripped
gas particles (large blue dots) from the ISM of an in-falling
galaxy
   simulated with VPH (top row) and SPH (bottom row) at $t=0.3 \, \rm{Gyr}$
   (left column), $t=0.6 \, \rm{Gyr}$ (middle column) and $t=0.9 \, \rm{Gyr}$
    (right column) (the cited
masses and times carry a factor of $h^{-1}$ in their units). The green
curve traces out the orbit of the galaxy
   through the cluster up to the time of each individual panel, as labelled.
   The gas particles of the target cluster's IGM are shown as yellow dots.  }
\label{RPS_particle_loss}
\end{center}
\end{figure*}

Both SPH and VPH rely on an artificial viscosity to capture shock
waves and to damp out small scale numerical noise. The choice of the
artificial viscosity parametrisation and the associated strength of
the viscous forces are crucial for the overall robustness and accuracy
of the scheme. Using a large viscosity leads to more robust shock
capturing, but on the other hand it may produce substantial viscous
effects in regions where the gas really ought to flow without
dissipation \citep{Cullen2010}. In particular, a too high viscosity
may suppress the growth of fluid instabilities that are important in
gas stripping processes. We have therefore checked how our wind tunnel
results for VPH depend on the viscosity setting. To this end we have
reduced the viscosity parameter from our default value of $\alpha=1$
(high viscosity) to $\alpha=0.5$ (intermediate) and $\alpha=0.25$ (low
artificial viscosity). In Figure~\ref{WT_AV}, we compare the
corresponding results for the stripping as a function of time in our
low-resolution wind tunnel set-up. We see that for reduced viscosity
the gas is actually stripped a bit faster in VPH than for the high
viscosity setting. This is consistent with our expectation that the
numerical viscosity will tend to damp small-scale fluid instabilities
and turbulence, which as a side effect reduces the stripping rate.

Another numerical nuisance parameter in VPH lies in the strength of
the shape correction forces that can be introduced into the technique
in order to regularise the mesh and encourage a quasi-regular particle
distribution. As explained in \citet{Hess2010}, this is accomplished
by adding a small term to the Lagrangian that penalises large
aspect-ratio distortions of cells and large offsets between the
generating point of a cell and its centre-of-mass.  While not strictly
necessary for VPH to work, we have found the technique to be
considerably less noisy in certain conditions when such weak shape
correction forces are used.  In our standard implementation, their
strength is controlled by two dimensionless parameters $\beta_0=0.2$
and $\beta_1=0.01$ \citep[see][for the exact definition of these
parameters]{Hess2010}.  In Figure~\ref{WT_SHP}, we show the dependence
of the stripping results for VPH when instead stronger shape
correction forces described by $\beta_0=0.6, \beta_1=0.03$ or even
$\beta_0=1.8, \beta_1=0.09$ are invoked. In both cases, this has only
a very weak influence on the results, as desired.

\section{Stripping of a galaxy during cluster in-fall} \label{Infall}

In order to complement our wind-tunnel experiments carried out in the
previous section with more realistic setups, we here want to conduct a
few simulations where galaxy models are in-falling into live models of
galaxy clusters. This approach includes a full treatment of gravity as
well as a modelling of cooling and star formation. In particular, it
accounts correctly for the tidal effects of a galaxy travelling within
the cluster potential. The latter inevitably changes the structure of
the galaxy, most prominently by reducing its dark matter mass through
tidal truncation, which in turn changes the conditions under which the
hydrodynamical processes occur.

Besides looking at the stripping of the gas, we will also compare how
the star formation rates in the in-falling galaxy decline in our
different simulations, since we expect that different numerical ram
pressure will strongly affect this quantity
\citep[e.g.][]{Kronberger2008,Kapferer2009}. We note that a number of
recent studies both with grid-based codes \citep{Roediger2007,
  Iapichino2008} and SPH codes \citep{Jachym2007,McCarthy2008} have
begun to investigate this question in detail.  Our focus here will be
much more limited though, and only be concerned with a comparison of our new
VPH scheme relative to SPH and {\small AREPO}.

\subsection{Setup of galaxy-cluster interaction simulations}

For definiteness, we consider a parabolic encounter of a
$10^{12}\,h^{-1}{\rm M}_\odot$ disk galaxy with a small $M_{200}=5\times
10^{13}\,h^{-1} {\rm M}_\odot$ galaxy cluster. The galaxy is constructed as
a compound system as described in Section~\ref{IsolatedGal} but scaled
down to fewer particles than in our earlier simulations. This was
necessary to reduce the computational cost to an acceptable level,
given that in this setup we need to represent the much larger cluster
with the same mass resolution as the galaxy in order to avoid
numerical problems in SPH \citep{Ott2003,Read2011}. In our default
set-up, we have therefore chosen $4000$ gas particles for the ISM of
the in-falling galaxy, and $10^6$ gas particles of identical mass for
the IGM of the galaxy cluster. This low resolution will of course
limit our ability to properly resolve hydrodynamical instabilities. We
note however that similar and often still lower resolution is
routinely used in cosmological simulations of galaxy formation, so our
results are directly representative of such studies. They in any case
allow us the extend our previous tests in Section~\ref{WindTunnel} to
lower resolution and investigate how reliably they can be assumed to
carry over to more complicated situations. 

\begin{table*}
\begin{tabular}{ l | c |  c | c |  c |  c } 
\hline 
Code & softening baryons & softening DM  & $T_{\rm SN}$ & $A_0$ &
regularisation $\beta_{0/1}$ \\
  \hline   
VPH & $0.25 \; h^{-1} \rm{kpc}$ & $0.5 \; h^{-1} \rm{kpc}$ & $3 \times
10^8$ & $3000$ & $\beta_0=0.6, \quad \beta_1=0.03$ \\
SPH & $0.25 \; h^{-1} \rm{kpc}$ & $0.5 \; h^{-1} \rm{kpc}$	& $3
\times 10^8$ & $3000$ \\
{\small AREPO} & $0.25 \; h^{-1} \rm{kpc}$ & $0.5 \; h^{-1} \rm{kpc}$
& $3 \times 10^8$ & $3000$ \\
  \hline  
\end{tabular}
\caption{Gravitational softening lengths for simulations in
  section~\ref{Infall}. The columns show the gravitational softening
  lengths for baryons, that is for gas and
  stellar particles. The next column shows these
  values for collisionless DM halo particles. 
  The following columns specify the parameters 
  of the supernova feedback model, where $T_{\rm SN}$ is the
  effective temperature and $A_0$ the evaporation efficiency
  \citep[using the notation of][]{SpringelHernquist2003}. The last column gives 
  the parameters used for the Voronoi cell shape
  correction scheme in VPH \citep[in the notation of][]{Hess2010}.}
\label{Infall_paramTable}
\end{table*}

For the cluster, we adopt a gas-fraction of $17\%$ and a NFW
concentration of $c=2.0$.  A detailed list of main numerical
parameters adopted for our simulations in this section is given in
Table~\ref{Infall_paramTable}.  The trajectory of the encounter starts
at a distance of $700\,h^{-1}{\rm kpc}$, just outside of the virial
radius, and follows a parabolic orbit where the minimal distance of
the two objects would be $50\,h^{-1}{\rm kpc}$ if they were point
masses. Relative to the orbital plane, the galaxy's disk is tilted by
$\theta=30^{\circ}$ and then turned by $\phi=30^{\circ}$ in the
azimuth \citep[see][for a sketch of the orbital geometry]{Duc2000}.
We note that we can expect to be able to draw quite general
conclusions from a single choice of inclination angles since the
orbital geometry has been shown to have no significant effect on the
gas stripping \citep{Roediger2006}.

\subsection{Properties of the head wind}

In Figure~\ref{GalStr_headwind}, we show the properties of the wind
encountered by the in-falling galaxy as a function of time. We give
the time evolution of the galaxy's environment as found in a sector
with opening angle $30^{\rm o}$ of a spherical shell in the radial range
$45\, h^{-1} {\rm kpc} < r <99\,h^{-1} {\rm kpc}$ centred around the
direction the galaxy is heading to. The velocity difference
$v_{\rm{headwind}}$ shown in the fist panel of
Figure~\ref{GalStr_headwind} is computed with respect to the centre of
mass of the galaxy. The figure shows that density, pressure and
temperature of the head wind reach their peak at the pericentre, as
expected. 

We include results for VPH, SPH and {\small AREPO} in
Fig.~\ref{GalStr_headwind}, and as the comparison shows, there is
generally very good agreement between the three different simulation
techniques. All three runs show a strong rise of density and pressure as
the galaxy approaches and passes pericentre, while the temperature is
roughly constant, reflecting the nearly isothermal conditions in the
cluster gas. There are no significant deviations in the orbit of the
galaxy between the different methods, hence any difference in the
evolution of the galaxies can only arise from differences in the
hydrodynamical treatment of the interaction of galaxy and cluster gas.

\subsection{Gas stripping and star formation truncation}

In order to define whether a gas particle or cell still belongs to the
galaxy we use the condition: 
\begin{equation}
\rho > 4 \times 10^{-26} \, {\rm g}\, {\rm cm}^{-3} \quad \mbox{and}  \quad 
T < 10^6 \,\rm{K}. 
\label{crit_1stpart}
\end{equation}
Furthermore, we additionally require that the separation $r$ of
particles from the centre-of-mass of the remaining gravitationally
bound dark matter of the galaxy is less than $20\,h^{-1}\rm{kpc}$.  We
note that with this definition the galaxy can in principle also
accrete new gas from the cluster which may then cool onto the ISM and
help to provide fuel for star formation.

\begin{figure}
\begin{center}
\includegraphics[width=0.5\textwidth]
{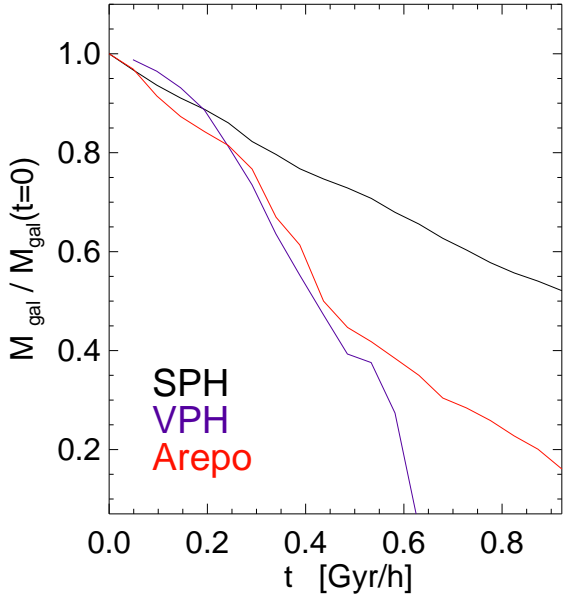}
\vspace{-7mm}
\caption{Gas mass as a function of time still bound to a galaxy that
  is in-falling into a galaxy cluster. We include  
simulation results for SPH (black), VPH (red) and {\small AREPO}
(blue). }
\label{GalStr_MassRatio}
\end{center}
\end{figure}

\begin{figure}
\begin{center}
\includegraphics[width=0.5\textwidth]
{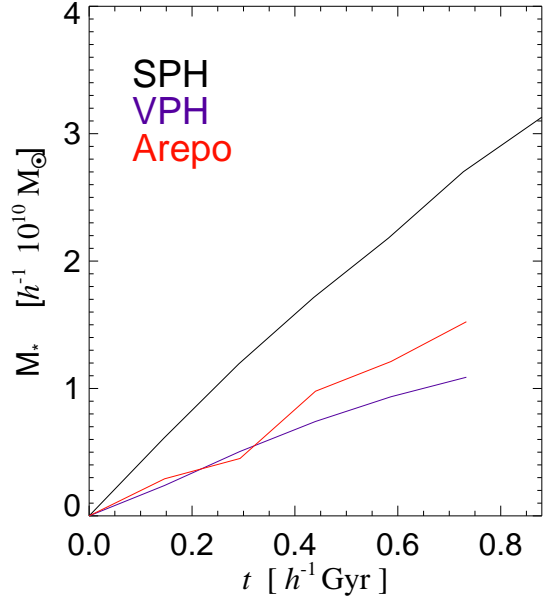}
\vspace{-7mm}
\caption{The stellar mass formed within a distance of $15\,h^{-1} 
  \mathrm{kpc}$ from the galaxy centre as a function of time.  We
  include simulation results for SPH (black), VPH (blue) and {\small
    AREPO} (red).  }
\label{GalStr_Sfrate}
\end{center}
\end{figure}

In Figure~\ref{RPS_particle_loss}, we show a visual comparison of the
gas stripping in VPH and SPH simulations, where it is readily apparent
that the gas mass loss proceeds much slower in SPH than in VPH.  This
is confirmed by quantitative measurements shown in
Figure~\ref{GalStr_MassRatio}, which include results for gas
stripping in three different simulations of the encounter of a galaxy
with the cluster ICM. Interestingly, VPH looses gas here even somewhat
faster than {\small AREPO}, but both methods yield a substantially
faster loss of gas than SPH. While at time $\sim 1\,{\rm Gyr}$ the
VPH simulation has lost {\em all} of the gas, the galaxy in SPH still
retains half of it.

This substantial difference is corroborated by the behaviour of the
star formation rate in the galaxies, shown in cumulative form in
Figure~\ref{GalStr_Sfrate}.  Here SPH shows the slowest decline
overall, while both VPH and {\small AREPO} lead to a rapid termination
of star formation, which implies a quick reddening of the galaxies. It
is a well known problem in galaxy formation to understand the colours
of cluster galaxies in detail. Semi-analytic models of galaxy
formation have commonly predicted a rather quick truncation of star
formation upon cluster in-fall, yielding cluster populations that are
actually too red \citep[e.g.][]{Weinmann2006}.  It can be expected,
due to the faster stripping, that our VPH and {\small AREPO}
simulations suffer from the same problem, potentially making it even
more severe. This may indicate that the stripping efficiencies in all
of the simulations are substantially too large, probably because the
ISM is still under-resolved and appears as a homogeneous dense phase
instead of being resolved into a true multi-phase medium. The latter
would make the medium more porous, allowing very dense clouds of gas
to resist the ICM wind for a longer time and to stay in the in-falling
galaxy.

\begin{figure}
\begin{center}
\vspace{4mm}
\includegraphics[width=0.43\textwidth]{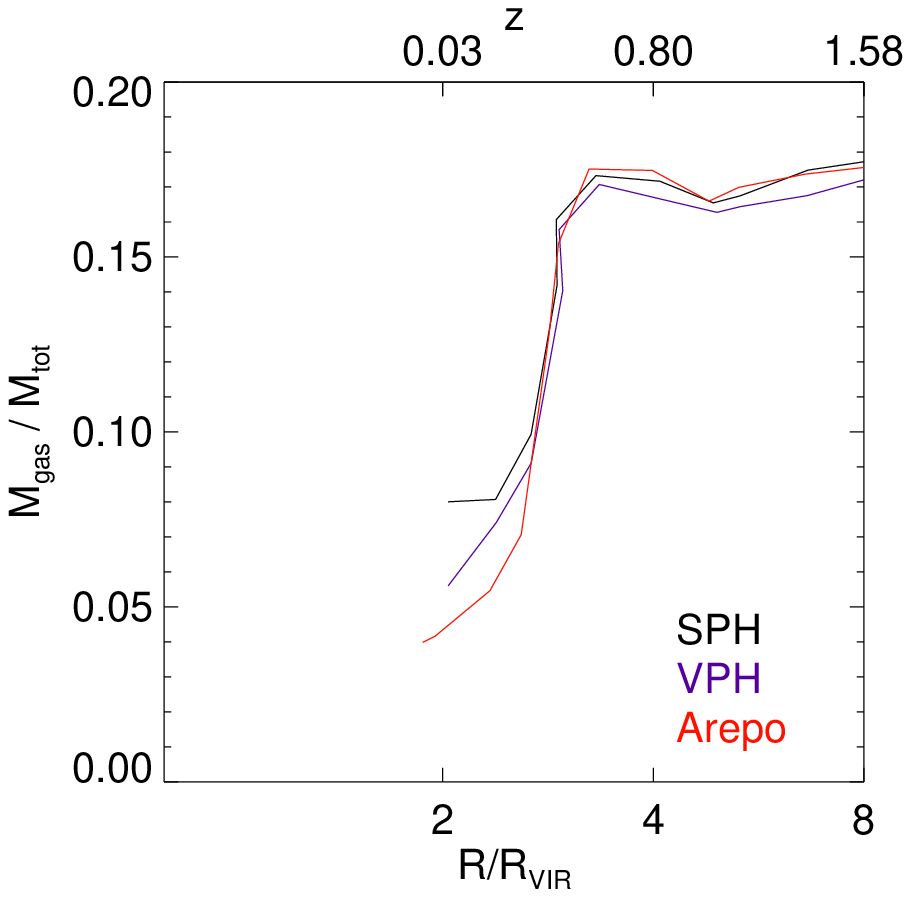}\\
\includegraphics[width=0.43\textwidth]{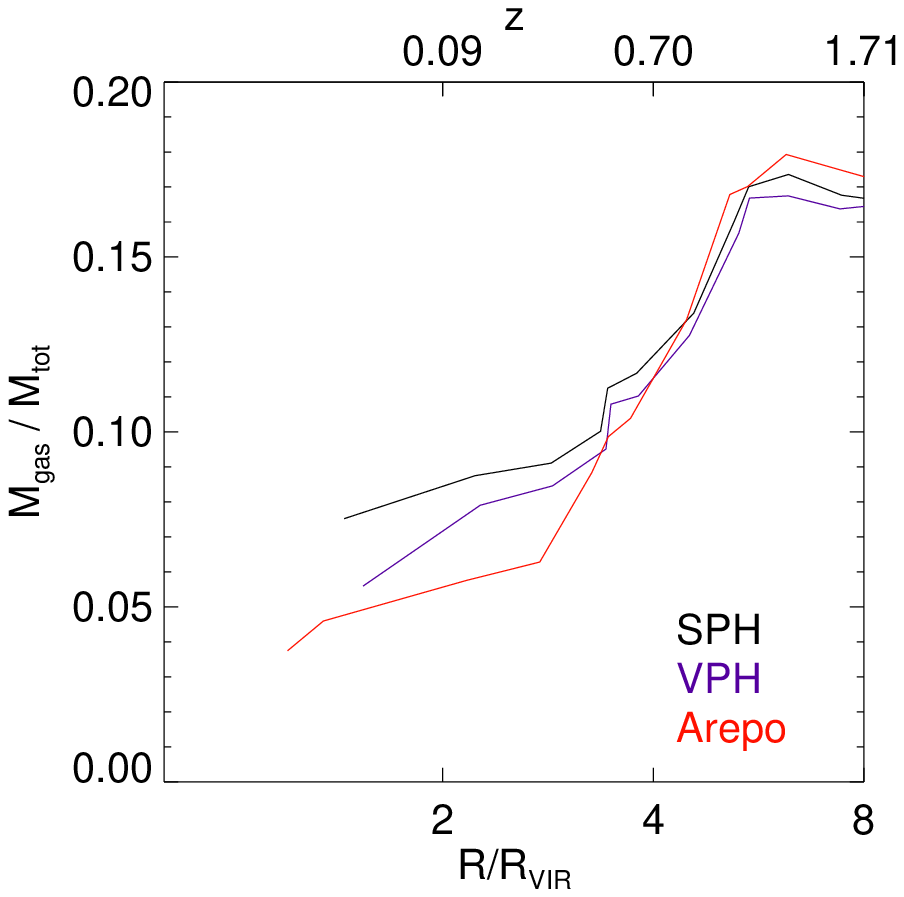}
\vspace{-7mm}
\caption{Gas fraction as a function of distance to the cluster centre
  for two representative (sub)halos that fall into the cluster. We
  show results for a substructure with $4.5\,{\rm k}$ DM-particles
  (top panel) and $1.8\,{\rm k}$ DM-particles (bottom) panel,
  comparing results for SPH (black), VPH (blue) and {\small AREPO}
  (red).  We find a clear difference in the final phase of the
  stripping process. In the SPH simulation, the substructure keeps a
  higher gas content until it is disrupted.}
\label{Mixing_Stripp}
\end{center}
\end{figure}

\begin{figure}
\begin{center}
\includegraphics[width=0.45\textwidth]{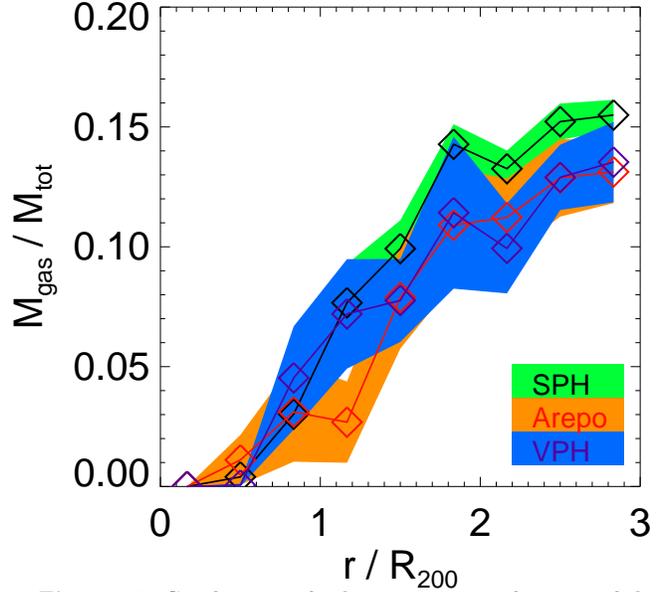}
\vspace{-4mm}
\caption{Gas fraction of substructures as a function of their distance
  to the cluster centre. We compare results for SPH (black with green shade),
  VPH (blue) and {\small AREPO} (red with orange shadow), including only
  subhalos more massive than $M > 3\times 10^{11} \,h^{-1}{\rm  M}_{\odot}$ to exclude
  poorly resolved small structures.  The shaded areas indicate one
  standard deviation from counting statistics. }
\label{Mixing_gasfrac}
\end{center}
\end{figure}

\section{Cosmological cluster simulations} \label{Cosmo}

We now turn to results for fully cosmological simulations of cluster
formation. We first carry out a comparison of the gas content of
satellite systems in `zoom' simulations of the formation of rich
galaxy clusters, carried out with VPH, SPH and {\small AREPO}. We here
primarily want to check whether the differences we have observed in
the stripping of dense ISM gas out of galaxies in our earlier more
idealised simulations manifest themselves also in non-radiative
simulations without star formation, where the density contrast is much
smaller. Secondly, we study the well known Santa Barbara cluster of
\citet{Frenk} in order to investigate whether VPH produces a higher
entropy in the cluster centre compared to SPH, which would then make
it closer to the results of mesh codes that have been applied to this
problem. We also use this cluster in order to assess the numerical
convergence of VPH for the properties of the intra-cluster gas.

\subsection{Gas stripping in non-radiative zoom simulations of galaxy clusters}
\label{Zoom}

In order to simulate the formation of rich galaxy clusters in the
$\Lambda$CDM cosmology, we extract a massive halo from the Millennium
Simulation \citep{SpringelMill}, and re-simulate this cluster with the
addition of gas.  To this end we trace the particles that make up the
cluster back to the original initial conditions at $z=127$, thereby
finding the Lagrangian region out of which the cluster has formed.
This region is then populated both with dark matter and gas particles,
which are perturbed with the original displacement field. We can also
increase the resolution compared to that in the original simulation if
desired, in which case additional small-scale fluctuation power is
added in the region between the old and new Nyquist frequencies.
Further away from the region that holds the cluster material and its
immediate surroundings, the resolution is reduced by combining
particles into progressively heavier `boundary' particles. In this
way, the resolution gradually declines with distance from the cluster
while the gravitational tidal field that influences its formation is
still accurately determined. With this standard `zoom' technique, the
computational cost is concentrated in the small region of interest,
allowing high resolution simulations of individual objects in
comparably short time.

For definiteness, we have picked a cluster of virial mass $M_{200}=
1.8\times 10^{15}\,h^{-1}M_{\odot}$, which we re-simulate with a
baryon fraction $\Omega_b=0.045$. The other cosmological parameters
are the same as in in the original Millennium run and are given by
$\Omega_{\rm m} = 0.25$, $\Omega_\Lambda=0.75$, $\sigma_8=0.9$,
$h=0.73$, and $n=1$.  We have initialised the re-simulations at
$z=127$ and evolved them with VPH, SPH and {\small AREPO} to the
present epoch. We employed an identical gravitational softening length
of $0.005 \; h^{-1} \rm{kpc}$ for all particle types and
hydrodynamical schemes, and treated the gas as a non-radiative mix of
hydrogen and helium.

\begin{table*}
\begin{tabular}{ l | r | r | r | c |  c  }  
\hline
Code & resolution & softening baryons & softening DM 
& $\rm{min}(T_{\rm gas})$  &  $\beta_{0/1}$ \\
  \hline   
VPH & $2\times 32^3$	& $25 \; h^{-1} \rm{kpc}$	&
$25 \; h^{-1} \rm{kpc}$	& $20\,{\rm K}$       &  
$\beta_0=0.8, \quad
\beta_1=0.04$\\
VPH & $2\times 64^3$	& $12 \; h^{-1} \rm{kpc}$	&
$12 \; h^{-1} \rm{kpc}$	& $20\,{\rm K}$	&  	 
$\beta_0=0.8, \quad \beta_1=0.04$\\
VPH & $2\times 128^3$	& $6 \; h^{-1} \rm{kpc}$	&
$6 \; h^{-1} \rm{kpc}$	& $20\,{\rm K}$	&  	 
$\beta_0=0.8, \quad \beta_1=0.04$\\
VPH & $2\times 256^3$	& $3 \; h^{-1} \rm{kpc}$	&
$3 \; h^{-1} \rm{kpc}$	& $20\,{\rm K}$	&  	
$\beta_0=0.8, \quad \beta_1=0.04$\\
  \hline  
\end{tabular}
\caption{List of parameters for simulations in
  section~\ref{Sec_SBcluster}. The column ``resolutions'' denotes the
  initial particle numbers. The following columns show the gravitational
  softening lengths for gas and collisionless DM halo particles. The
  columns on the right specify parameters deviating from our standard
  choice, such as minimal gas temperature
  and slightly more aggressive values for VPH's mesh regularisation
  scheme.}
\label{SB_paramTable}
\end{table*}

We identify halos in the simulations by applying the FOF algorithm to
the high-resolution dark matter particles. Each gas particle is
assigned to the group in which its closest dark matter particle lies.
We then apply the {\small SUBFIND} algorithm \citep{Springel2001} to
the particle groups we found in this way in order to decompose them
into gravitationally bound (sub)groups. Using the IDs attached to each
particle we can track individual halos/subhalos as a function of time,
and, in particular, study how the gas content of subhalos declines as
a halo falls into the cluster. In order to reduce numerical noise in
our simulation comparison, we however restrict our substructure
selection to a sample where more than $60\%$ of the DM particles can
be found in every studied simulation run.

As an example, we show in Figure~\ref{Mixing_Stripp} the evolution of
the gas fraction of two different substructures as a function of their
distance to the cluster centre (the corresponding times are indicated
by the redshift labels on the upper axis), comparing results for VPH,
SPH and {\small AREPO} simulations.  For both substructures, we find
good agreement in the early in-fall phase, for distances larger than
about $\sim 3\,R_{\rm vir}$. At smaller separations, the influence of
the cluster is however very noticeable already, and the substructures
loose gas quickly. In both of these examples, we identify a
considerably larger gas loss in the VPH run than in SPH at the last
time when we can still find the substructures before they are
disrupted. This is consistent with our earlier findings for the
stripping of the dense ISM gas.

In Figure~\ref{Mixing_gasfrac}, we now consider all of the
substructures around the cluster at the final time, and simply compare
the gas mass fraction in substructures as a function of
cluster-centric radius. Even though the results are a bit noisy, we
find a clear statistical trend of a smaller gas mass fraction in the
VPH run relative to SPH, especially in the region $\sim 1-3\,R_{\rm
  vir}$, where the gas fraction declines rapidly. Interestingly, a
simulation with {\small AREPO} for the same cluster initial conditions
shows a bound gas fraction that is still smaller in-falling dark
matter halos. We hence conclude that even at the level of
non-radiative simulations there are already significant differences in
the stripping of gas out of in-falling substructure, which also
implies that the mixing of gas in massive halos can be expected to
differ in the three examined numerical techniques.

\begin{figure*}
\begin{center}
\resizebox{8.4cm}{7cm}{\includegraphics{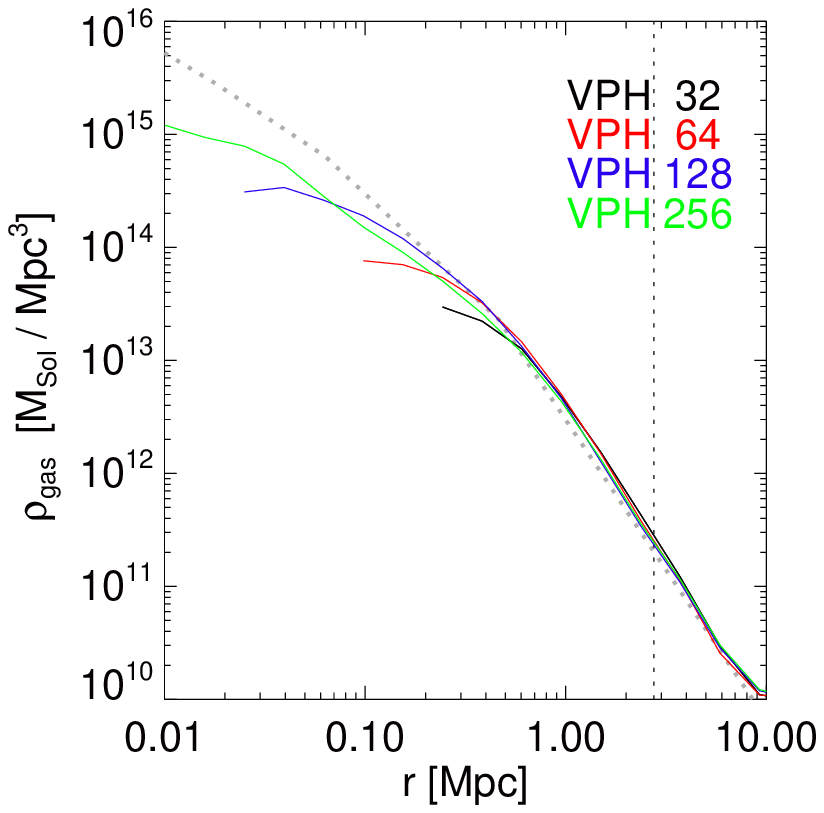}}%
\resizebox{8.4cm}{7cm}{\includegraphics{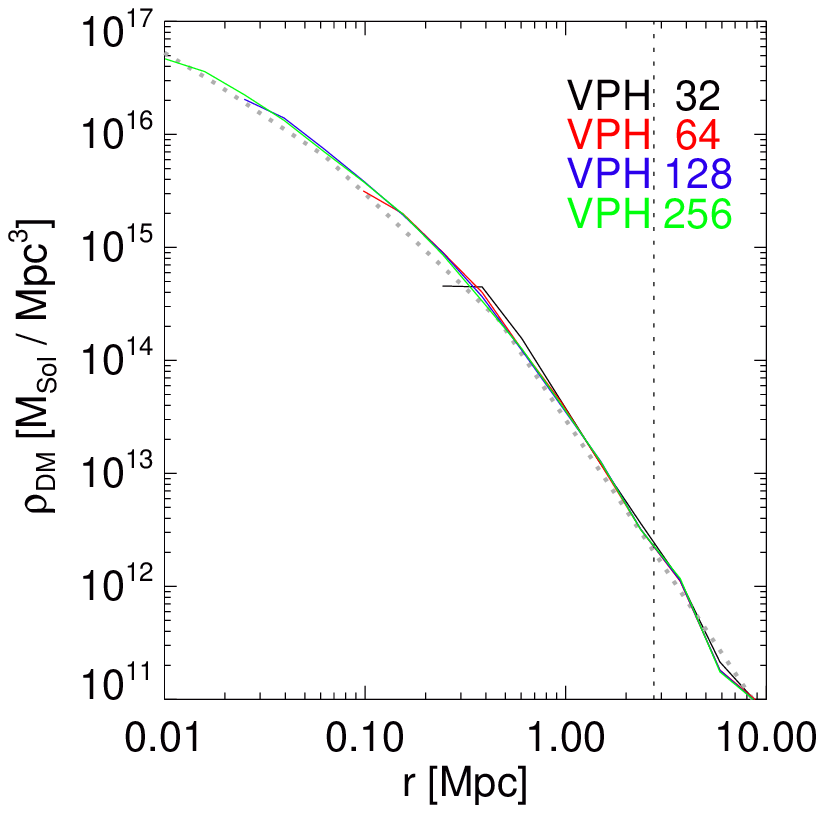}}\\
\vspace*{-5mm}%
\resizebox{8.4cm}{7cm}{\includegraphics{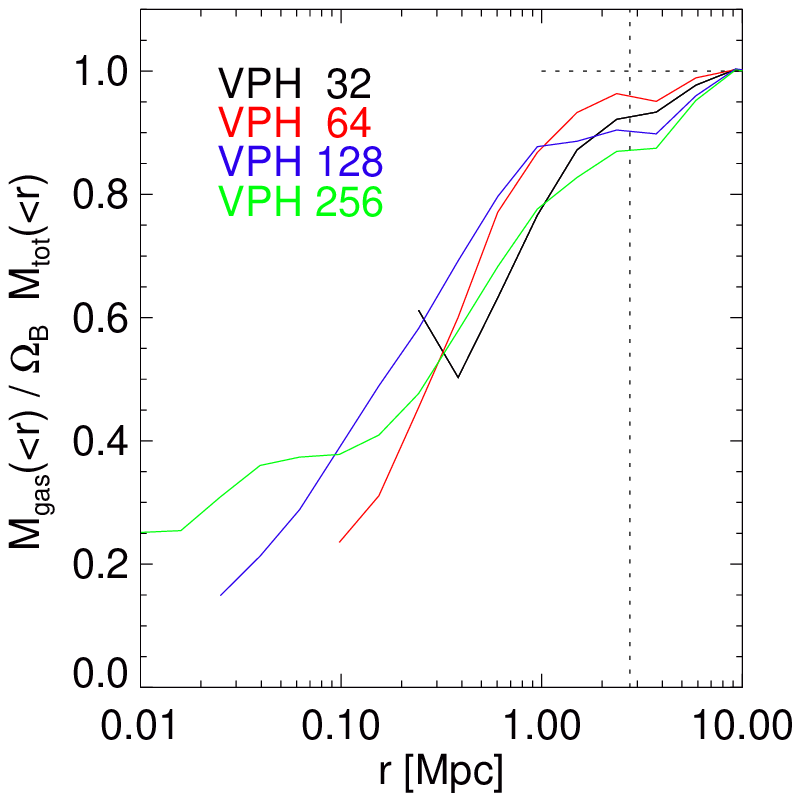}}%
\resizebox{8.4cm}{7cm}{\includegraphics{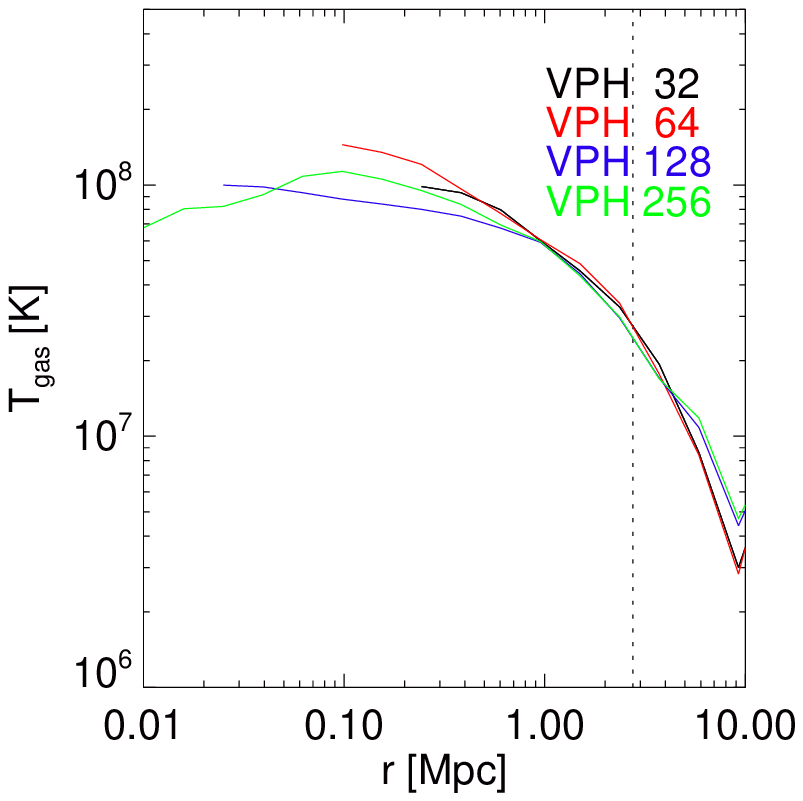}}\\
\vspace*{-5mm}%
\resizebox{8.4cm}{7cm}{\includegraphics{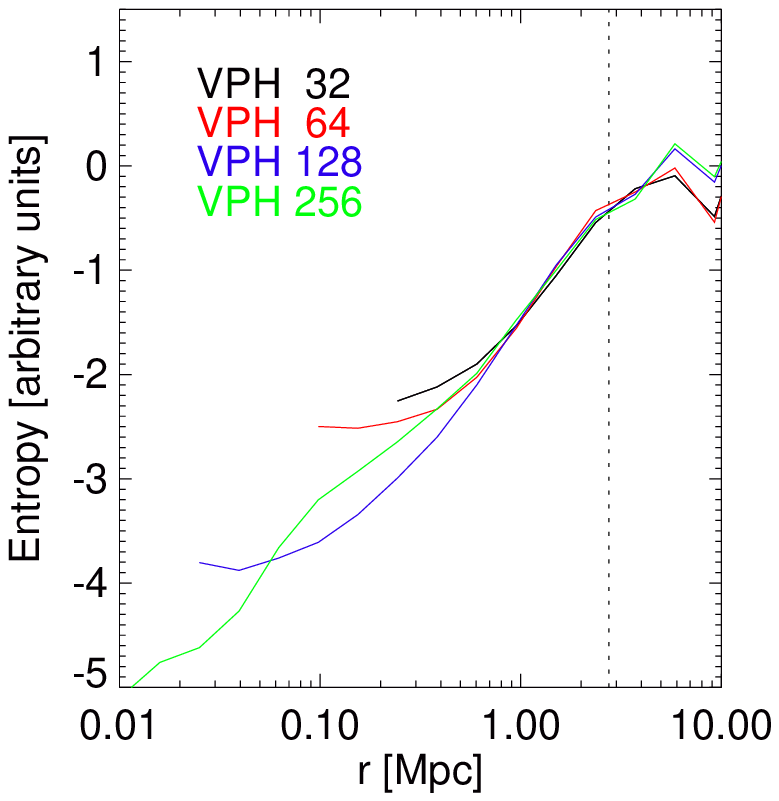}}%
\resizebox{8.4cm}{7cm}{\includegraphics{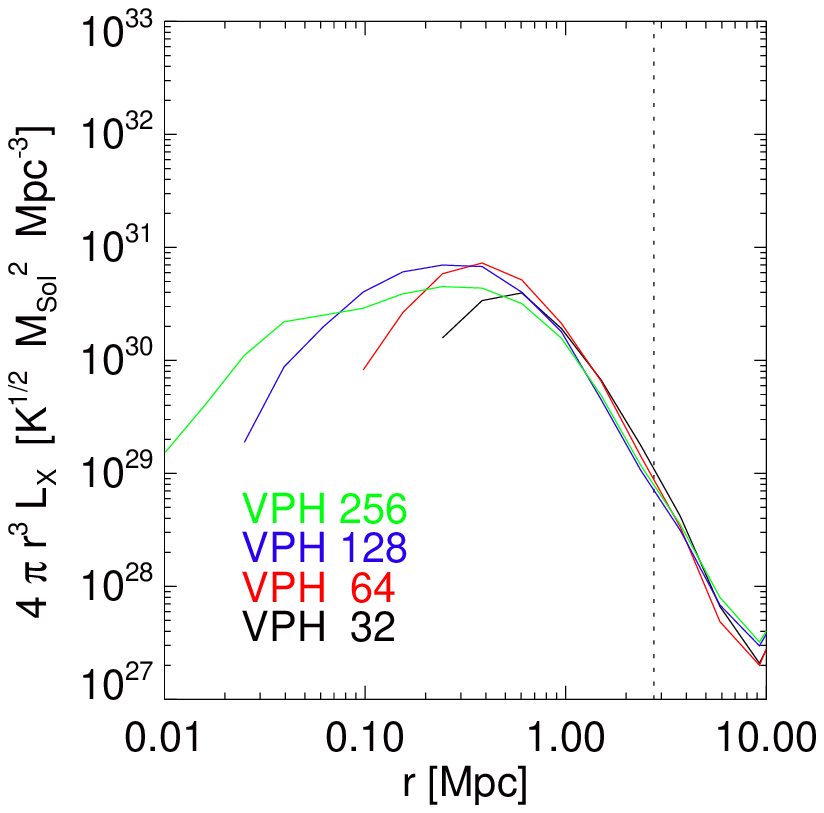}}\\ 
\caption{Resolution study with VPH for the cluster of the cosmological
  ``Santa Barbara Cluster Comparison project``. Radial profiles of
  (from top left to bottom right) give gas-density, dark matter
  density, cumulative gas-fraction, gas-temperature, specific entropy
  and X-ray luminosity. All properties have been obtained as
  mass-weighted arithmetic averages within logarithmically spaced
  bins (the cited masses and times carry a factor of $h^{-1}$ in
  their units). The different colours give VPH results with initial
  gas particle numbers of $32^3$ (black), $64^3$ (red), $128^3$ (blue)
  and $256^3$ (green), respectively.}
\label{SBcluster_1_RES}
\end{center}
\end{figure*}

\subsection{The Santa Barbara Cluster} 
\label{Sec_SBcluster}

The `Santa Barbara cluster comparison project' of \citet{Frenk}
analysed the results of a large number of different cosmological
hydrodynamical codes for the formation of a rich cluster of galaxies
with non-radiative gas. The comparison involved both SPH codes and
hydrodynamical mesh codes, and focused, in particular, on the
resulting thermodynamic properties of the intra-cluster gas. An
important result that emerged from the study was that the different
methods systematically disagree in the amount of entropy predicted for
the central regions of the cluster, with the mesh-based approaches
yielding consistently higher central entropy and correspondingly lower
density than the SPH codes.

The Santa Barbara (SB) cluster has become a standard test problem for
cosmological hydrodynamic codes, with results reported in numerous
studies. Recently, some studies have suggested that the difference
seen between the various techniques is primarily associated with
differences in the treatment of mixing \citep{Mitchell2009,Vazza2011},
which is suppressed in SPH by construction, and this may artificially
lower the central entropy. Specifically, it has been suggested that
this problem may be related to a suppression of Raleigh-Taylor fluid
instability in SPH \citep{Wadsley2008}. If the difference is caused by the
lack of mixing at the particle level \citep{Tasker2008,Wadsley2008}
the VPH results should in principle agree with SPH.

In the spirit of the original project of \citet{Frenk}, we here re-run
the SB cluster with our new hydrodynamical VPH method. We are
especially interested in the question whether VPH differs in its
predictions for the central cluster region compared with SPH, which
perhaps could arise due to the different stripping efficiency of this
technique.  We use the same initial conditions that have been used in
the original SB cluster comparison project, where an
Einstein-de-Sitter cosmological model with mean density $\Omega_{\rm
  m} =1$, Hubble constant $H_0 = 50\, \rm{km} \, \rm{s^{-1}
  Mpc^{-1}}$, and baryon fraction of $\Omega_b = 0.1$ was used. The
initial conditions were constructed as a constrained realisation where
a rich cluster corresponding to a $3\sigma$ peak was imposed to form
in the centre of a cubic box with a side-length of $64\, \rm{Mpc}$, in
an otherwise random realisation.

We have simulated the SB cluster at a variety of different resolutions
using VPH, ranging from $2\times 32^3$ to $2\times 256^3$ particles
with parameters as denoted in Table~\ref{SB_paramTable}, in order to
see how well the method converges for the primary cluster properties.
In Figure~\ref{SBcluster_1_RES}, we show the results of this
convergence test, in terms of radial profiles for gas density, dark
matter density, enclosed gas fraction, temperature, entropy, and
specific X-ray emissivity. We find that the convergence is in general
good for the outer profiles and the dark matter properties. Only in
the very centre some interesting differences can be observed. For the
most part they can be simply understood as an imprint of the varying
spatial and mass resolution across the sequence, where shortly before
the resolution limit is reached the lower resolution runs peel away
from the converged result seen in the higher resolution
simulations. This is for example the case for the entropy and gas
density profiles. 

It is clear however that VPH does not predict the formation of a large
constant entropy core region, unlike what is typically seen in mesh
codes and consistent with the falling entropy profiles that are
typically observed with SPH all the way to the centre of halos. In
fact, in general our results obtained with VPH for the Santa Barbara
cluster are in good agreement with SPH results reported in the
literature \citep[e.g.][]{Frenk,Ascasibar2003,gadget2}, suggesting
that it makes comparatively little difference for the bulk
thermodynamic structures which particle-based method is used in
non-radiative cosmological simulations. In particular, the different
stripping efficiency alone and the slightly higher spatial resolving
power of the Voronoi-based density estimate do not yield an entropy
core as generally found in hydrodynamical mesh-codes.

\section{Discussion and Conclusions}		\label{SecConclusions}

In this study, we have carried out a systematic comparison of the
properties of our new Voronoi particle hydrodynamics (VPH) method
\citep{Hess2010} with respect to standard SPH and moving-mesh
hydrodynamics as implemented in the {\small AREPO} code. We have
focused on stripping processes in galaxies and halos upon in-fall into
galaxy clusters. Here, it is expected that the outer parts of gaseous
disks are quickly removed due to ram pressure stripping
\citep{GunnGott1972}, but the subsequent more gradual gas loss
sensitively depends on the ability of hydrodynamical codes to capture
fluid instabilities occurring in shear flows around the galaxies. The
recent findings that SPH appears to exhibit severe inaccuracies in
this regime has prompted us to develop the alternative VPH method. In
the present paper, we have studied how this new technique compares
with traditional SPH and the new moving-mesh code technique in a
number of basic problems relevant for galaxy formation.

\begin{figure}
\begin{center}
\resizebox{8.5cm}{!}{\includegraphics{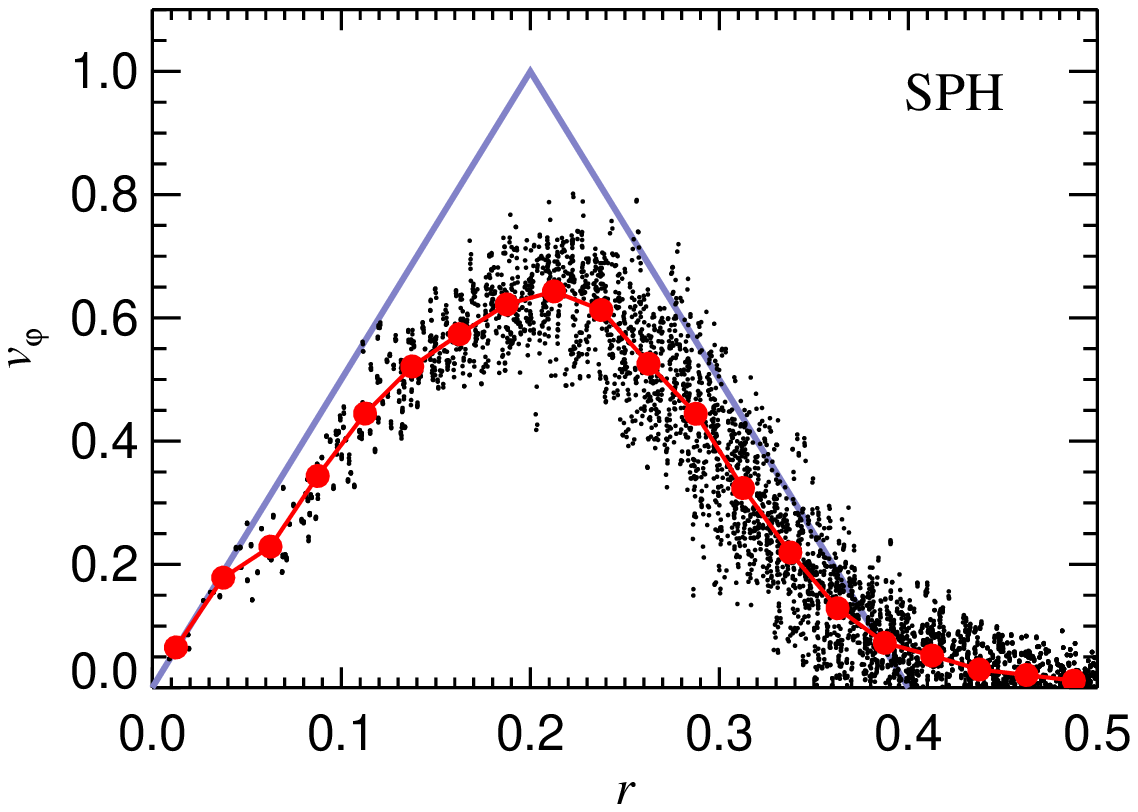}}\\
\resizebox{8.5cm}{!}{\includegraphics{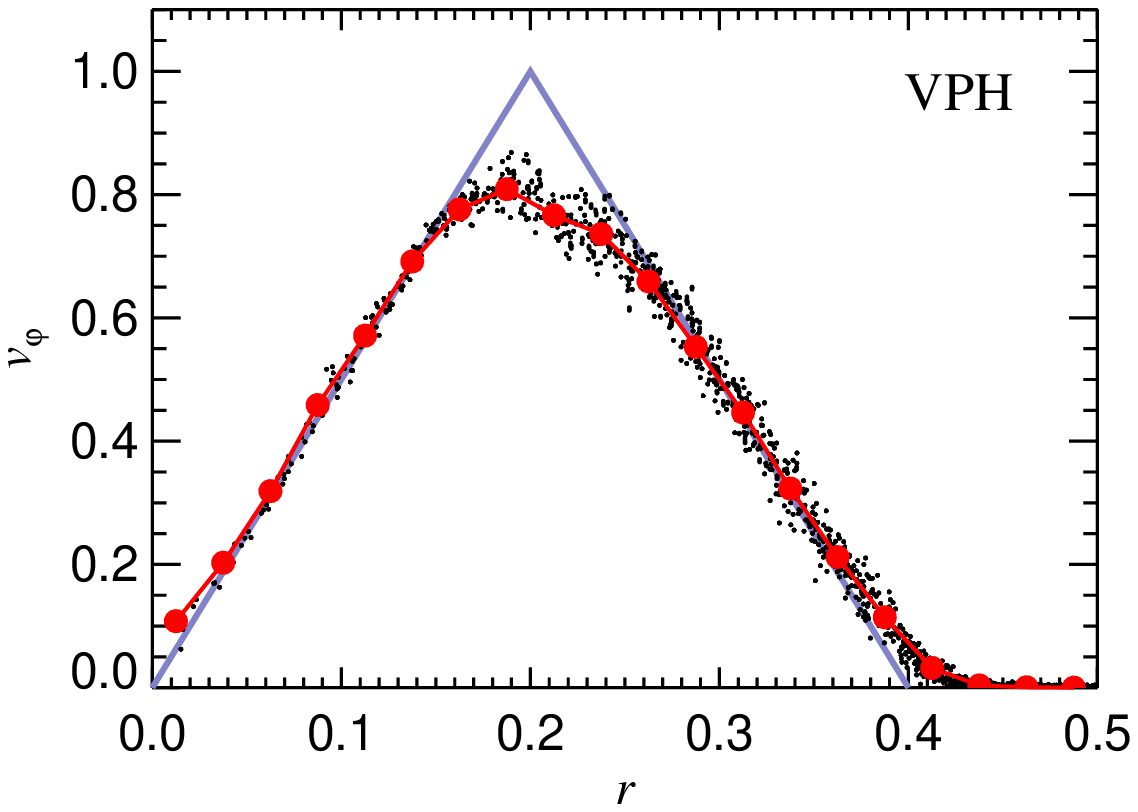}}\\
\resizebox{8.5cm}{!}{\includegraphics{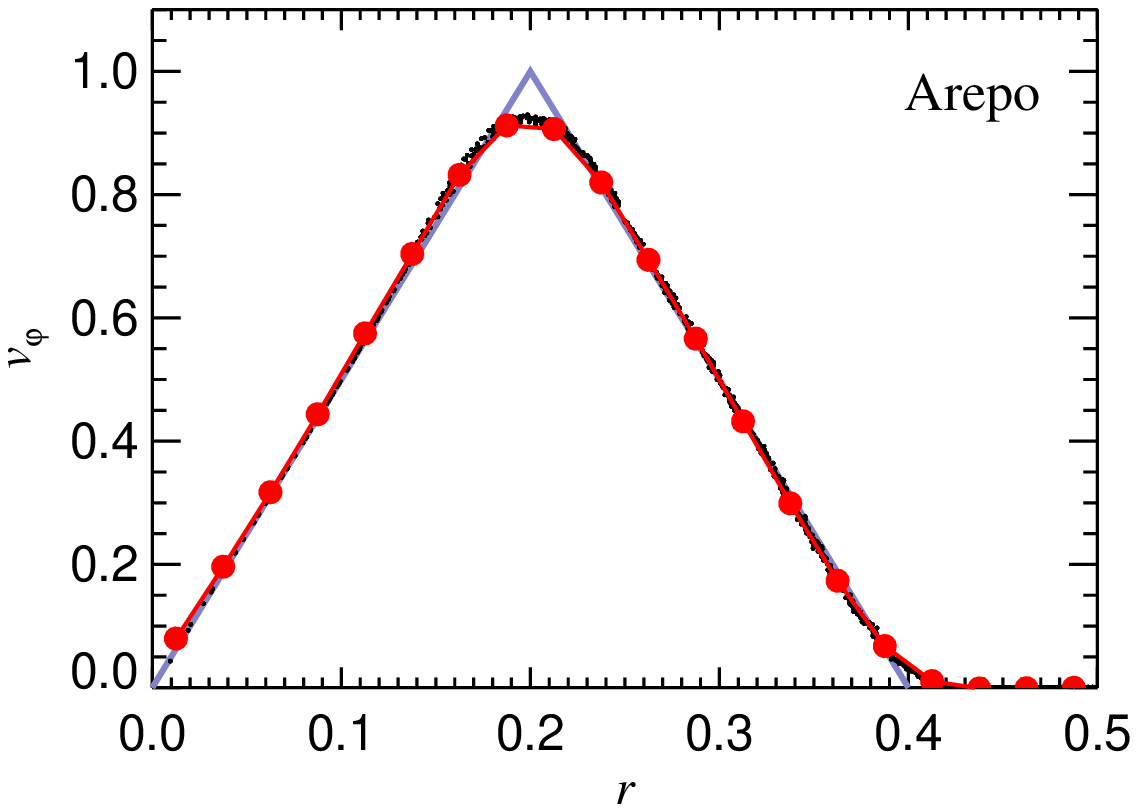}}
\caption{Radial profiles of the azimuthal velocity in evolved simulations
  ($t=1.0$) of the Gresho vortex test, calculated at resolution
  $80\times 80$ with different numerical techniques (SPH, VPH, and
  {\small AREPO}), as labelled. The small black dotes show individual
  velocities of simulation particles/cells, while the red dots give
  the averaged solution binned in radial annuli.  The blue thick lines give
  the analytic solution that is realised in the initial conditions.}
\label{Gresho_profiles}
\end{center}
\end{figure}

\begin{figure}
\begin{center}
\resizebox{8.5cm}{!}{\includegraphics{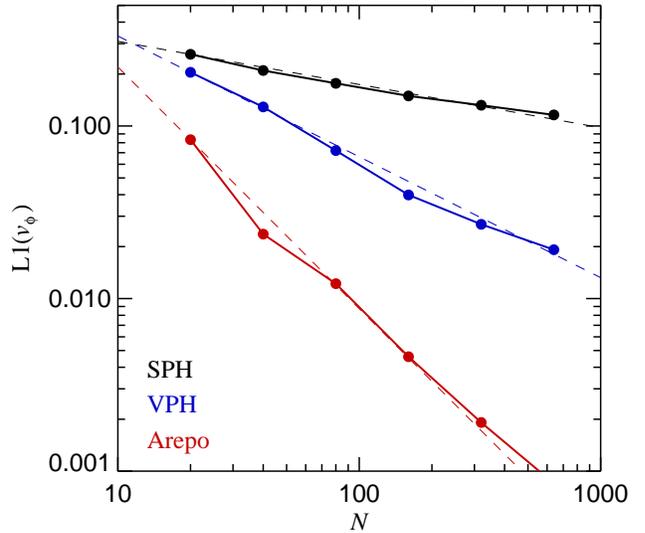}}
\caption{Convergence rate of three different numerical techniques for
  the Gresho vortex test. The black points refer to measurements of
  the L1 error norm at $t=2.0$ for the SPH method, with resolutions
  ranging from $20\times 20$ to $640\times 640$. The L1 error has been
  calculated on the basis of all single particles/cells. The blue dots
  give the corresponding measurements for our VPH method, while the
  red points are for the {\small AREPO} moving-mesh code. The dashed
  lines indicate power-laws with slopes $-0.3$, $-0.7$ and $-1.4$ for
  SPH, VPH and {\small AREPO}, respectively.}
\label{GreshoConvRate}
\end{center}
\end{figure}

To this end, we first compared results for isolated compound galaxy
models, both in isolation and in wind tunnels where they were exposed
to a supersonic head wind. This set-up allowed relatively
high-resolution simulations of wind--ISM interactions. Our simulations
have revealed non-negligible differences in the rate at which dense
ISM gas is stripped and dispersed, and in the appearance of this gas
in the downstream part of the flow. SPH showed a lower stripping rate
than both VPH and the mesh-code {\small AREPO}. As a result, the SPH
galaxy also experienced the largest displacement due to the ram
pressure of the impinging wind. Also, we were able to show that
essentially none of the ISM gas in SPH and VPH could ever be
transferred to much lower density. Instead, if gas was stripped, it
stayed in coherent dense blobs, where even star formation could
continue. Furthermore due to elevated pressure caused by the surface
tension effect in SPH, the star formation in these stripped blobs
remained at an unphysically high level. {\small AREPO}, in contrast,
showed a rapid loss of gas out of the disk, which was furthermore
efficiently mixed with other gas, so that lower densities were reached
quickly by the stripped gas and star formation was stopped.

We followed up these simulations with numerical experiments where we
dropped galaxy models in live cluster models. Even though here the
resolution was substantially lower, we obtained results in good
qualitative agreement with our wind tunnel runs. Likewise, in
non-radiative cosmological simulations of galaxy cluster formation, we
followed individual subhalos as they fell into the forming cluster,
finding again that the gas content of satellite systems declined
slowest in SPH, while the stripping in VPH and especially in {\small
  AREPO} proceeded noticeably faster.

Finally, we considered simulations of the Santa Barbara cluster, which
has become an important test problem for evaluating cosmological
hydrodynamical codes. While the VPH runs revealed a slightly elevated
entropy compared to SPH at the smallest radii, they in general agreed
quite well with SPH and in particular did not provide evidence at the
highest resolution that an entropy core similar to those found in
mesh-codes such as {\small AREPO} is formed. This is perhaps to be
expected if the entropy core indeed primarily arises from mixing
processes \citep{Mitchell2009} that are largely absent in SPH and VPH,
by construction.

Overall, it thus appears that VPH offers some improvements over SPH
without however changing its fundamental character. We argue that the
most important origin of these differences lies in an improved
gradient estimate in VPH compared to SPH. VPH is second-order accurate
in the gradient estimates, i.e.~a linear gradient is always reproduced
exactly, {\em independent} of the particle distribution
\citep{Springel2010}. In contrast, SPH has a so-called zero-th order
error in its gradient estimate \citep[e.g.][]{Read2010}. This in
particular means that even for equal pressures for all particles the
pressure force not necessarily vanishes \citep{Abel2011}, and
furthermore, the absolute size of the gradient error grows with the
pressure itself. The gradient errors in SPH have also been linked to
the creation of small-scale velocity noise in studies of subsonic
turbulence \citep{Bauer2011}.

\label{gresho_in_conclusion}

A good test problem for appreciating this difference in the gradient
accuracy is the flow of a stable vortex as suggested by
\citet{Gresho1990}.  In this `triangular vortex problem' a fluid is
set up with an azimuthal velocity profile (see
Appendix~\ref{gresho_appendix} for details), such that the pressure
gradients counter the centrifugal force, and the vortex evolves in a
time-independent, stable fashion.  Figure~\ref{Gresho_profiles} shows
the radial velocity profile after the vortex has been evolved for a
time $t=1$ with the VPH, SPH and {\small AREPO} codes in 2D, using a
$80\times 80$ Cartesian grid for the initial conditions in the domain
$[-0.5,0.5]^2$.  It can be seen clearly that SPH shows a much larger
velocity scatter than the other two codes, and its solution has
already degraded quite noticeably relative to VPH and {\small
  AREPO}. Especially in the inner domain, where the fluid rotates like
a solid body, SPH deviates systematically from the analytic solution.
We note that part of this degradation can be influenced by the
artificial viscosity setting \citep{Springel2010b,Read2011}, but if a
higher viscosity is used to suppress the velocity noise it typically
also leads to a faster viscous erosion of the vortex. In VPH, some
velocity scatter is seen as well, but it is appreciably smaller than
in SPH, which we interpret as a consequence of the more accurate
gradient estimates in VPH.

Arguably one of the best ways to quantify the accuracy of the
different numerical techniques for this analytic test problem is to
look at an objective error measure as a function of resolution. In
Figure~\ref{GreshoConvRate}, we compare the L1-error norm for the
azimuthal streaming velocity of the Gresho test as a function of
resolution for all three techniques. It is evident that ordinary SPH
converges only very slowly, whereas VPH shows a considerably improved
convergence rate. This directly demonstrates an important improvement
brought about by VPH, which we can directly trace to better gradient
estimates. The latter appears also as the primary reason for the
better results for stripping we obtained with VPH compared to SPH,
which is due to the more accurate treatment of contact discontinuities
and the avoidance of the `gap' phenomenon of SPH.

Note that the difference in convergence rate also means that VPH will
outperform SPH in terms of computational cost if very high accuracy
needs to be achieved, provided its computational cost differs only by
a constant factor of order unity which is indeed the case in our
present implementation. However, the Voronoi mesh construction adds
substantial computational cost compared to ordinary ``standard'' SPH,
making the VPH code about a factor 3-4 slower for pure hydrodynamics
when the same number of particles is used. This is mitigated to some
extent if self-gravity is included (which is typically about as
expensive or slightly more expensive than SPH-based hydrodynamics),
reducing the difference to less than a factor of 2. We note that some
alternative suggestions to improve standard SPH, such as the scheme by
\citet{Read2010} which involves a dramatic increase of the number of
smoothing neighbours, also require an increased computational cost per
resolution element. Which of these schemes is ultimately the most
efficient one (in the sense of reaching a given accuracy goal with the
smallest computational cost) is difficult to answer in general, and is
in any case a problem-dependent and implementation-dependent question.

According to Fig.~\ref{GreshoConvRate}, VPH still falls short of the
better convergence rate obtained with the moving-mesh code {\small
  AREPO} \citep[and similarly with fixed grid mesh codes such as
{\small ATHENA}, see][]{Stone2008,Springel2010}.  The above discussion
suggests that higher order density estimates combined with at least
equally accurate gradient estimates are needed to improve on SPH and
VPH in this respect. Some suggestions in this direction have recently
been made \citep[e.g.][]{Read2011,Maron2011,McNally2011}, and it will
be interesting to see whether they can successfully yield significant
accuracy improvements in cosmological applications such as those
discussed here.

\section*{ACKNOWLEDGEMENTS}

The authors thank the anonymous referee for insightful comments about
the paper. VS acknowledges support by the DFG Research Centre SFB-881
`The Milky Way System' through project A1.

\bibliography{paper}
\bibliographystyle{mn2e.bst}

\appendix

\section{Gresho Vortex Test}

\begin{figure}
\begin{center}
\resizebox{4.2cm}{!}
{\includegraphics{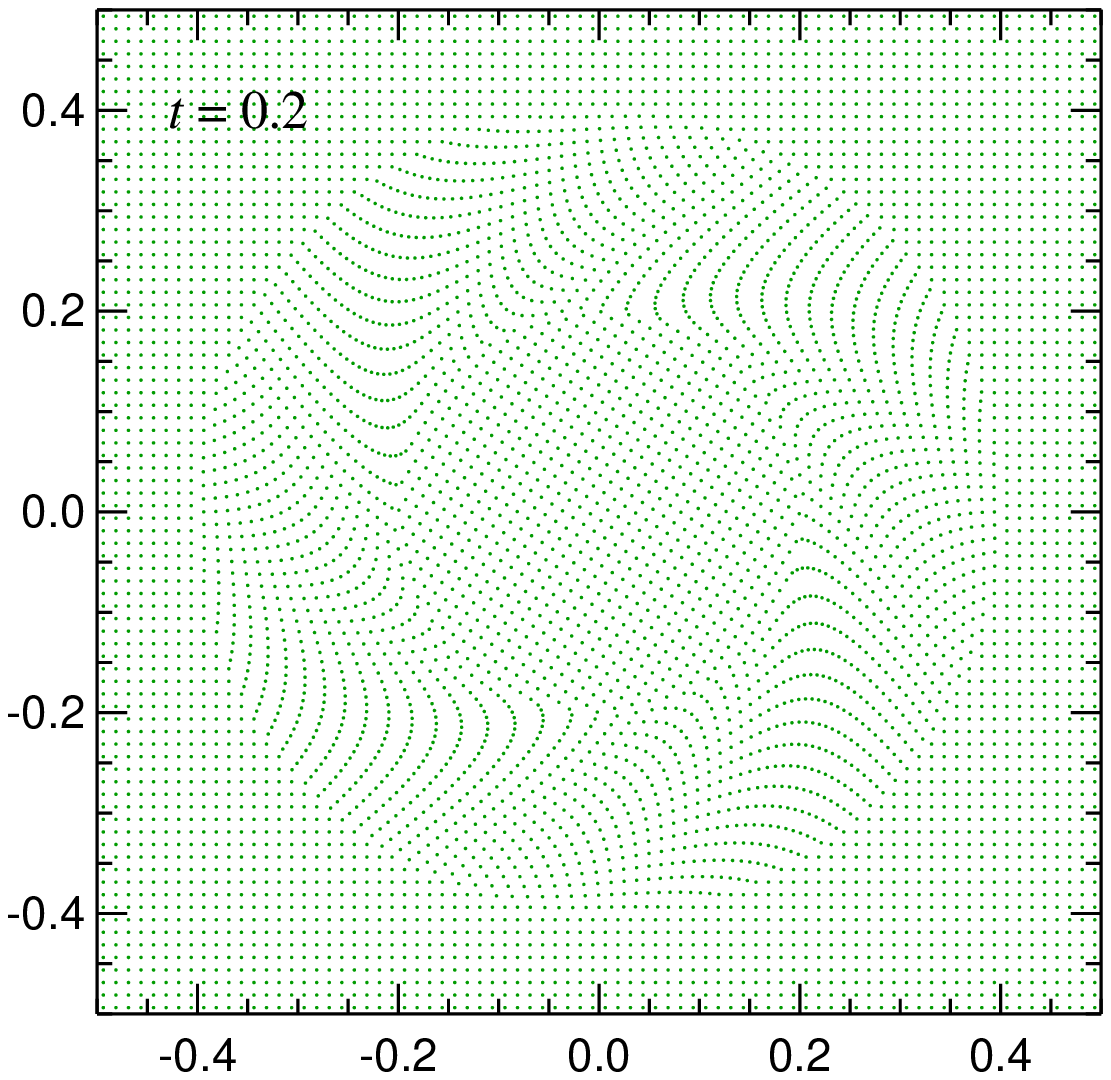}}%
\resizebox{4.2cm}{!}
{\includegraphics{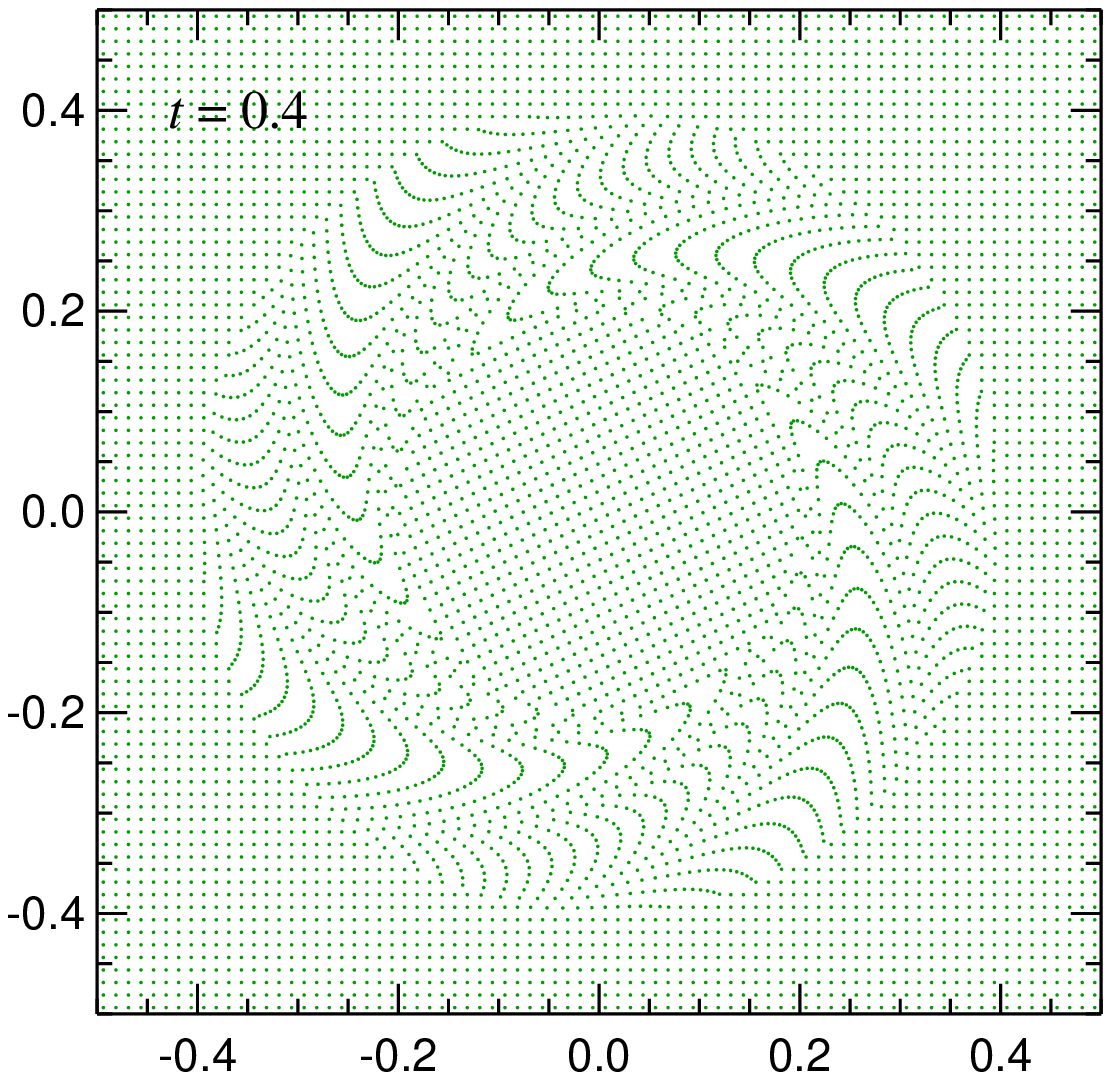}}\\
\resizebox{4.2cm}{!}
{\includegraphics{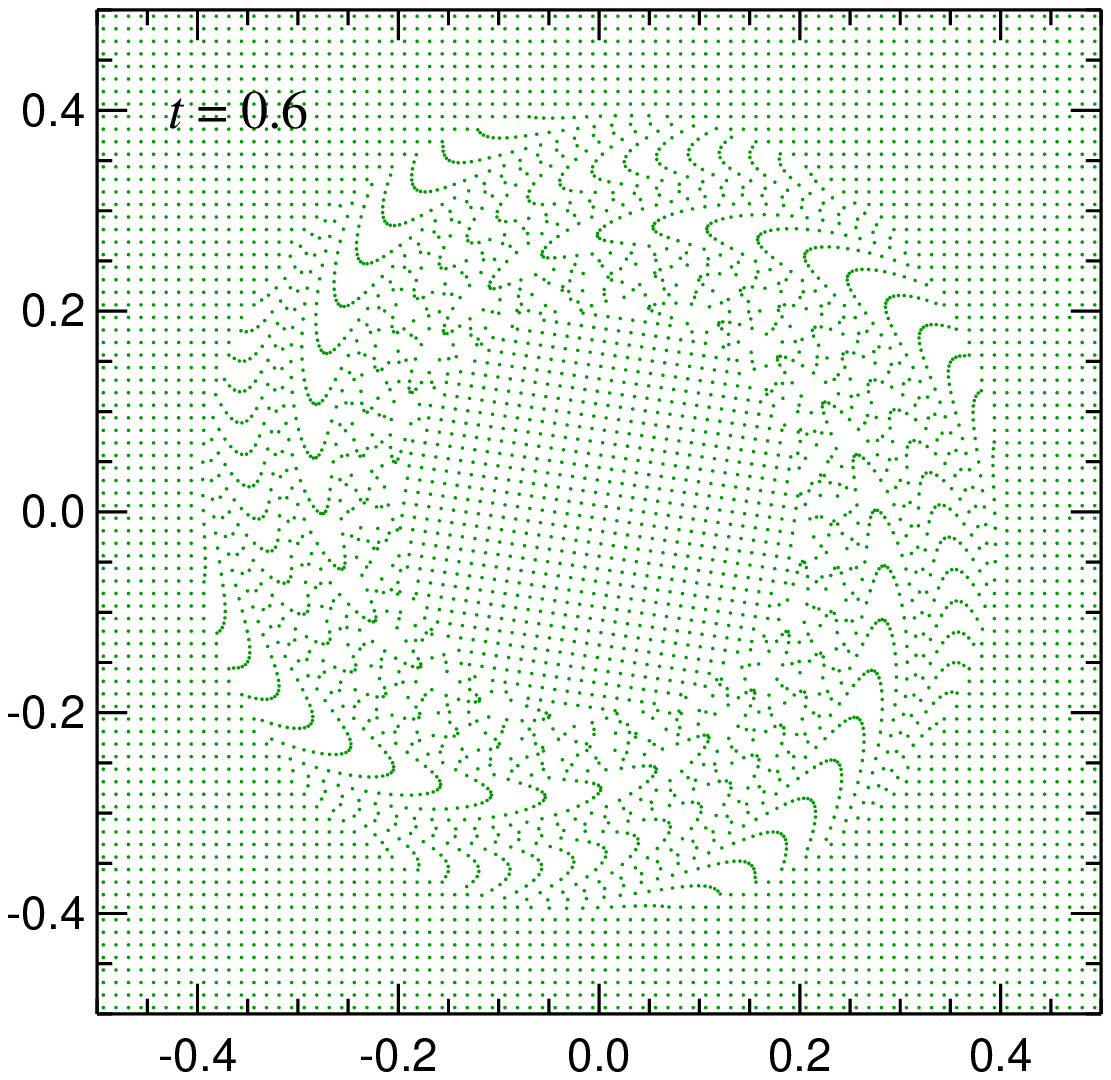}}%
\resizebox{4.2cm}{!}
{\includegraphics{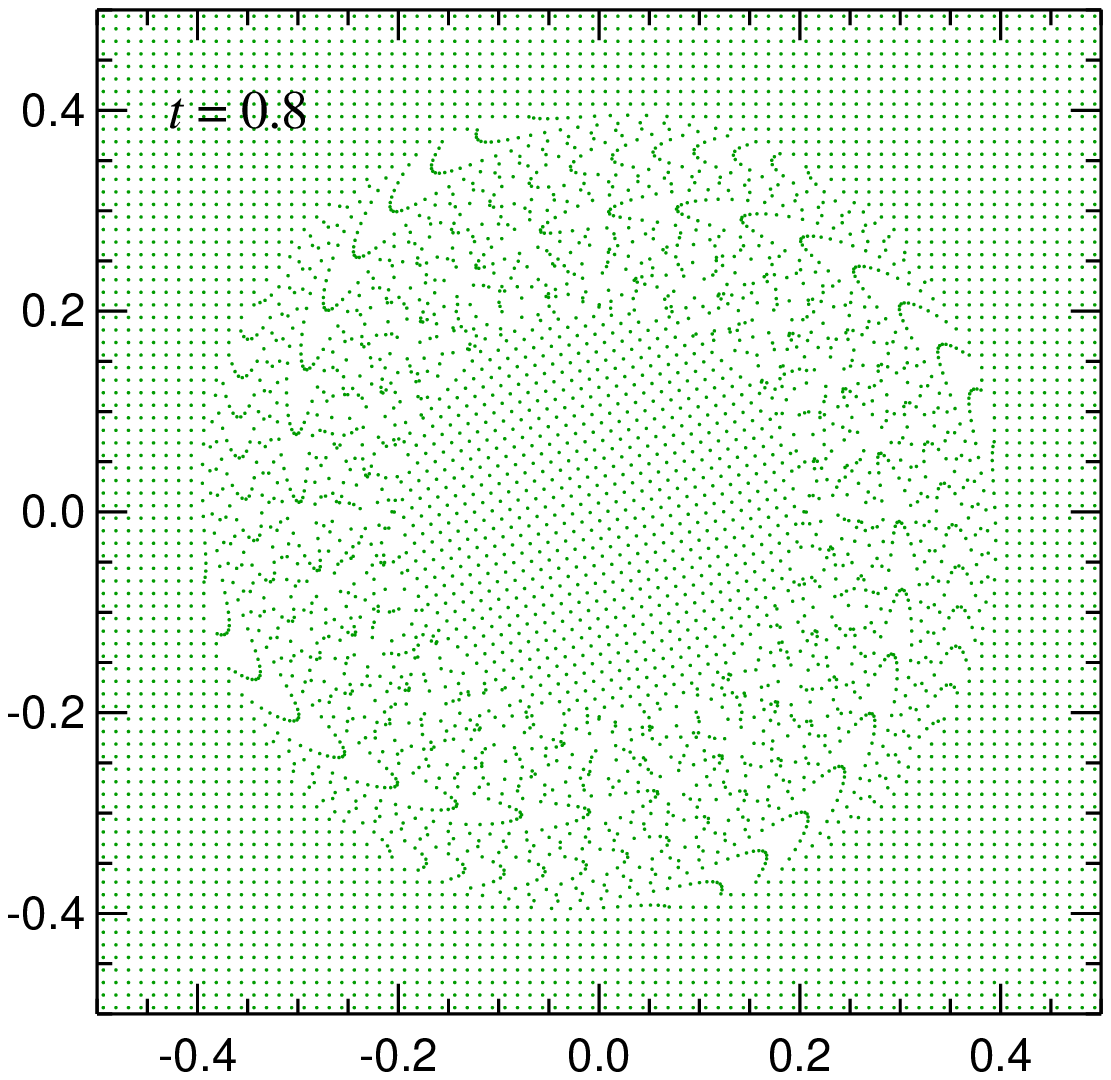}}\\
\resizebox{8.4cm}{!}
{\includegraphics{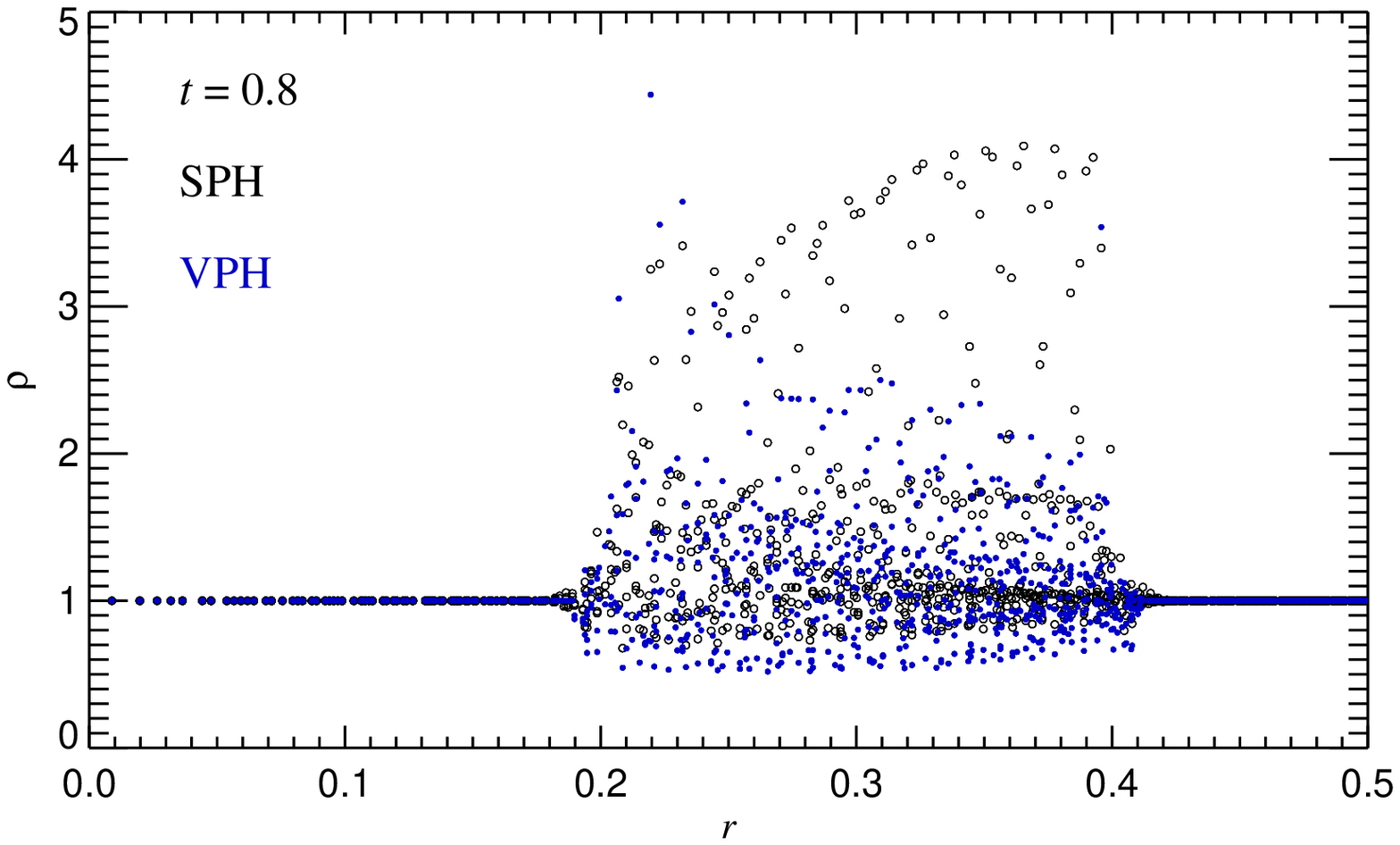}}\\
\caption{Particle distribution and resulting density measurements in
  the case of perfectly circular particle motions. The four panels on
  top illustrate the distortion and shearing of the initial Cartesian
  grid if the points would move on perfectly circular paths with their
  initial azimuthal velocity. Such orbits would be obtained if the
  numerical pressure gradient would always agree with the analytically
  expected gradient. The bottom panel gives the density estimates
  obtained with VPH (blue points) and SPH (black circles) for the
  particle distribution obtained at time $t=0.8$ (middle right
  panel). }
\label{Gresho_grid}
\end{center}
\end{figure}

\label{gresho_appendix}
In the test problem of a stable vortex as suggested by
\citet{Gresho1990} the azimuthal velocity profile has the form
\begin{equation}
v_\phi(r) = \left\{
\begin{array}{ll}
5r & {\rm for}\;\;\; 0\le r <0.2 \\
2-5r & {\rm for}\;\;\; 0.2\le r <0.4 \\
0 & {\rm for}\;\;\; r \ge 0.4 \\
\end{array}
\right.
\end{equation}
in a gas of constant density equal to $\rho=1$ and adiabatic index
of $\gamma=5/3$. The pressure is
chosen as
\begin{equation}
P(r) = \left\{
\begin{array}{ll}
5 + 25/2 r^2 & {\rm for}\;\;\; 0\le r <0.2 \\
9 + 25/2 r^2 - \\\;\;\;\;\;\;20 r + 4 \ln (r/0.2) & {\rm for}\;\;\;
0.2\le r <0.4 \\
3+4\ln 2 & {\rm for}\;\;\; r \ge 0.4 \\
\end{array}
\right.
\end{equation}
so that the pressure gradients balance the centrifugal force.

In Figure~\ref{Gresho_profiles}, we showed the radial velocity profile after
the vortex has been evolved for a time $t=1$ with the VPH, SPH and
{\small AREPO} codes in 2D, using a $80\times 80$ Cartesian grid for
the initial conditions in the domain $[-0.5,0.5]^2$.  
Interestingly, one can also clearly see that the velocity noise in VPH
is still lower in the solid body rotation part of the vortex, for
$r<0.2$. Here the initial pressure gradients are correct in VPH, and
they will in principle stay that way during the motion of the solid
body part. However, the strongly shearing part between $0.2< r <0.4$
will eventually disturb the evolution -- and this can in some sense be
blamed on the density estimate.

To see this, let us imagine for the moment that the fluid particles
would all move under the exact analytic pressure gradient, then they
would move on circular orbits with constant angular velocity. The
resulting time evolution of the point set is shown in the top four
panels of Figure~\ref{Gresho_grid}, together with the resulting
density estimates at time $t=0.8$ for VPH and SPH in the lower
panel. The shearing motion creates substantial irregularities in the
particle distribution, manifesting itself in significant fluctuations
of the density estimates around the desired background density of
$\rho=1$. These fluctuations tend to be roughly of comparable
magnitude for the kernel-based density estimates in SPH and the
Voronoi-based density estimates in VPH. In any case, the density
fluctuations of course cause pressure fluctuation that {\em
  necessarily} prevent that the particles move on exactly circular
orbits. Instead, additional motions are created that locally restore
the pressure equilibrium. Motions that do not agree with the analytic
characteristics are here needed to mitigate pressure fluctuations and
to prevent that density errors as large as implied by the analytic
trajectories occur in the first place.  In contrast to SPH and VPH, in
{\small AREPO}, the density values can stay correct because this code
calculates mass exchanges between cells in such a way that the density
of cells stays (nearly exactly) constant, {\em despite} the strong
shearing of the tessellation. Combined with the accurate gradient
operator in this code (which is the same as in VPH), this avoids
spurious contributions to the pressure gradients, leading to higher
accuracy in the final solution.

A further layer of complexity is added when the ambiguous role of the
shape correction forces in VPH is considered. On one hand they induce
deviations of particle motions from circular orbits in order to make
cells more regular, which can degrade the accuracy with which the
analytic solution is followed.  On the other hand they maintain
regularity of the cell configuration, which improves the robustness of
the scheme and helps to ensure accurate density and gradient
estimates \citep{Hess2010}. It is well possible that an improved cell
regularisation scheme is able to improve VPH's performance for the
Gresho problem, something we leave for future investigations.

\label{lastpage}
\end{document}